\documentclass[12pt, reqno]{amsart}

\usepackage[margin=.9in, a4paper, foot=.3in]{geometry}

\let\oldtocsection=\tocsection

\let\oldtocsubsection=\tocsubsection

\let\oldtocsubsubsection=\tocsubsubsection

\renewcommand{\tocsection}[2]{\hspace{0em}\oldtocsection{#1}{#2}}
\renewcommand{\tocsubsection}[2]{\hspace{1em}\oldtocsubsection{#1}{#2}}
\renewcommand{\tocsubsubsection}[2]{\hspace{2em}\oldtocsubsubsection{#1}{#2}}

\usepackage{cite}

\usepackage{graphicx, color}

\definecolor{darkblue}{rgb}{0,0,.5}
\definecolor{darkred}{rgb}{.5,0,0}
\definecolor{darkgreen}{rgb}{0,0.5,0}

\usepackage{hyperref}
\hypersetup {
    pdfstartview=FitH,
    colorlinks,
    citecolor=darkgreen,
    linkcolor=darkred,
    urlcolor=darkblue
}

\usepackage[noBBpl,slantedGreek]{mathpazo}
\usepackage[mathcal]{euscript}

\usepackage{amssymb}

\makeatletter
\def\section{\@startsection{section}{1}%
  \z@{-.7\linespacing\@plus -\linespacing}{.5\linespacing}%
  {\normalfont\scshape\centering}}
\def\subsection{\@startsection{subsection}{2}%
  \z@{-.5\linespacing\@plus -.7\linespacing}{.5em}%
  {\normalfont\bfseries\mathversion{bold}}}
\makeatother

\allowdisplaybreaks

\numberwithin{equation}{section}

\newcommand {\id}{\mathrm{id}}
\newcommand {\Osc}{\mathrm{Osc}}
\newcommand {\rme}{\mathrm e}

\newcommand {\bbC}{\mathbb C}
\newcommand {\bbD}{\mathbb D}
\newcommand {\bbL}{\mathbb L}
\newcommand {\bbM}{\mathbb M}
\newcommand {\bbN}{\mathbb N}
\newcommand {\bbO}{\mathbb O}
\newcommand {\bbP}{\mathbb P}
\newcommand {\bbQ}{\mathbb Q}
\newcommand {\bbT}{\mathbb T}
\newcommand {\bbX}{\mathbb X}
\newcommand {\bbY}{\mathbb Y}
\newcommand {\bbZ}{\mathbb Z}

\newcommand {\calB}{\mathcal B}
\newcommand {\calC}{\mathcal C}
\newcommand {\calD}{\mathcal D}
\newcommand {\calE}{\mathcal E}
\newcommand {\calH}{\mathcal H}
\newcommand {\calK}{\mathcal K}
\newcommand {\calL}{\mathcal L}
\newcommand {\calM}{\mathcal M}

\newcommand {\calP}{\mathcal P}
\newcommand {\calQ}{\mathcal Q}
\newcommand {\calR}{\mathcal R}
\newcommand {\calT}{\mathcal T}
\newcommand {\calX}{\mathcal X}
\newcommand {\calY}{\mathcal Y}

\newcommand {\mbar}[3]{\hskip #2 \overline{\hskip -#2 #1 \hskip -#3} \hskip #3}

\newcommand {\oM}{\mbar{M}{.2em}{.1em}}
\newcommand {\oQ}{\mbar{Q}{.07em}{.07em}}
\newcommand {\oT}{\mbar{T}{.07em}{.03em}}
\newcommand {\oV}{\mbar{V}{.07em}{.07em}}
\newcommand {\oW}{\mbar{W}{.03em}{.03em}}
\newcommand {\orho}{\mbar{\rho}{.03em}{.03em}}
\newcommand {\ovarphi}{\mbar{\varphi}{.07em}{.07em}}

\newcommand {\obbL}{\mbar{\mathbb L}{.0em}{.1em}}
\newcommand {\obbM}{\mbar{\mathbb M}{.05em}{.1em}}
\newcommand {\obbT}{\mbar{\mathbb T}{.05em}{.05em}}
\newcommand {\obbQ}{\mbar{\mathbb Q}{.05em}{.05em}}

\newcommand {\ocalL}{\mbar{\mathcal L}{.07em}{.07em}}
\newcommand {\ocalM}{\mbar{\mathcal M}{.07em}{.07em}}
\newcommand {\ocalQ}{\mbar{\mathcal Q}{.07em}{.07em}}
\newcommand {\ocalT}{\mbar{\mathcal T}{.03em}{.03em}}

\newcommand {\gothh}{\mathfrak h}
\newcommand {\gothg}{\mathfrak g}
\newcommand {\gothgl}{\mathfrak{gl}}
\newcommand {\gothsl}{\mathfrak{sl}}

\newcommand {\tlsliii}{\widetilde{\mathcal L}(\mathfrak{sl}_3)}
\newcommand {\uqbp}{\mathrm U_q(\mathfrak b_+)}
\newcommand {\uqbm}{\mathrm U_q(\mathfrak b_-)}
\newcommand {\uqnp}{\mathrm U_q(\mathfrak n_+)}
\newcommand {\uqnm}{\mathrm U_q(\mathfrak n_-)}
\newcommand {\uqgliii}{\mathrm U_q(\mathfrak{gl}_3)}

\newcommand {\uqlslii}{\mathrm U_q(\mathcal L(\mathfrak{sl}_2))}
\newcommand {\uqlsliii}{\mathrm U_q(\mathcal L(\mathfrak{sl}_3))}
\newcommand {\uqtlsliii}{\mathrm U_q(\widetilde{\mathcal L}(\mathfrak{sl}_3))}

\DeclareMathOperator {\End}{End}
\DeclareMathOperator {\im}{im}
\DeclareMathOperator {\sgn}{sgn}
\DeclareMathOperator {\tr}{tr}

\title{Quantum groups and functional relations for higher rank}

\author[H. Boos]{Hermann Boos}
\address{Fachbereich C -- Physik, Bergische Universit\"at Wuppertal, 42097 Wuppertal, Germany}
\email{hboos@uni-wuppertal.de}

\author[F. G\"ohmann]{Frank G\"ohmann}
\address{Fachbereich C -- Physik, Bergische Universit\"at Wuppertal, 42097 Wuppertal, Germany}
\email{goehmann@uni-wuppertal.de}

\author[A. Kl\"umper]{Andreas Kl\"umper}
\address{Fachbereich C -- Physik, Bergische Universit\"at Wuppertal, 42097 Wuppertal, Germany}
\email{kluemper@uni-wuppertal.de}

\author[Kh. S. Nirov]{\vskip .2em Khazret S. Nirov}
\address{Institute for Nuclear Research of the Russian Academy of Sciences, 60th October Ave 7a, 117312 Moscow, Russia}
\curraddr{Fachbereich C -- Physik, Bergische Universit\"at Wuppertal, 42097 Wuppertal, Germany}
\email{nirov@uni-wuppertal.de}

\author[A. V. Razumov]{Alexander V. Razumov}
\address{Institute for High Energy Physics, 142281 Protvino, Moscow region, Russia}
\email{Alexander.Razumov@ihep.ru}

\begin{document}

\addtolength {\jot}{3pt}

\begin{abstract}
A detailed construction of the universal integrability objects related to the integrable systems associated with the quantum group $\uqlsliii$ is given. The full proof of the functional relations in the form independent of the representation of the quantum group on the quantum space is presented. The case of the general gradation and general twisting is treated. The specialization of the universal functional relations to the case when the quantum space is the state space of a discrete spin chain is described.
\end{abstract}

\maketitle

\tableofcontents

\section{Introduction}

The method of functional relations was proposed by Baxter to solve statistical models which can or cannot be treated with the help of the Bethe ansatz, see, for example, \cite{Bax82, Bax04}. It appears that its main ingredients, transfer matrices and $Q$-operators, are essential not only for the integration of the corresponding quantum statistical models in the sense of calculating the partition function in the thermodynamic limit. One of the remarkable recent applications is their usage in the construction of the fermionic basis \cite{BooJimMiwSmiTak07, BooJimMiwSmiTak09, JimMiwSmi09, BooJimMiwSmi10} for the observables of the XXZ spin chain.

It seems that the most productive, although not comprehensive, modern approach to the theory of quantum integrable systems is the approach based on the concept of quantum group invented by Drinfeld and Jimbo \cite{Dri87, Jim85}. In accordance with this approach, the transfer matrices and $Q$-operators are constructed by choosing first the representations for the factors of the tensor product of two copies of the underlying quantum group. Then these representations are applied to the universal $R$-matrix, and finally the trace over one of the representation spaces is taken. Here the functional relations are consequences of the properties of the used representations of the quantum group. For the first time, it was conceived by Bazhanov, Lukyanov and Zamolodchikov \cite{BazLukZam96, BazLukZam97, BazLukZam99}.

Following the physical tradition, we call the representation space corresponding to the first factor of the tensor product the auxiliary space, and the representation space corresponding to the second factor the quantum space. The representation on the auxiliary space defines the integrability object, while the representation on the quantum space defines the concrete physical integrable system. It appears convenient for our purposes to fix the representation for the first factor only, see, for example, \cite{AntFei97, RosWes02, BazTsu08, BooGoeKluNirRaz14a, BooGoeKluNirRaz13}. We use for the objects obtained in such a way the term ``universal''. The relations obtained for them can be used for any physical model related to the quantum group under consideration.

In the papers \cite{BazLukZam96, BazLukZam97, BazLukZam99} the case of the quantum group $\uqlslii$ was considered and as the quantum space the state space of a conformally invariant two dimensional field theory was taken. We reconsidered the case of this quantum group in the papers \cite{BooGoeKluNirRaz14a, BooGoeKluNirRaz13} where we obtained the functional relations in the universal form and then chose as the quantum space the state space of the XXZ-spin chain.

A two dimensional field theory with extended conformal symmetry related to the quantum group $\uqlsliii$ was analysed in the paper \cite{BazHibKho02}. Here as the quantum space the corresponding state space of the quantum continuous field theory under consideration  was taken again.  In the present paper we define for the case of the quantum group $\uqlsliii$ universal integrability objects and derive the corresponding functional relations. Here to define the integrability objects we use the general gradation and general twisting. However, the main difference from \cite{BazHibKho02} is that we give a new and detailed proof of
the functional relations. In the paper \cite{BazHibKho02} proofs are often skipped, or given in schematic form. This was one of the reasons for writing our paper. Another reason was the desire to have the specialization of the universal functional relations to the case when the quantum space is the state space of a discrete spin chain.

Below we use the notation
\begin{equation*}
\kappa_q = q - q^{-1}
\end{equation*}
so the definition of the $q$-deformed number can be written as 
\begin{equation*}
[\nu]_q = \frac{q^\nu - q^{- \nu}}{q - q^{-1}} = \kappa_q^{-1}(q^\nu - q^{-\nu}), \qquad \nu \in \bbC.
\end{equation*}
We denote by $\calL(\gothg)$ the loop Lie algebra of a finite dimensional simple Lie algebra $\gothg$ and by $\widetilde \calL(\gothg)$ its standard central extension, see, for example, the book by Kac \cite{Kac90}. The symbol~$\bbN$ means the set of natural numbers and the symbol $\bbZ_+$ the set of non-negative integers.

To construct integrability objects one uses spectral parameters. They are introduced by defining a $\bbZ$-gradation of the quantum group. In the case under consideration a $\bbZ$-gradation of $\uqlsliii$ is determined by three integers $s_0$, $s_1$ and $s_2$. We use the notation $s = s_0 + s_1 + s_2$ and denote by $r_s$ some fixed $s$th root of $-1$, so that $(r_s)^s = -1$.

\section{Integrability objects}

In this paper we consider integrable systems related to the quantum group $\uqlsliii$. Depending on the sense of the ``deformation parameter'' $q$, there are at least three definitions of a quantum group. According to the first definition, $q = \exp \hbar$, where $\hbar$ is an indeterminate, according to the second one, $q$ is indeterminate, and according to the third one, $q = \exp \hbar$, where $\hbar$ is a complex number. In the first case a quantum group is a $\bbC[[\hbar]]$-algebra, in the second case a $\bbC(q)$-algebra, and in the third case it is just a complex algebra. We define the quantum group as a $\bbC$-algebra, see, for example, the books \cite{JimMiw95, EtiFreKir98}.

To construct representations of the quantum group $\uqlsliii$ we use the Jimbo's homomorphism from $\uqlsliii$ to the quantum group $\uqgliii$. Therefore, we first remind the definition of $\uqgliii$ and then discuss $\uqlsliii$. 

\subsection{\texorpdfstring{Quantum group $\uqgliii$}{Quantum group Uq(gl3)}} \label{ss:qguqslii}

Denote by $\gothg$ the standard Cartan subalgebra of the Lie algebra $\gothgl_3$ and by $G_i = E_{ii}$, $i = 1, 2, 3$, the elements forming the standard basis of $\gothg$.\footnote{We use the usual notation $E_{ij}$ for the matrix units.} The root system of $\gothgl_3$ relative to $\gothg$ is generated by the simple roots $\alpha_i \in \gothg^*$, $i = 1, 2$, given by the relations
\begin{equation}
\alpha_j(G_i) = c_{ij}, \label{alphah}
\end{equation}
where
\begin{equation*}
(c_{ij}) = \left( \begin{array}{rr}
1 & 0 \\
-1 & 1 \\
0 & -1
\end{array} \right).
\end{equation*}
The Lie algebra $\gothsl_3$ is a subalgebra of $\gothgl_3$, and the standard Cartan subalgebra $\gothh$ of $\gothsl_3$ is a subalgebra of $\gothg$. Here the standard Cartan generators $H_i$, $i = 1, 2$, of $\gothsl_3$ are
\begin{equation}
H_1 = G_1 - G_2, \qquad H_2 = G_2 - G_3, \label{hk}
\end{equation}
and we have
\begin{equation*}
\alpha_j(H_i) = a_{ij},
\end{equation*}
where
\begin{equation}
(a_{ij}) = \left( \begin{array}{rr}
2 & -1 \\
-1 & 2
\end{array} \right) \label{cm}
\end{equation}
is the Cartan matrix of $\gothsl_3$.

Let $\hbar$ be a complex number and $q = \exp \hbar$. We define the quantum group $\uqgliii$ as a unital associative $\bbC$-algebra generated by the elements $E_i$, $F_i$, $i = 1, 2$, and $q^X$, $X \in \gothg$, with the relations\footnote{It is necessary to assume that $q^2 \ne 1$. In fact, we assume that $q$ is not any root of unity.}
\begin{gather}
q^0 = 1, \qquad q^{X_1} q^{X_2} = q^{X_1 + X_2}, \label{xx} \\
q^X E_i \, q^{-X} = q^{\alpha_i(X)} E_i, \qquad q^X F_i \, q^{-X} = q^{- \alpha_i(X)} F_i, \label{xexf} \\
[E_i, F_j] = \kappa_q^{-1} \delta_{ij} \, (q^{H_i} - q^{-H_i}) \label{ef}
\end{gather}
satisfied for any $i$ and $j$, and the Serre relations
\begin{equation}
E_i^2 E_j^{} - [2]_q E_i^{} E_j^{} E_i^{} + E_j^{} E_i^2 = 0, \qquad
F_i^2 F_j^{} - [2]_q F_i^{} F_j^{} F_i^{} + F_j^{} F_i^2 = 0 \label{sr}
\end{equation}
satisfied for any distinct $i$ and $j$. Note that $q^X$ is just a convenient notation. There are no elements of $\uqgliii$ corresponding to the elements of $\gothg$. In fact, this notation means a set of elements of $\uqgliii$ parametrized by $\gothg$. It is convenient to use the notations
\begin{equation*}
q^{X + \nu} = q^\nu q^X
\end{equation*}
and
\begin{equation}
[X + \nu]_q = \kappa_q^{-1} \, (q^{X + \nu}- q^{ -X -\nu}) = \kappa_q^{-1} \, (q^{\nu} q^X - q^{-\nu} q^{-X}) \label{xnq}
\end{equation}
for any $X \in \gothg$ and $\nu \in \bbC$. Here equation (\ref{ef}) takes the form
\begin{equation*}
[E_i, F_j] = \delta_{ij} [H_i]_q.
\end{equation*}
Similar notations are used below for the case of the quantum groups $\uqtlsliii$ and $\uqlsliii$.

With respect to the properly defined coproduct, counit and antipode the quantum group $\uqgliii$ is a Hopf algebra.

Looking at (\ref{xexf}) one can say that the generators $E_i$ and $F_i$ are related to the roots $\alpha_i$ and $-\alpha_i$ respectively. Define the elements related to the roots $\alpha_1 + \alpha_2$ and $-(\alpha_1 + \alpha_2)$ as
\begin{equation}
E_3 = E_1 E_2 - q^{-1} E_2 E_1, \qquad F_3 = F_2 F_1 - q \, F_1 F_2. \label{e3f3}
\end{equation}
The Serre relations (\ref{sr}) give
\begin{align}
& E_3 E_1 = q^{-1} E_1 E_3, && E_3 E_2 = q \, E_2 E_3, \label{e3e} \\
& F_3 F_1 = q^{-1} F_1 F_3, && F_3 F_2 = q \, F_2 F_3. \label{f3f}
\end{align}
One can also verify that
\begin{equation*}
[E_3, F_3] = [H_1 + H_2]_q,
\end{equation*}
and, besides,
\begin{align}
& [E_1, F_3] = - q \, F_2 \, q^{H_1}, && [E_2, F_3] = F_1 \, q^{- H_2}, \label{f3e} \\
& [E_3, F_1] = - E_2 \, q^{- H_1}, && [E_3, F_2] = q^{-1} E_1 \, q^{H_2}. \label{e3f}
\end{align}

Using the above relations, one can find explicit expressions for the action of the generators of $\uqgliii$ on Verma $\uqgliii$-modules, see appendix \ref{a:vmr}.

\subsection{\texorpdfstring{Quantum group $\uqlsliii$}{Quantum group Uq(L(sl3))}}

\subsubsection{Definition}

We start with the quantum group $\uqtlsliii$. Recall that the Cartan subalgebra of $\tlsliii$ is
\begin{equation*}
\widetilde \gothh = \gothh \oplus \bbC c,
\end{equation*}
where $\gothh = \bbC H_1 \oplus \bbC H_2$ is the standard Cartan subalgebra of $\gothsl_3$ and $c$ is the central element \cite{Kac90}. Define the Cartan elements
\begin{equation*}
h_0 = c - H_1 - H_2, \qquad h_1 = H_1, \qquad h_2 = H_2,
\end{equation*}
so that one has
\begin{equation}
c = h_0 + h_1 + h_2 \label{c}
\end{equation}
and
\begin{equation*}
\widetilde \gothh = \bbC h_0 \oplus \bbC h_1 \oplus \bbC h_2.
\end{equation*}
The simple roots $\alpha_i \in \widetilde{\gothh}^*$, $i = 0$, $1$, $2$, are given by the equation
\begin{equation*}
\alpha_j(h_i) = \tilde a_{ij},
\end{equation*}
where
\begin{equation*}
(\tilde a_{ij}) = \left(\begin{array}{rrr}
2 & -1 & -1 \\
-1 & 2 & -1 \\
-1 & -1 & 2
\end{array} \right)
\end{equation*}
is the Cartan matrix of the Lie algebra $\tlsliii$.

As before, let $\hbar$ be a complex number and $q = \exp \hbar$. The quantum group $\uqtlsliii$  is a unital associative $\bbC$-algebra generated by the elements $e_i$, $f_i$, $i = 0, 1, 2$, and $q^x$, $x
\in \widetilde \gothh$, with the relations
\begin{gather}
q^0 = 1, \qquad q^{x_1} q^{x_2} = q^{x_1 + x_2}, \label{lxx} \\
q^x e_i q^{-x} = q^{\alpha_i(x)} e_i, \qquad q^x f_i q^{-x} = q^{-\alpha_i(x)} f_i, \\
[e_i, f_j] = \delta_{ij} [h_i]_q
\end{gather}
satisfied for all $i$ and $j$, and the Serre relations
\begin{equation}
e_i^2 e_j^{\mathstrut} - [2]_q  e_i^{\mathstrut} e_j^{\mathstrut} e_i^{\mathstrut}
+ e_j^{\mathstrut} e_i^2 = 0, \qquad f_i^2 f_j^{\mathstrut} - [2]_q  f_i^{\mathstrut} f_j^{\mathstrut} f_i^{\mathstrut} + f_j^{\mathstrut} f_i^2 = 0 \label{lsr}
\end{equation}
satisfied for all distinct $i$ and $j$.

The quantum group $\uqtlsliii$ is a Hopf algebra with the comultiplication $\Delta$, the antipode $S$, and the counit $\varepsilon$ defined by the
relations
\begin{gather}
\Delta(q^x) = q^x \otimes q^x, \qquad \Delta(e_i) = e_i \otimes 1 + q^{- h_i} \otimes e_i, \qquad \Delta(f_i) = f_i \otimes q^{h_i} + 1 \otimes f_i, \label{cmul} \\
S(q^x) = q^{- x}, \qquad S(e_i) = - q^{h_i} e_i, \qquad S(f_i) = - f_i \, q^{- h_i}, \\
\varepsilon(q^x) = 1, \qquad \varepsilon(e_i) = 0, \qquad \varepsilon(f_i) = 0. \label{cu}
\end{gather}

The quantum group $\uqlsliii$ can be defined as the quotient algebra of $\uqtlsliii$ by the two-sided ideal generated by the elements of the form $q^{\nu c} - 1$, $\nu \in \bbC^\times$. In terms of generators and relations the quantum group $\uqlsliii$ is a $\bbC$-algebra generated by the elements $e_i$, $f_i$, $i = 0, 1, 2$, and $q^x$, $x \in \widetilde{\gothh}$, with relations (\ref{lxx})--(\ref{lsr}) and an additional relation
\begin{equation}
q^{\nu c} = 1, \label{qh0h1}
\end{equation}
where $\nu \in \bbC^\times$. It is a Hopf algebra with the Hopf structure defined by (\ref{cmul})--(\ref{cu}). One of the reasons to use the quantum group $\uqlsliii$ instead of $\uqtlsliii$ is that in the case of $\uqtlsliii$ we have no expression for the universal $R$-matrix.

\subsubsection{Universal $R$-matrix}

As any Hopf algebra the quantum group $\uqlsliii$ has another comultiplication called the opposite comultiplication. It can be defined explicitly by the equations
\begin{gather}
\Delta^{\mathrm{op}}(q^x) = q^x \otimes q^x, \label{doqx} \\
\Delta^{\mathrm{op}}(e_i) = e_i \otimes q^{- h_i} + 1 \otimes e_i, \qquad \Delta^{\mathrm{op}}(f_i) = f_i \otimes 1 + q^{h_i} \otimes f_i. \label{doefi}
\end{gather}
When the quantum group $\uqlsliii$ is defined as a $\bbC[[\hbar]]$-algebra it is a quasitriangular Hopf algebra. It means that there exists an element $\calR \in \uqlsliii \otimes \uqlsliii$, called the universal $R$-matrix, such that
\begin{equation*}
\Delta^{\mathrm{op}}(a) = \calR \, \Delta(a) \, \calR^{-1}
\end{equation*}
for all $a \in \uqlsliii$, and\footnote{For the explanation of the notation see, for example, the book \cite{ChaPre94} or the papers \cite{BooGoeKluNirRaz10, BooGoeKluNirRaz14a}.}
\begin{equation}
(\Delta \otimes \id) (\calR) = \calR^{13} \calR^{23}, \qquad (\id \otimes \Delta) (\calR) = \calR^{13} \calR^{12}. \label{urm}
\end{equation}
The expression for the universal $R$-matrix of $\uqlsliii$ considered as a $\bbC[[\hbar]]$-algebra can be constructed using the procedure proposed by Khoroshkin and Tolstoy \cite{TolKho92}. Note that here the universal $R$-matrix is an element of $\uqbp \otimes \uqbm$, where $\uqbp$ is the Borel subalgebra of $\uqlsliii$ generated by $e_i$, $i = 0, 1, 2$, and $q^x$, $x \in \widetilde{\gothh}$, and $\uqbm$ is the Borel subalgebra of $\uqlsliii$ generated by $f_i$, $i = 0, 1, 2$, and $q^x$, $x \in \widetilde{\gothh}$.

In fact, one can use the expression for the universal $R$-matrix from the paper \cite{TolKho92} also for the case of the quantum group $\uqlsliii$ defined as a $\bbC$-algebra having in mind that in this case the quantum group is quasitriangular only in some restricted sense. Namely, all the relations involving the universal $R$-matrix should be considered as valid only for the weight representations of $\uqlsliii$, see in this respect the paper \cite{Tan92} and the discussion below.

Recall that a representation $\varphi$ of $\uqlsliii$ on the vector space $V$ is a weight representation if
\begin{equation*}
V = \bigoplus_{\lambda \in \widetilde \gothh^*} V_\lambda,
\end{equation*}
where
\begin{equation*}
V_\lambda = \{v \in V \mid q^x v = q^{\lambda(x)} v \mbox{ for any } x \in \widetilde \gothh \}.
\end{equation*}
Taking into account relations (\ref{qh0h1}) and (\ref{c}), we conclude that $V_\lambda \ne \{0\}$ only if
\begin{equation}
\lambda(h_0 + h_1 + h_2) = 0. \label{lambdah0h1h2}
\end{equation}
Let $\varphi_1$ and $\varphi_2$ be weight representations of $\uqlsliii$ on the vector spaces $V_1$ and $V_2$ with the weight decompositions
\begin{equation*}
V_1 = \bigoplus_{\lambda \in \widetilde \gothh^*} (V_1)_\lambda, \qquad V_2 = \bigoplus_{\lambda \in \widetilde \gothh^*} (V_2)_\lambda.
\end{equation*}
In the tensor product $V_1 \otimes V_2$ the role of the universal $R$-matrix is played by the operator
\begin{equation}
\calR_{\varphi_1, \varphi_2} = (\varphi_1 \otimes \varphi_2)(\calB) \, \calK_{\varphi_1, \varphi_2}. \label{rpipi}
\end{equation}
Here $\calB$ is an element of $\uqnp \otimes \uqnm$, where $\uqnp$ and $\uqnm$ are the subalgebras of $\uqlsliii$ generated by $e_i$, $i = 0, 1, 2$, and  $f_i$, $i = 0, 1, 2$, respectively. The operator $\calK_{\varphi_1, \varphi_2}$ acts on a vector $v \in (V_1)_{\lambda_1} \otimes (V_2)_{\lambda_2}$ in accordance with the equation
\begin{equation}
\calK_{\varphi_1, \varphi_2} \, v = q^{\sum_{i, j = 1}^2 b_{i j} \lambda_1(h_i) \lambda_2(h_j)} \, v, \label{kpipii}
\end{equation}
where
\begin{equation*}
(b_{ij}) = \frac{1}{3} \left( \begin{array}{cc}
2 & 1 \\
1 & 2
\end{array} \right)
\end{equation*}
is the inverse matrix of the Cartan matrix (\ref{cm}) of the Lie algebra $\gothsl_3$. It follows from (\ref{lambdah0h1h2}) that (\ref{kpipii}) can be written in a simpler form
\begin{equation}
\calK_{\varphi_1, \varphi_2} \, v = q^{\sum_{i = 0}^2 \lambda_1(h_i) \lambda_2(h_i) / 3} \, v. \label{kpipi}
\end{equation}

\subsection{\texorpdfstring{$R$-operators}{R-operators}}

To obtain $R$-operators, or, as they also called, $R$-matrices, one uses for both factors of the tensor product of the two copies of the quantum group one and the same finite dimensional representation. We do not use in this paper the explicit form of the $R$-operators for the quantum group $\uqlsliii$. The corresponding calculations for this case and for some other quantum groups can be found in the papers \cite{KhoTol92, LevSoiStu93, ZhaGou94, BraGouZhaDel94, BraGouZha95, BooGoeKluNirRaz10, BooGoeKluNirRaz11}.

\subsection{Universal monodromy and universal transfer operators}

\subsubsection{General remarks} \label{ss:gr}

To construct universal monodromy and transfer operators we endow $\uqlsliii$ with a $\bbZ$-gradation, see, for example, \cite{BooGoeKluNirRaz14a, BooGoeKluNirRaz13}. The usual way to do it is as follows.

Given $\zeta \in \bbC^\times$, we define an automorphism $\Gamma_\zeta$ of $\uqlsliii$ by its action on the generators of $\uqlsliii$ as
\begin{equation*}
\Gamma_\zeta(q^x) = q^x, \qquad \Gamma_\zeta(e_i) = \zeta^{s_i} e_i, \qquad \Gamma_\zeta(f_i) = \zeta^{-s_i} f_i,
\end{equation*}
where $s_i$ are arbitrary integers. The family of automorphisms $\Gamma_\zeta$, $\zeta \in \bbC^\times$, generates the $\bbZ$-gradation with the grading subspaces
\begin{equation*}
\uqlsliii_m = \{ a \in \uqlsliii \mid \Gamma_\zeta(a) = \zeta^m a \}.
\end{equation*}
Taking into account (\ref{cmul}) we see that
\begin{equation}
(\Gamma_\zeta \otimes \Gamma_\zeta) \circ \Delta = \Delta \circ \Gamma_\zeta. \label{ggd}
\end{equation}
It also follows from the explicit form of the universal $R$-matrix obtained with the help of the Tolstoy--Khoroshkin construction, see \cite{TolKho92, BooGoeKluNirRaz10}, that for any $\zeta \in \bbC^\times$ we have
\begin{equation}
(\Gamma_\zeta \otimes \Gamma_\zeta)(\calR) = \calR.  \label{ggr}
\end{equation}
Following the physical tradition, we call $\zeta$ the spectral parameter.

If $\pi$ is a representation of $\uqlsliii$, then  for any $\zeta \in \bbC^\times$ the mapping $\pi \circ \Gamma_\zeta$ is also a representation of $\uqlsliii$ . Below, for any homomorphism $\varphi$ from $\uqlsliii$ to some algebra we use the notation
\begin{equation}
\varphi_\zeta = \varphi \circ \Gamma_\zeta. \label{vpz}
\end{equation}
If $V$ is a $\uqlsliii$-module corresponding to the representation $\pi$, we denote by $V_\zeta$ the $\uqlsliii$-module corresponding to the representation $\pi_\zeta$. Certainly, as vector spaces $V$ and $V_\zeta$ coincide.

Now let $\pi$ be a representation of the quantum group $\uqlsliii$ on a vector space $V$. The universal monodromy operator $\calM_\pi(\zeta)$ corresponding to the representation $\pi$ is defined by the relation
\begin{equation*}
\calM_\pi(\zeta) = (\pi_\zeta \otimes \id)(\calR).
\end{equation*}
It is clear that $\calM_\pi(\zeta)$ is an element of $\End(V) \otimes \uqlsliii$.

Universal monodromy operators are auxiliary objects needed for construction of universal transfer operators. The universal transfer operator $\calT_\pi(\zeta)$ corresponding to the universal monodromy operator $\calM_\pi(\zeta)$ is defined as
\begin{equation*}
\calT_\pi(\zeta) = (\tr \otimes \id)(\calM_\pi(\zeta) (\pi_\zeta(t) \otimes 1)) = ((\tr \circ \pi_\zeta) \otimes \id)(\calR (t \otimes 1)),
\end{equation*}
where $t$ is a group-like element of $\uqlsliii$ called a twist element. Note that $\calT_\pi(\zeta)$ is an element of $\uqlsliii$. An important property of the universal transfer operators $\calT_\pi(\zeta)$ is that they commute for all representations $\pi$ and all values of $\zeta$. They also commute with all generators $q^x$, $x \in \widetilde \gothh$, see, for example, our papers \cite{BooGoeKluNirRaz14a, BooGoeKluNirRaz13}. 

As we noted above, to construct representations of $\uqlsliii$ we are going to use Jimbo's homomorphism. It is a homomorphism $\varphi: \uqlsliii \to \uqgliii$ defined by the equations\footnote{Recall that the Cartan generators of $\gothsl_3$ are related to those of $\gothgl_3$ by relation (\ref{hk}).}
\begin{align}
& \varphi(q^{\nu h_0}) = q^{\nu(G_3 - G_1)}, && \varphi(q^{\nu h_1}) = q^{\nu (G_1 - G_2)}, && \varphi(q^{\nu h_2}) = q^{\nu (G_2 - G_3)}, \label{jha} \\
& \varphi(e_0) = F_3 \, q^{- G_1 - G_3}, && \varphi(e_1) = E_1, && \varphi(e_2) = E_2, \\
& \varphi(f_0) = E_3 \, q^{G_1 + G_3} , && \varphi(f_1) = F_1, && \varphi(f_2) = F_2, \label{jhc}
\end{align}
see the paper \cite{Jim86a}. If $\pi$ is a representation of $\uqgliii$, then $\pi \circ \varphi$ is a representation of $\uqlsliii$. Define the universal monodromy operator
\begin{equation*}
\calM_\varphi(\zeta) = (\varphi_\zeta \otimes \id)(\calR)
\end{equation*}
being an element of $\uqgliii \otimes \uqlsliii$. It is evident that
\begin{equation*}
\calM_{\pi \circ \varphi}(\zeta) = (\pi \otimes \id)(\calM_\varphi(\zeta)) = ((\pi \circ \varphi_\zeta) \otimes \id)(\calR).
\end{equation*}
For the corresponding transfer operator we have
\begin{equation*}
\calT_{\pi \circ \varphi}(\zeta) = (\tr \otimes \id)(\calM_{\pi \circ \varphi}(\zeta)((\pi \circ \varphi_\zeta)(t) \otimes 1)) = ((\tr \circ \pi \circ \varphi_\zeta) \otimes \id)(\calR (t \otimes 1)).
\end{equation*}
Introduce the notation
\begin{equation*}
\tr_\pi = \tr \circ \pi.
\end{equation*}
Note that $\tr_\pi$ is a trace on $\uqgliii$. This means that it is a linear mapping from $\uqgliii$ to $\bbC$ satisfying the cyclicity condition
\begin{equation*}
\tr_\pi(a b) = \tr_\pi(b a)
\end{equation*}
for any $a, b \in\uqgliii$. One can write
\begin{equation*}
\calT_{\pi \circ \varphi}(\zeta) = (\tr \otimes \id)(\calM_{\pi \circ \varphi}(\zeta) ((\pi \circ \varphi _\zeta)(t) \otimes 1)) = (\tr_\pi \otimes \id)(\calM_\varphi(\zeta) (\varphi_\zeta(t) \otimes 1)).
\end{equation*}
Thus, to obtain the universal transfer operators $\calT_{\pi \circ \varphi}(\zeta)$, one can use different universal monodromy operators $\calM_{\pi \circ \varphi}(\zeta)$ corresponding to different representations $\pi$, or use one and the same universal monodromy operator $\calM_\varphi(\zeta)$ but different traces $\tr_\pi$ corresponding to different representations $\pi$.

\subsubsection{More universal monodromy operators}

Additional universal monodromy operators can be defined with the help of automorphisms of $\uqlsliii$. There are two special automorphisms of $\uqlsliii$. The first one is defined by the relations
\begin{align}
& \sigma(e_0) = e_1, && \sigma(e_1) = e_2, && \sigma(e_2) = e_0, \label{sigmae} \\*
& \sigma(f_0) = f_1, && \sigma(f_1) = f_2, && \sigma(f_2) = f_0, \\*
& \sigma(q^{\nu h_0}) = q^{\nu h_1}, && \sigma(q^{\nu h_1}) = q^{\nu h_2}, && \sigma(q^{\nu h_2}) = q^{\nu h_0}, \label{sigmah}
\end{align}
and the second one is given by
\begin{align}
& \tau(e_0) = e_0, && \tau(e_1) = e_2, && \tau(e_2) = e_1, \label{taue} \\
& \tau(f_0) = f_0, && \tau(f_1) = f_2, && \tau(f_2) = f_1, \\
& \tau(q^{\nu h_0}) = q^{\nu h_0}, && \tau(q^{\nu h_1}) = q^{\nu h_2}, && \tau(q^{\nu h_2}) = q^{\nu h_1}. \label{tauh}
\end{align}
These automorphisms generate a subgroup of the automorphism group of $\uqlsliii$ isomorphic to the dihedral group $\mathrm D_3$. Using the automorphisms $\sigma$ and $\tau$, we define two families of homomorphisms from $\uqlsliii$ to $\uqgliii$ generalizing the Jimbo's homomorphism as
\begin{equation*}
\varphi_i = \varphi \circ \sigma^{- i + 1}, \qquad \ovarphi{}'_i = \varphi \circ \tau \circ \sigma^{- i + 1},
\end{equation*}
and the corresponding universal monodromy operators as
\begin{equation*}
\calM_i(\zeta) = ((\varphi_i)_\zeta \otimes \id)(\calR), \qquad \ocalM{}'_i(\zeta) = ((\ovarphi{}'_i)_\zeta \otimes \id)(\calR).
\end{equation*}
The prime means that the corresponding homomorphisms and the objects related to them are redefined below to have simpler form of the functional relations. Below we often define objects with the help of the powers of the automorphism $\sigma$. Different powers are marked by the values of the corresponding index. We assume that if the index is omitted it means that the object is taken for the index value $1$, in particular, we have $\varphi = \varphi_1$. Since $\sigma^3$ is the identity automorphism of $\uqlsliii$, we have
\begin{equation*}
\calM_{i + 3}(\zeta) = \calM_i(\zeta), \qquad \ocalM'_{i + 3}(\zeta) = \ocalM{}'_i(\zeta).
\end{equation*}
Therefore, there are only six different universal monodromy operators of such kind.

It follows from (\ref{cmul}) that
\begin{equation*}
(\sigma \otimes \sigma) \circ \Delta = \Delta \circ \sigma.
\end{equation*}
Similarly, (\ref{doqx}) and (\ref{doefi}) give
\begin{equation*}
(\sigma \otimes \sigma) \circ \Delta^{\mathrm{op}} = \Delta^{\mathrm{op}} \circ \sigma.
\end{equation*}
Using the definition of the universal $R$-matrix (\ref{urm}), we obtain the equation
\begin{equation*}
((\sigma \otimes \sigma)(\calR)) \Delta(\sigma(a)) ((\sigma \otimes \sigma)(\calR))^{-1} = \Delta^{\mathrm{op}}(\sigma(a)).
\end{equation*}
Taking into account the uniqueness theorem for the universal $R$-matrix \cite{KhoTol92}, we conclude that
\begin{equation}
(\sigma \otimes \sigma)(\calR) = \calR. \label{ssr}
\end{equation}
Using this relation, it is not difficult to demonstrate that
\begin{equation}
\calM_{i + 1}(\zeta) = (\id \otimes \sigma)(\calM_i(\zeta))|_{s \to \sigma(s)}, \qquad \ocalM{}'_{i + 1}(\zeta) = (\id \otimes \sigma)(\ocalM{}'_i(\zeta))|_{s \to \sigma(s)}, \label{mtom}
\end{equation}
where $s \to \sigma(s)$ stands for
\begin{equation*}
s_0 \to s_1, \ s_1 \to s_2, \ s_2 \to s_0.
\end{equation*}
One can also show that
\begin{equation}
(\tau \otimes \tau)(\calR) = \calR. \label{ttr}
\end{equation}
This relation, together with the equation
\begin{equation}
\sigma \circ \tau \circ \sigma = \tau, \label{sts}
\end{equation}
gives
\begin{equation}
\ocalM{}'_i(\zeta) = (\id \otimes \tau)(\calM_{- i + 2}(\zeta))|_{s \to \tau(s)}, \label{bmtom}
\end{equation}
where $s \to \tau(s)$ stands for
\begin{equation*}
s_0 \to s_0, \ s_1 \to s_2, \ s_2 \to s_1,
\end{equation*}
and we take into account that $\tau^2 = \id$.

Starting with the infinite dimensional representation $\widetilde \pi^{\lambda}$ of the quantum group $\uqgliii$ described in appendix \ref{a:vmr}, we define the infinite dimensional representations
\begin{equation*}
\widetilde \varphi_i^{\lambda} = \widetilde \pi^{\lambda} \circ \varphi_i, \qquad \widetilde{\ovarphi}{}'^\lambda_i = \widetilde \pi{}^{\lambda} \circ \ovarphi{}'_i
\end{equation*}
of the quantum group $\uqlsliii$. Slightly abusing notation we denote the corresponding $\uqlsliii$-modules by $\widetilde V_i^\lambda$ and $\widetilde{\oV}{}'^\lambda_i$. We define two families of universal monodromy operators:
\begin{equation*}
\widetilde \calM_i^\lambda(\zeta) = ((\widetilde \varphi_i^\lambda)_\zeta \otimes \id)(\calR), \qquad \widetilde \ocalM{}'^\lambda_i(\zeta) = ((\widetilde {\ovarphi}{}'^\lambda_i)_\zeta \otimes \id)(\calR).
\end{equation*}
In the same way one defines two families of universal monodromy operators associated with the finite dimensional representation $\pi^\lambda$ of $\uqgliii$:
\begin{equation*}
\calM_i^\lambda(\zeta) = ((\varphi_i^\lambda)_\zeta \otimes \id)(\calR), \qquad \ocalM{}'^\lambda_i(\zeta) = ((\ovarphi{}'^\lambda_i)_\zeta \otimes \id)(\calR),
\end{equation*}
where
\begin{equation*}
\varphi{}_i^\lambda = \pi^\lambda \circ \varphi_i, \qquad \ovarphi{}'^\lambda_i = \pi^\lambda \circ \ovarphi{}'_i.
\end{equation*}
The newly defined universal monodromy operators satisfy relations similar to $(\ref{mtom})$ and $(\ref{bmtom})$.

\subsubsection{Universal transfer operators} \label{sss:uto}

Recall that a universal transfer operator is constructed by taking the trace over the auxiliary space. In the case of an infinite dimensional representation there is a problem of convergence which can be solved with the help of a nontrivial twist element. We use a twist element of the form
\begin{equation}
t = q^{(\phi_0 h_0 + \phi_1 h_1 + \phi_2 h_2) / 3}, \label{t}
\end{equation}
where $\phi_0$, $\phi_1$ and $\phi_2$ are complex numbers. Taking into account (\ref{qh0h1})and (\ref{c}), we assume that
\begin{equation*}
\phi_0 + \phi_1 + \phi_2 = 0.
\end{equation*}

We define two families of universal transfer operators associated with the infinite dimensional representations $\widetilde \pi^\lambda$ of $\uqgliii$ as
\begin{align*}
& \widetilde \calT_i^\lambda(\zeta) = (\tr \otimes \id)(\widetilde \calM_i^\lambda(\zeta) ((\widetilde \varphi_i^\lambda)_\zeta(t) \otimes 1)) = (\widetilde \tr^\lambda \otimes \id)(\calM_i(\zeta) ((\varphi_i)_\zeta(t) \otimes 1)), \\
& \widetilde \ocalT{}'^\lambda_i(\zeta) = (\tr \otimes \id)(\widetilde \ocalM{}'^\lambda_i(\zeta) ((\widetilde {\ovarphi}_i^\lambda)_\zeta(t) \otimes 1)) = (\widetilde \tr^\lambda \otimes \id)(\ocalM{}'_i(\zeta) (({\ovarphi}_i)_\zeta(t) \otimes 1)),
\end{align*}
where
\begin{equation*}
\widetilde \tr{}^\lambda = \tr \circ \widetilde \pi^\lambda,
\end{equation*}
and two families of universal transfer operators associated with the finite dimensional representations $\pi^\lambda$ of $\uqgliii$ as
\begin{align*}
& \calT_i^\lambda(\zeta) = (\tr \otimes \id)(\calM_i^\lambda(\zeta) ((\varphi_i^\lambda)_\zeta(t) \otimes 1)) = (\tr^\lambda \otimes \id)(\calM_i(\zeta) ((\varphi_i)_\zeta(t) \otimes 1)), \\
& \ocalT{}'^\lambda_i(\zeta) = (\tr \otimes \id)(\ocalM{}'^\lambda_i(\zeta) ((\ovarphi{}'^\lambda_i)_\zeta(t) \otimes 1)) = (\tr^\lambda \otimes \id)(\ocalM{}'_i(\zeta) ((\ovarphi{}'_i)_\zeta(t) \otimes 1)),
\end{align*}
where
\begin{equation*}
\tr^\lambda = \tr \circ \pi^\lambda.
\end{equation*}
Note that the mappings $\widetilde \tr{}^\lambda$ and $\tr^\lambda$ are traces on the algebra $\uqgliii$.

Let us discuss the dependence of the universal transfer operators on the spectral parameter $\zeta$. Consider, for example, the universal transfer operator $\widetilde \calT^\lambda(\zeta)$. From the structure of the universal $R$-matrix, it follows that the dependence on $\zeta$ is determined by the dependence on $\zeta$ of the elements of the form $\varphi_\zeta(a)$, where $a \in \uqnp$. Any such element is a linear combination of monomials each of which is a product of $E_1$, $E_2$, $F_3$ and $q^X$ for some $X \in \gothg$. Let $A$ be such a monomial. We have
\begin{equation*}
q^{H_1} A q^{- H_1} = q^{2 n_1 - n_2 - n_3} A, \qquad q^{H_2} A q^{- H_2} = q^{- n_1 +2 n_2 - n_3} A,
\end{equation*}
where $n_1$, $n_2$ and $n_3$ are the numbers of $E_1$, $E_2$ and $F_3$ in $A$. Hence $\widetilde \tr{}^\lambda(A)$ can be non-zero only if
\begin{equation*}
n_1 = n_2 = n_3 = n.
\end{equation*}
Each $E_1$ enters $A$ with the factor $\zeta^{s_1}$, each $E_2$ with the factor $\zeta^{s_2}$, and each $F_3$ with the factor $\zeta^{s_0}$. Thus, for a monomial with non-zero trace we have a dependence on $\zeta$ of the form $\zeta^{n s}$. Therefore, the universal transfer operator $\widetilde \calT^\lambda(\zeta)$ depends on $\zeta$ only via $\zeta^s$, where $s$ is the sum of $s_0$, $s_1$ and $s_2$. The same is evidently true for all other universal transfer operators defined above. Using this fact, we obtain from (\ref{mtom}) and (\ref{bmtom}) the relations
\begin{gather}
\widetilde \calT{}^\lambda_{i + 1}(\zeta) = \sigma(\widetilde \calT^\lambda_i(\zeta))|_{\phi \to \sigma(\phi)}, \qquad
\widetilde{\ocalT}{}'^\lambda_{i + 1}(\zeta) = \sigma(\widetilde{\ocalT}{}'^\lambda_i(\zeta))|_{\phi \to \sigma(\phi)}, \label{tst} \\
\widetilde{\ocalT}{}'^\lambda_i(\zeta) = \tau(\widetilde{\calT}{}^\lambda_{- i + 2}(\zeta))|_{\phi \to \tau(\phi)}, \label{ttt}
\end{gather}
where $\phi \to \sigma(\phi)$ stands for
\begin{equation*}
\phi_0 \to \phi_1, \quad \phi_1 \to \phi_2, \quad \phi_2 \to \phi_0
\end{equation*}
and $\phi \to \tau(\phi)$ for
\begin{equation*}
\phi_0 \to \phi_0, \quad \phi_1 \to \phi_2, \quad \phi_2 \to \phi_1.
\end{equation*}
Similarly, for the universal transfer operators corresponding to the finite dimensional representations $\pi^\lambda$ we have
\begin{gather}
\calT{}^\lambda_{i + 1}(\zeta) = \sigma(\calT^\lambda_i(\zeta))|_{\phi \to \sigma(\phi)}, \qquad \ocalT{}'^\lambda_{i + 1}(\zeta) = \sigma(\ocalT{}'^\lambda_i(\zeta))|_{\phi \to \sigma(\phi)}, \label{tsta} \\
\ocalT{}'^\lambda_i(\zeta) = \tau(\calT{}^\lambda_{- i + 2}(\zeta))|_{\phi \to \tau(\phi)}.
\end{gather} 

\subsubsection{Application of BGG resolution}

Recall first some definitions and properties of the necessary objects. We have denoted the standard Cartan subalgebra of the Lie algebra $\gothgl_3$ by $\gothg$. The Weyl group $W$ of the root system of $\gothgl_3$ is generated by the reflections $r_i: \gothg^* \to \gothg^*$, $i = 1, 2$, defined by the equation
\begin{equation*}
r_i(\lambda) = \lambda - \lambda(H_i) \alpha_i.
\end{equation*}
The minimal number of generators $r_i$ necessary to represent an element $w \in W$ is said to be the length of $w$ and is denoted by $\ell(w)$. It is assumed that the identity element has the length equal to $0$.

Let $\{\gamma_i\}$ be a dual basis of the standard basis $\{G_i\}$ of $\gothg$. Using (\ref{alphah}), it is easy to see that
\begin{equation*}
\alpha_1 = \gamma_1 - \gamma_2, \qquad \alpha_2 = \gamma_2 - \gamma_3.
\end{equation*}
One can get convinced that
\begin{align*}
r_1 (\gamma_1) & = \gamma_2, & r_1 (\gamma_2) & = \gamma_1, & r_1(\gamma_3) = \gamma_3, \\
r_2 (\gamma_1) & = \gamma_1, & r_2 (\gamma_2) & = \gamma_3, & r_2(\gamma_3) = \gamma_2.
\end{align*}
Identifying an element of $\gothg^*$ with the set of its components respective to the basis $\{\gamma_i\}$, we see that the reflection $r_1$ transposes the first and second components, and $r_2$ transposes the second and third components. One can verify that the whole Weyl group $W$ can be identified with the symmetric group $\mathrm S_3$. Here $(-1)^{\ell(w)}$ is evidently the sign of the permutation corresponding to the element $w \in W$.  The order of $W$ is equal to six. There are one element of length $0$, two elements of length $1$, two elements of length $2$, and one element of length $3$.

Consider now a finite dimensional $\uqgliii$-module $V^\lambda$, see appendix \ref{a:vmr}. Introduce the following direct sums of infinite dimensional  $\uqgliii$-modules:
\begin{equation*}
U_k = \bigoplus_{\substack{w \in W \\ \ell(w) = k}} \widetilde V^{w \cdot \lambda},
\end{equation*}
where $w \cdot \lambda$ means the affine action of $w$ defined as
\begin{equation*}
w \cdot \lambda = w(\lambda + \rho) - \rho
\end{equation*}
with $\rho = \alpha_1 + \alpha_2$. The quantum version of the Bernstein--Gelfand--Gelfand (BGG) resolution \cite{Ros91} for the quantum group $\uqgliii$ is the following exact sequence of $\uqgliii$-modules and $\uqgliii$-homomorphisms:
\begin{equation*}
0 \longrightarrow U_3 \overset{\varphi_3}{\longrightarrow} U_2 \overset{\varphi_2}{\longrightarrow} U_1 \overset{\varphi_1}{\longrightarrow} U_0 \overset{\varphi_0}{\longrightarrow} U_{-1} \longrightarrow 0,
\end{equation*}
where $U_{-1} = V^\lambda$.

Let $\rho_k$ be the representation of $\uqgliii$ corresponding to the $\uqgliii$-module $U_k$. Note that $\rho_{-1} = \pi^\lambda$. The subspaces $\ker \varphi_k$ and $\im \varphi_k$ are $\uqgliii$-submodules of $U_k$ and $U_{k - 1}$ respectively. For each $k = 0, 1, 2, 3$ we have
\begin{equation*}
\tr \circ \rho_k = \tr \circ \rho_k|_{\ker \varphi_k} + \tr \circ \rho_{k - 1} |_{\im \varphi_k}
\end{equation*}
and
\begin{equation*}
\im \varphi_k = \ker \varphi_{k - 1}.
\end{equation*}
Hence,
\begin{equation*}
\sum_{k = 0}^3 (-1)^k \tr \circ \rho_k = \tr \circ \rho_{-1}|_{\im \varphi_0} - \tr \circ \rho_3|_{\ker \varphi_3}.
\end{equation*}
Finally, having in mind that
\begin{equation*}
\im \varphi_0 = V^\lambda, \qquad \ker \varphi_3 = 0,
\end{equation*}
we obtain
\begin{equation*}
\sum_{k = 0}^3 (-1)^k \tr \circ \rho_k = \tr \circ \pi^\lambda = \tr^\lambda.
\end{equation*}
From the definition of $U_k$ it follows that
\begin{equation*}
\sum_{k = 0}^3 (-1)^k \tr \circ \rho_k = \sum_{w \in W} (-1)^{\ell(w)} \widetilde \tr{}^{\, w \cdot \lambda}.
\end{equation*}
Thus, we have
\begin{equation*}
\tr^\lambda = \sum_{w \in W} (-1)^{\ell(w)} \widetilde \tr{}^{\, w \cdot \lambda}.
\end{equation*}
This gives
\begin{equation*}
\calT^\lambda_i(\zeta) = \sum_{w \in W} (-1)^{\ell(w)} \widetilde \calT^{\, w \cdot \lambda}_i(\zeta), \qquad \ocalT{}'^\lambda_i(\zeta) = \sum_{w \in W} (-1)^{\ell(w)} \widetilde \ocalT{}'^{\, w \cdot \lambda}_i(\zeta).
\end{equation*}
Using the identification of $W$ with $\mathrm S_3$ described above, we write
\begin{equation}
\calT^\lambda_i(\zeta) = \sum_{p \in \mathrm S_3} \sgn(p) \, \widetilde \calT^{\, p(\lambda + \rho) - \rho}_i(\zeta), \qquad \ocalT{}'^\lambda_i(\zeta) = \sum_{p \in \mathrm S_3} \sgn(p) \, \widetilde \ocalT{}'^{\, p(\lambda + \rho) - \rho}_i(\zeta), \label{twtt}
\end{equation}
where $\sgn(p)$ is the sign of the permutation $p$, and $p$ acts on an element of $\gothg^*$ appropriately permuting its components with respect to the basis $\{\gamma_i\}$.

\subsection{\texorpdfstring{Universal $L$-operators and universal $Q$-operators}{Universal L-operators and universal Q-operators}}

\subsubsection{General remarks}

It follows from (\ref{rpipi}) and (\ref{kpipi}) that to construct universal monodromy operators and transfer operators it suffices to have a representation of the Borel subalgebra $\uqbp$. This observation is used to construct universal $L$-operators and $Q$-operators. In distinction to the case of universal transfer operators, we use here representations of $\uqbp$ which cannot be obtained by restriction of representations of $\uqlsliii$, or equivalently, representations of $\uqbp$ which cannot be extended to representations of $\uqlsliii$. It is clear that to have interesting functional relations one should use for the construction of universal $L$-operators and $Q$-operators representations which are connected in some way to the representations used for the construction of universal monodromy operators and transfer operators.

In general, a universal $L$-operator associated with a representation $\rho$ of $\uqbp$ is defined as
\begin{equation*}
\calL_\rho(\zeta) = (\rho_\zeta \otimes \id)(\calR).
\end{equation*}
As the universal monodromy operators are auxiliary objects needed for the construction of the universal transfer operators, the universal $L$-operators are needed for the construction of the universal $Q$-operators. The universal $Q$-operator $\calQ_\rho(\zeta)$ corresponding to the universal $L$-operator $\calL_\rho(\zeta)$ is defined as
\begin{equation*}
\calQ_\rho(\zeta) = (\tr \otimes \id)(\calL_\rho(\zeta) (\rho_\zeta(t) \otimes 1)) = ((\tr \circ \rho_\zeta) \otimes \id)(\calR (t \otimes 1)),
\end{equation*}
where $t$ is a twist element.

\subsubsection{Basic representation} \label{sss:br}

We start the construction of universal $L$-operators and $Q$-ope\-ra\-tors with the construction of the basic representation of $\uqbp$. The initial point is the representations $\widetilde \varphi^\lambda$ of $\uqlsliii$ described above.

First define the notion of a shifted representation. Let $\xi$ be an element of $\widetilde \gothh^*$ satisfying the equation
\begin{equation*}
\xi(h_0 + h_1 + h_2) = 0.
\end{equation*}
If $\rho$ is a representation of $\uqbp$, then the representation $\rho[\xi]$ defined by the relations
\begin{equation*}
\rho[\xi](e_i) = \rho(e_i), \qquad \rho[\xi](q^x) = q^{\xi(x)} \rho (q^x)
\end{equation*}
is a representation of $\uqbp$ called a shifted representation. If $W$ is a $\uqbp$-module corresponding to the representation $\rho$, then $W[\xi]$ denotes the $\uqbp$-module corresponding to the representation $\rho[\xi]$.

Consider the restriction of the representation $\widetilde \varphi^\lambda$ to $\uqbp$ and a shifted representation $\widetilde \varphi^\lambda[\xi]$ of $\uqbp$. One can show that for $\xi \ne 0$ this representation cannot be extended to a representation of $\uqlsliii$ and we can use it to construct a universal $Q$-operator. However, it follows from (\ref{rpipi}) and (\ref{kpipi}) that the universal $Q$-operator defined with the help of the representation $\widetilde \varphi^\lambda[\xi]$ is connected with the universal transfer operator defined with the help of the representation $\widetilde \varphi^\lambda$ by the relation
\begin{equation*}
\calQ_{\widetilde \varphi^\lambda[\xi]}(\zeta) = \calT_{\widetilde \varphi^\lambda}(\zeta) \, q^{\sum_{i = 0}^2 \xi(h^{\mathstrut}_i) h'_i / 3},
\end{equation*}
where
\begin{equation*}
h'_i = h_i + \phi_i.
\end{equation*}
Here we assume that the twist element is of the form (\ref{t}). We see that the use of shifted representations does not give anything really new.

Consider again the restriction of the representation $(\widetilde \varphi^\lambda)_\zeta$ to $\uqbp$. We have for this representation\footnote{Recall that $\mu_1 = \lambda_1 - \lambda_2$ and $\mu_2 = \lambda_2 - \lambda_3$.}
\begin{align}
& q^{\nu h_0} v_n = q^{\nu(- \mu_1 - \mu_2 + n_1 + 2 n_2 + n_3)} v_n, \label{ah0vn} \\*
& q^{\nu h_1} v_n = q^{\nu(\mu_1 - 2 n_1 - n_2 + n_3)} v_n, \\*
& q^{\nu h_2} v_n = q^{\nu(\mu_2 + n_1 - n_2 - 2 n_3)} v_n, \label{ah2vn} \\
& e_0 v_n = \zeta^{s_0} q^{- \lambda_1 - \lambda_3 - n_3} v_{n + \varepsilon_2}, \label{ae0vn} \\*
& e_1 v_n = \zeta^{s_1} \bigl( [\mu_1 - n_1 - n_2 + n_3 + 1]_q [n_1]_q v_{n - \varepsilon_1} - q^{\mu_1 - n_2 + n_3 + 2} [n_2]_q v_{n - \varepsilon_2 + \varepsilon_3} \bigr), \\*
& e_2 v_n = \zeta^{s_2} \bigl( [\mu_2 - n_3 + 1]_q [n_3]_q v_{n - \varepsilon_3} + q^{- \mu_2 + 2 n_3} [n_2]_q v_{n + \varepsilon_1 - \varepsilon_2} \bigr), \label{ae2vn}
\end{align}
where the basis vectors $v_n$ are defined by (\ref{vn}). Let us try to go to the limit $\mu_1, \mu_2 \to \infty$. Looking at relations (\ref{ah0vn})--(\ref{ae2vn}), we see that we cannot perform this limit directly. Therefore, we consider first a shifted representation $(\widetilde \varphi^\lambda)_\zeta[\xi]$ with $\xi$ defined by the relations
\begin{equation*}
\xi(h_0) = \mu_1 + \mu_2, \qquad \xi(h_1) = - \mu_1, \qquad \xi(h_2) = - \mu_2.
\end{equation*}
Then we introduce a new basis
\begin{equation*}
w_n = c_1^{n_1 + n_2} c_2^{n_2 + n_3} v_n,
\end{equation*}
where
\begin{equation*}
c_1 = q^{- \mu_1 - 1 - 2 (\lambda_3 - 1) s_1 / s}, \qquad c_2 = q^{- \mu_2 - 1 - 2 (\lambda_3 - 1) s_2 / s}.
\end{equation*}
Now relations (\ref{ah0vn})--(\ref{ah2vn}) take the form
\begin{gather}
q^{\nu h_0} w_n = q^{\nu(n_1 + 2 n_2 + n_3)} w_n, \label{bh01wn} \qquad q^{\nu h_1} w_n = q^{\nu(- 2 n_1 - n_2 + n_3)} w_n, \\*
q^{\nu h_2} w_n = q^{\nu(n_1 - n_2 - 2 n_3)} w_n, \label{bh2wn}
\end{gather}
and instead of relations (\ref{ae0vn})--(\ref{ae2vn}) we have
\begin{align}
e_0 w_n &= \widetilde \zeta^{s_0} q^{- n_3} w_{n + \varepsilon_2}, \label{be0wn} \\*
e_1 w_n &= \widetilde \zeta^{s_1} \bigl( \kappa_q^{-1} (q^{- n_1 - n_2 + n_3} - q^{- 2 \mu_1 + n_1 + n_2 - n_3 - 2}) [n_1]_q w_{n - \varepsilon_1} \notag \\*
& \hspace{18em} {} - q^{- n_2 + n_3 + 1}[n_2]_q w_{n - \varepsilon_2 + \varepsilon_3} \bigr), \\
e_2 w_n &= \widetilde \zeta^{s_2} \bigl( \kappa_q^{-1} (q^{- n_3} - q^{- 2 \mu_2 + n_3 - 2}) [n_3]_q w_{n - \varepsilon_3} + q^{- 2 \mu_2 + 2 n_3 - 1} [n_2]_q w_{n + \varepsilon_1 - \varepsilon_2} \bigr), \label{be3wn}
\end{align}
where
\begin{equation*}
\widetilde \zeta = q^{- 2 (\lambda_3 - 1) / s} \zeta.
\end{equation*}
It is possible now to consider the limit $\mu_1, \mu_2 \to \infty$. Relations (\ref{bh01wn}), (\ref{bh2wn}) retain their form, while (\ref{be0wn})--(\ref{be3wn}) go to
\begin{align*}
e_0 w_n &= \widetilde \zeta^{s_0} q^{- n_3} w_{n + \varepsilon_2}, \\*
e_1 w_n &= \widetilde \zeta^{s_1} \bigl( \kappa_q^{-1} q^{- n_1 - n_2 + n_3} [n_1]_q w_{n - \varepsilon_1} - q^{- n_2 + n_3 + 1} [n_2]_q w_{n - \varepsilon_2 + \varepsilon_3} \bigr), \\*
e_2 w_n &= \widetilde \zeta^{s_2} \kappa_q^{-1} q^{- n_3} [n_3]_q w_{n - \varepsilon_3}.
\end{align*}

Denote by $\rho''$ the representation of $\uqbp$ defined by the relations
\begin{align*}
q^{\nu h_0} v_n &= q^{\nu(n_1 + 2 n_2 + n_3)} v_n, \\
q^{\nu h_1} v_n &= q^{\nu(- 2 n_1 - n_2 + n_3)} v_n, \\
q^{\nu h_2} v_n &= q^{\nu(n_1 - n_2 - 2 n_3)} v_n, \\
e_0 v_n &= q^{- n_3} v_{n + \varepsilon_2}, \\*
e_1 v_n &= \bigl( \kappa_q^{-1} q^{- n_1 - n_2 + n_3} [n_1]_q v_{n - \varepsilon_1} - q^{- n_2 + n_3 + 1} [n_2]_q v_{n - \varepsilon_2 + \varepsilon_3} \bigr), \\*
e_2 v_n &= \kappa_q^{-1} q^{- n_3} [n_3]_q v_{n - \varepsilon_3},
\end{align*}
and by $W''$ the corresponding $\uqbp$-module. It is clear that if we define the universal $Q$-operator $\calQ''(\zeta)$ by the relation
\begin{equation*}
\calQ''(\zeta) = ((\tr \circ (\rho'')_\zeta) \otimes \id)(\calR (t \otimes 1)),
\end{equation*}
then we have
\begin{equation*}
\calQ''(\zeta) = \lim_{\mu_1, \mu_2 \to \infty} \left( \widetilde \calT^{(\mu_1 + \mu_2, \mu_2, 0)}(q^{- 2 / s} \zeta) \, q^{((\mu_1 + \mu_2) h'_0 - \mu_1 h'_1 - \mu_2 h'_2) / 3} \right).
\end{equation*}
We use the double prime having in mind that we will redefine $Q$-operators twice. It follows from the above relation that the universal $Q$-operators $\calQ''(\zeta)$ for all values of $\zeta$ commute. In addition, they commute with all universal transfer operators $\calT^\lambda_i(\zeta)$, $\ocalT^\lambda_i(\zeta)$ and with all generators $q^x$, $x \in \widetilde \gothh$,  see, for example, our papers \cite{BooGoeKluNirRaz14a, BooGoeKluNirRaz13}.

In fact, the representation $\rho''$ is not irreducible. Certainly, the same is true for the representation $(\rho'')_\zeta$. In fact, for any $\zeta \in \bbC^\times$ there is a filtration
\begin{equation*}
\{0\} = ((W'')_\zeta)_{-1} \subset ((W'')_\zeta)_0 \subset ((W'')_\zeta)_1 \subset \cdots
\end{equation*}
formed by the $\uqbp$-submodules
\begin{equation*}
((W'')_\zeta)_k = \bigoplus_{n_1 = 0}^k \bigoplus_{n_2, n_3 = 0}^\infty \bbC \, v_n.
\end{equation*}
Denote by $\rho'$ the representation of $\uqbp$ defined by the relations
\begin{gather}
q^{\nu h_0} v_n = q^{\nu(2 n_1 + n_2)} v_n, \quad
q^{\nu h_1} v_n = q^{\nu(- n_1 + n_2)} v_n, \quad
q^{\nu h_2} v_n = q^{\nu(- n_1 - 2 n_2)} v_n, \label{qhv} \\
e_0 v_n = q^{- n_2} v_{n + \varepsilon_1}, \label{eva} \\*
e_1 v_n = - q^{- n_1 + n_2 + 1} [n_1]_q v_{n - \varepsilon_1 + \varepsilon_2}, \qquad
e_2 v_n = \kappa_q^{-1} q^{- n_2} [n_2]_q v_{n - \varepsilon_2}, \label{evb}
\end{gather}
and by $W'$ the corresponding $\uqbp$-module. Here we assume that $n$ stands for the pair $(n_1, n_2) \in \bbZ_+^2$. It is easy to see that
\begin{equation*}
((W'')_\zeta)_k / ((W'')_\zeta)_{k - 1} \cong (W'[\xi_k])_\zeta,
\end{equation*}
where
\begin{equation*}
\xi_k(h_0) = k, \qquad \xi_k(h_1) = - 2 k, \qquad \xi_k(h_2) = k.
\end{equation*}
Hence, the universal $Q$-operator $\calQ'(\zeta)$ defined by the relation
\begin{equation*}
\calQ'(\zeta) = ((\tr \circ (\rho')_\zeta) \otimes \id)(\calR (t \otimes 1)),
\end{equation*}
is related to the universal $Q$-operator $\calQ''(\zeta)$ as
\begin{equation*}
\calQ''(\zeta) = \sum_{k = 0}^\infty \calQ'(\zeta) \, q^{k (h'_0 - 2 h'_1 + h'_2) / 3} = \calQ'(\zeta) \bigl( 1 - q^{(h'_0 - 2 h'_1 + h'_2) / 3} \bigr)^{-1},
\end{equation*}
Here we use the fact that
\begin{equation}
\calQ_{\rho[\xi]}(\zeta) = \calQ_\rho(\zeta) \, q^{\sum_{i = 0}^2 \xi(h^{\mathstrut}_i) h'_i / 3} \label{qshq}
\end{equation}
for any representation $\rho$ of $\uqbp$ and any $\xi \in \widetilde \gothh^*$.
We use the representation $\rho'$ as the basic representation for the construction of all necessary universal $L$-operators and $Q$-operators. In fact, it is an asymptotic, or prefundamental, representation of $\uqbp$, see the papers  \cite{HerJim12, FreHer13}.

\subsubsection{Interpretation in terms of $q$-oscillators}

It is useful to give an interpretation of relations (\ref{qhv})--(\ref{evb}) in terms of $q$-oscillators. Let us remind the necessary definitions, see, for example, the book \cite{KliSch97}.

Let $\hbar$ be a complex number and $q = \exp \hbar$.\footnote{We again assume that $q$ is not a root of unity.} The $q$-oscillator algebra $\Osc_q$ is a unital associative $\bbC$-algebra with generators $b^\dagger$, $b$, $q^{\nu N}$, $\nu \in \bbC$, and relations
\begin{gather*}
q^0 = 1, \qquad q^{\nu_1 N} q^{\nu_2 N} = q^{(\nu_1 + \nu_2)N}, \\
q^{\nu N} b^\dagger q^{-\nu N} = q^\nu b^\dagger, \qquad q^{\nu N} b q^{-\nu N} = q^{-\nu} b, \\
b^\dagger b = [N]_q, \qquad b b^\dagger = [N + 1]_q,
\end{gather*}
where we use the notation similar to (\ref{xnq}). It is easy to understand that the monomials $(b^\dagger)^{k + 1} q^{\nu N}$, $b^{k + 1} q^{\nu N}$
and $q^{\nu N}$ for $k \in \bbZ_+$ and $\nu \in \bbC$ form a basis of $\Osc_q$.

Two representations of $\Osc_q$ are interesting for us. First, let $W^{\scriptscriptstyle +}$ be a free vector space generated by the set $\{ v_0, v_1, \ldots \}$. One can show that the relations
\begin{gather*}
q^{\nu N} v_n = q^{\nu n} v_n, \\*
b^\dagger v_n = v_{n + 1}, \qquad b \, v_n = [n]_q v_{n - 1},
\end{gather*}
where we assume that $v_{-1} = 0$, endow $W^{\scriptscriptstyle +}$ with the structure of an $\Osc_q$-module. We denote the corresponding representation of the algebra $\Osc_q$ by $\chi^{\scriptscriptstyle +}$. Further, let $W^{\scriptscriptstyle -}$ be a free vector space generated again by the set $\{ v_0, v_1, \ldots \}$. The relations
\begin{gather*}
q^{\nu N} v_n = q^{- \nu (n + 1)} v_n, \\
b \, v_n = v_{n + 1}, \qquad b^\dagger v_n = - [n]_q v_{n - 1},
\end{gather*}
where we again assume that $v_{-1} = 0$, endow the vector space $W^{\scriptscriptstyle -}$ with the structure of an $\Osc_q$-module. We denote the corresponding representation of $\Osc_q$ by $\chi^{\scriptscriptstyle -}$.

Consider the algebra $\Osc_q \otimes \Osc_q$. As is usual, define
\begin{gather*}
b^{}_1 = b \otimes 1, \qquad b^\dagger_1 = b^\dagger \otimes 1, \qquad b^{}_2 = 1 \otimes b, \qquad b^\dagger_2 = 1 \otimes b^\dagger, \\
q^{\nu N_1} = q^{\nu N} \otimes 1, \qquad q^{\nu N_2} = 1 \otimes q^{\nu N},
\end{gather*}
and denote
\begin{equation*}
q^{\nu_1 N_1 + \nu_2 N_2 + \nu} = q^\nu q^{\nu_1 N_1} q^{\nu_2 N_2}.
\end{equation*}
Assume that the generators of $\Osc_q \otimes \Osc_q$ act on the module $W'$ defined by equations (\ref{qhv})--(\ref{evb}) as on the module $W^{\scriptscriptstyle +} \otimes W^{\scriptscriptstyle +}$. This allows us to write (\ref{qhv})--(\ref{evb})  as
\begin{align*}
& q^{\nu h_0} v_n = q^{\nu(2 N_1 + N_2)} v_n, &&
q^{\nu h_1} v_n = q^{\nu(- N_1 + N_2)} v_n, &&
q^{\nu h_2} v_n = q^{\nu(- N_1 - 2 N_2)} v_n, \\*
& e_0 v_n = b^\dagger_1 q^{- N_2} v_n, && e_1 v_n = - b_1^{\mathstrut} b_2^\dagger q^{- N_1 + N_2 + 1} v_n, && e_2 v_n = \kappa_q^{-1} b_2^{\mathstrut} q^{- N_2}  v_n.
\end{align*}
These equations suggest defining a homomorphism $\rho: \uqbp \to \Osc_q \otimes \Osc_q$ by
\begin{align}
& \rho(q^{\nu h_0}) = q^{\nu(2 N_1 + N_2)}, && \rho(q^{\nu h_1}) = q^{\nu(- N_1 + N_2)}, &&
\rho(q^{\nu h_2}) = q^{\nu(- N_1 - 2 N_2)}, \label{rhoh} \\*
& \rho(e_0) = b^\dagger_1 q^{- N_2}, && \rho(e_1) = - b_1^{\mathstrut} b_2^\dagger q^{- N_1 + N_2 + 1}, && \rho(e_2) = \kappa_q^{-1} b_2^{\mathstrut} q^{- N_2}. \label{rhoe}
\end{align}
Using this homomorphism, we can write for the representation $\rho'$ the equation
\begin{equation*}
\rho' = (\chi^{\scriptscriptstyle +} \otimes \chi^{\scriptscriptstyle +}) \circ \rho.
\end{equation*}

Define the universal $L$-operator\footnote{The prime here and below means that the corresponding universal $L$-operators will be used to define primed universal $Q$-operators.}
\begin{equation*}
\calL'_\rho(\zeta) = (\rho_\zeta \otimes \id)(\calR)
\end{equation*}
being an element of $(\Osc_q \otimes \Osc_q) \otimes \uqlsliii$. Then if $\chi$ is a representation of $\Osc_q \otimes \Osc_q$, we have the universal $L$-operator
\begin{equation*}
\calL'_{\chi \circ \rho}(\zeta) = (\chi \otimes \id)(\calL'_\rho(\zeta)) = ((\chi \circ \rho_\zeta) \otimes \id)(\calR),
\end{equation*}
and the corresponding universal $Q$-operator is
\begin{equation*}
\calQ'_{\chi \circ \rho}(\zeta) = ((\tr \circ \chi \circ \rho_\zeta) \otimes \id)(\calR (t \otimes 1)).
\end{equation*}
One can write
\begin{equation*}
\calQ'_{\chi \circ \rho}(\zeta) = (\tr \otimes \id)(\calL'_{\chi \circ \rho}(\zeta) ((\chi \circ \rho_\zeta)(t) \otimes 1)) = (\tr_\chi \otimes \id)(\calL'_\rho(\zeta) (\rho_\zeta(t) \otimes 1)),
\end{equation*}
where $\tr_\chi = \tr \circ \chi$. Thus, to obtain the universal $Q$-operators $\calQ'_{\chi \circ \rho}(\zeta)$, one can use different $L$-operators $\calL'_{\chi \circ \rho}(\zeta)$ corresponding to different representations $\chi$, or use one and the same universal $L$-operator $\calL'_\rho(\zeta)$ but different traces $\tr_\chi$ corresponding to different representations $\chi$.

\subsubsection{More universal $L$-operators and $Q$-operators}

Using the homomorphism $\rho$ defined by (\ref{rhoh})--(\ref{rhoe}) and the automorphisms $\sigma$ and $\tau$ defined by (\ref{sigmae})--(\ref{sigmah}) and (\ref{taue})--(\ref{tauh}), we define the homomorphisms
\begin{equation*}
\rho_i = \rho \circ \sigma^{- i}, \qquad \orho_i = \rho \circ \tau \circ \sigma^{- i + 1},
\end{equation*}
where $i = 1, 2, 3$, and the universal $L$-operators
\begin{equation*}
\calL'_i(\zeta) = ((\rho_i)_\zeta \otimes \id)(\calR), \qquad \ocalL{}'_i(\zeta) = ((\orho_i)_\zeta \otimes \id)(\calR).
\end{equation*}
These universal $L$-operators are again elements of $(\Osc_q \otimes \Osc_q) \otimes \uqlsliii$. In the same way as for the universal monodromy operators, using relations (\ref{ssr}), (\ref{ttr}) and (\ref{sts}), we obtain the equations
\begin{gather}
\calL'_{i + 1}(\zeta) = (\id \otimes \sigma)(\calL'_i(\zeta))|_{s \to \sigma(s)}, \qquad
\ocalL{}'_{i + 1}(\zeta) = (\id \otimes \sigma)(\ocalL'_i(\zeta))|_{s \to \sigma(s)}, \label{lipo} \\
\ocalL{}'_i(\zeta) = (\id \otimes \tau)(\calL{}'_{- i + 1}(\zeta))|_{s \to \tau(s)}. \label{bli}
\end{gather}

Two standard representations $\chi^{\scriptscriptstyle +}$ and $\chi^{\scriptscriptstyle -}$ of the algebra $\Osc_q$ generate two traces on $\Osc_q$. We denote
\begin{equation*}
\tr^{\scriptscriptstyle +} = \tr \circ \chi^{\scriptscriptstyle +}, \qquad \tr^{\scriptscriptstyle -} = \tr \circ \chi^{\scriptscriptstyle -}.
\end{equation*}
We see that
\begin{equation*}
\tr^{\scriptscriptstyle +} ((b^\dagger)^{k + 1} q^{\nu N}) = 0, \qquad \tr^{\scriptscriptstyle +} (b^{k + 1} q^{\nu N}) = 0,
\end{equation*}
and that
\begin{equation*}
\tr^{\scriptscriptstyle +}(q^{\nu N}) = (1 - q^\nu)^{-1}
\end{equation*}
for $|q| < 1$. For $|q| > 1$ we define the trace $\tr^{\scriptscriptstyle +}$ by analytic continuation. Since the monomials $(b^\dagger)^{k + 1} q^{\nu N}$, $b^{k + 1} q^{\nu N}$ and $q^{\nu N}$ for $k \in \bbZ_+$ and $\nu \in \bbC$ form a basis of $\Osc_q$, the above relations are enough to determine the trace of any element of $\Osc_q$. It appears that
\begin{equation}
\tr^{\scriptscriptstyle -} = - \tr^{\scriptscriptstyle +}. \label{trtr}
\end{equation}
Fixing the representations, $\chi^{\scriptscriptstyle +}$ or $\chi^{\scriptscriptstyle -}$, for the factors of the tensor product $\Osc_q \otimes \Osc_q$, we obtain a trace on the algebra $\Osc_q \otimes \Osc_q$. We use for such traces the convention
\begin{equation*}
\tr^{\epsilon_1 \epsilon_2} = \tr \circ (\chi^{\epsilon_1} \otimes \chi^{\epsilon_2}),
\end{equation*}
where $\epsilon_1, \epsilon_2 = +, -$. It follows from (\ref{trtr}) that different traces $\tr^{\epsilon_1 \epsilon_2}$ differ at most by a sign. Therefore, since $\sigma^3 = \id$, we have only six really different universal $Q$-operators which can be obtained from the universal $L$-operators $\calL'_i(\zeta)$ and $\ocalL'_i(\zeta)$. We use the following definition
\begin{gather*}
\calQ'_i(\zeta) = \varsigma^{\epsilon_1 \epsilon_2} (\tr^{\epsilon_1 \epsilon_2} \otimes \id) (\calL'_i(\zeta) ((\rho_i)_\zeta(t) \otimes 1)), \\
\ocalQ'_i(\zeta) = \varsigma^{\epsilon_1 \epsilon_2} (\tr^{\epsilon_1 \epsilon_2} \otimes \id) (\ocalL{}'_i(\zeta) ((\orho_i)_\zeta(t) \otimes 1)),
\end{gather*}
where
\begin{equation*}
\varsigma^{\scriptscriptstyle ++} = 1, \qquad \varsigma^{\scriptscriptstyle +-} = -1, \qquad \varsigma^{\scriptscriptstyle -+} = -1, \qquad \varsigma^{\scriptscriptstyle --} = 1. 
\end{equation*}
Certainly, we can use the same trace but different representations and define the universal $Q$-operators as
\begin{gather*}
\calQ'_i(\zeta) = \varsigma^{\epsilon_1 \epsilon_2} ((\tr \circ (\rho^{\epsilon_1 \epsilon_2}_i)_\zeta) \otimes \id)(\calR (t \otimes 1)), \\
\ocalQ'_i(\zeta) = \varsigma^{\epsilon_1 \epsilon_2} ((\tr \circ (\orho_i^{\epsilon_1 \epsilon_2})_\zeta) \otimes \id)(\calR (t \otimes 1)),
\end{gather*}
where
\begin{equation*}
\rho_i^{\epsilon_1 \epsilon_2} = (\chi^{\epsilon_1} \otimes \chi^{\epsilon_2}) \circ \rho_i, \qquad \orho_i^{\epsilon_1 \epsilon_2} = (\chi^{\epsilon_1} \otimes \chi^{\epsilon_2}) \circ \orho_i
\end{equation*}
are representations of $\uqbp$. We denote by $W^{\epsilon_1 \epsilon_2}$ the corresponding $\uqbp$-mo\-dules.

Similarly as for the case of the universal transfer operators, one can demonstrate that the universal $Q$-operators $\calQ'_i(\zeta)$ and $\ocalQ{}'_i(\zeta)$ depend on $\zeta$ via $\zeta^s$. Therefore, equations (\ref{lipo}) and (\ref{bli}) lead to the relations
\begin{gather}
\calQ'_{i + 1}(\zeta) = \sigma(\calQ'_i(\zeta))|_{\phi \to \sigma(\phi)}, \qquad \ocalQ{}'_{i + 1}(\zeta) = \sigma(\ocalQ{}'_i(\zeta))|_{\phi \to \sigma(\phi)}, \label{qsq} \\
\ocalQ{}'_i(\zeta) = \tau(\calQ{}'_{- i + 1}(\zeta))|_{\phi \to \tau(\phi)}. \label{bqtq}
\end{gather}

Based on section \ref{sss:br}, all universal $Q$-operators $\calQ'_i(\zeta)$ and $\ocalQ{}'_i(\zeta)$ can be considered as limits of the corresponding universal transfer operators. Therefore, they commute for all values of $i$ and $\zeta$. They commute also with all universal transfer operators $\calT^\lambda_i(\zeta)$, $\ocalT^\lambda_i(\zeta)$ and with all generators $q^x$, $x \in \widetilde \gothh$.

\section{Functional relations}

\subsection{\texorpdfstring{Product of two universal $Q$-operators}{Product of two universal Q-operators}}

The functional relations are some equations for products of transfer operators and $Q$-operators. Consider the product of two universal $Q$-operators. In fact, one can consider any pair of the universal $Q$-operators. The functional relations for other pairs are simple consequences of the functional relations for the initial pair. It appears that the pair $\calQ'_3(\zeta_3)$ and $\calQ'_2(\zeta_2)$ is the most convenient choice. To simplify the proof we use for the definition of $\calQ'_3(\zeta_3)$ the tensor product of two copies of the representation $\chi^{\scriptscriptstyle +}$, and for the definition of $\calQ'_2(\zeta_2)$ the tensor product of the representations $\chi^{\scriptscriptstyle -}$ and $\chi^{\scriptscriptstyle +}$. Concretely speaking, we use the equations
\begin{align}
& \calQ'_3(\zeta_3) =  ((\tr \circ (\rho^{\scriptscriptstyle ++}_3)_{\zeta_3}) \otimes \id)(\calR (t \otimes 1)), \label{qp3} \\
& \calQ'_2(\zeta_2) =  - ((\tr \circ (\rho^{\scriptscriptstyle -+}_2)_{\zeta_2}) \otimes \id)(\calR (t \otimes 1)). \label{qp2}
\end{align}
One can show that
\begin{equation*}
\calQ'_3(\zeta_3) \calQ'_2(\zeta_2) = - ((\tr \circ ((\rho^{\scriptscriptstyle ++}_3)_{\zeta_3} \otimes_\Delta (\rho^{\scriptscriptstyle -+}_2)_{\zeta_2})) \otimes \id) (\calR (t \otimes 1)),
\end{equation*}
where\footnote{We use the notation $\otimes_\Delta$ to distinguish between the tensor product of representations and the usual tensor product of mappings.}
\begin{equation*}
(\rho^{\scriptscriptstyle ++}_3)_{\zeta_3} \otimes_\Delta (\rho^{\scriptscriptstyle -+}_2)_{\zeta_2} = ((\rho^{\scriptscriptstyle ++}_3)_{\zeta_3} \otimes (\rho^{\scriptscriptstyle -+}_2)_{\zeta_2})\circ \Delta,
\end{equation*}
see, for example, our paper~\cite{BooGoeKluNirRaz14a}.

It is demonstrated in appendix \ref{a:tprhorho} that the $\uqbp$-module$(W^{\scriptscriptstyle ++}_3)_{\zeta_3} \otimes (W^{\scriptscriptstyle -+}_2)_{\zeta_2}$ has a basis $\{w^k_n\}$ such that
\begin{align*}
q^{\nu h_0} w^k_n &= q^{\nu (n_1 + 2 n_2 + n_3 + k + 1)} w^k_n, \\*
q^{\nu h_1} w^k_n &= q^{\nu (- 2 n_1 - n_2 + n_3 + k + 1)} w^k_n, \\*
q^{\nu h_2} w^k_n &= q^{\nu (n_1 - n_2 - 2 n_3 - 2 k - 2)} w^k_n, \\
e_0 w_n^k &= q^{- n_1} w_{n + \varepsilon_2}^k, \\
e_1 w^k_n &= - q [n_2]_q w^k_{n - \varepsilon_2 + \varepsilon_3} + \zeta_2^s \kappa_q^{-1} q^{- n_1 + 2 n_3 + 1} [n_1]_q w^k_{n - \varepsilon_1} \\*
& \hspace{14em} {} + \zeta_2^{s_2} \kappa_q \, q^{- n_1 + 2 n_3 + 2} [n_1]_q [k]_q w^{k - 1}_{n -\varepsilon_1 + \varepsilon_3}, \\
e_2 w^k_n &= (q^{- n_3} \zeta_3^s - q^{n_3} \zeta_2^s) \kappa_q^{-1} q^{n_1 - n_2 - 1} [n_3]_q w^k_{n - \varepsilon_3} \\*
& \hspace{8em} {} + q^{n_1 - n_2 + 1} [n_2]_q w^k_{n + \varepsilon_1 - \varepsilon_2} - \zeta_2^{s_2} q^{n_1 - n_2 + 2 n_3}[k]_q w^{k - 1}_n.
\end{align*}
Here $k \in \bbZ_+$ and $n$ means a triple $(n_1, n_2, n_3) \in \bbZ_+^3$. We see that there is an increasing filtration of $\uqbp$-submodules
\begin{multline*}
\{0\} = ((W^{\scriptscriptstyle ++}_3)_{\zeta_3} \otimes (W^{\scriptscriptstyle -+}_2)_{\zeta_2})_{-1} \subset ((W^{\scriptscriptstyle ++}_3)_{\zeta_3} \otimes (W^{\scriptscriptstyle -+}_2)_{\zeta_2})_{0} \\ \subset ((W^{\scriptscriptstyle ++}_3)_{\zeta_3} \otimes (W^{\scriptscriptstyle -+}_2)_{\zeta_2})_{1} \subset \cdots,
\end{multline*}
where
\begin{equation*}
((W^{\scriptscriptstyle ++}_3)_{\zeta_3} \otimes (W^{\scriptscriptstyle -+}_2)_{\zeta_2})_k = \bigoplus_{\ell = 0}^ k \bigoplus_{n_1, n_2, n_3 = 0}^\infty \bbC w_n^\ell.
\end{equation*}
It is worth to note here that
\begin{equation*}
\bigcup_{k = - 1}^\infty ((W^{\scriptscriptstyle ++}_3)_{\zeta_3} \otimes (W^{\scriptscriptstyle -+}_2)_{\zeta_2})_k = (W^{\scriptscriptstyle ++}_3)_{\zeta_3} \otimes (W^{\scriptscriptstyle -+}_2)_{\zeta_2}.
\end{equation*}

Define now a $\uqbp$-module $(W_{32})_{\zeta_3, \zeta_2}$ as a free vector space generated by the set $\{v_n\}$, $n = (n_1, n_2, n_3) \in \bbZ_+^3$, with the following action of the generators
\begin{align*}
q^{\nu h_0} v_n &= q^{\nu (n_1 + 2 n_2 + n_3 + 1)} v_n, \\*
q^{\nu h_1} v_n &= q^{\nu (- 2 n_1 - n_2 + n_3 + 1)} v_n, \\*
q^{\nu h_2} v_n &= q^{\nu (n_1 - n_2 - 2 n_3 - 2)} v_n, \\
e_0 v_n &= q^{- n_1} v_{n + \varepsilon_2}, \\*
e_1 v_n &= - q [n_2]_q v_{n - \varepsilon_2 + \varepsilon_3} + \zeta_2^s \kappa_q^{-1} q^{- n_1 + 2 n_3 + 1} [n_1]_q v_{n - \varepsilon_1}, \\*
e_2 v_n &= (q^{- n_3} \zeta_3^s - q^{n_3} \zeta_2^s) \kappa_q^{-1} q^{n_1 - n_2 - 1} [n_3]_q v_{n - \varepsilon_3} + q^{n_1 - n_2 + 1} [n_2]_q v_{n + \varepsilon_1 - \varepsilon_2},
\end{align*}
and denote by $(\rho_{32})_{\zeta_3, \zeta_2}$ the corresponding representation of $\uqbp$. It is clear that
\begin{equation*}
((W^{\scriptscriptstyle ++}_3)_{\zeta_3} \otimes (W^{\scriptscriptstyle -+}_2)_{\zeta_2})_k / ((W^{\scriptscriptstyle ++}_3)_{\zeta_3} \otimes (W^{\scriptscriptstyle -+}_2)_{\zeta_2})_{k - 1} \cong (W_{32})_{\zeta_3, \zeta_2}[\xi_k],
\end{equation*}
where the elements $\xi_k \in \widetilde \gothh^*$ are defined by the relations
\begin{equation*}
\xi_k(h_0) = k, \qquad \xi_k(h_1) = k, \qquad \xi_k(h_2) = - 2 k.
\end{equation*}
Hence, we have
\begin{equation*}
\calQ'_3(\zeta_3) \calQ'_2(\zeta_2) = - \sum_{k = 0}^\infty ((\tr \circ (\rho_{32})_{\zeta_3, \zeta_2}[\xi_k]) \otimes \id)(\calR(t \otimes 1)),
\end{equation*}
and a relation similar to (\ref{qshq}) gives
\begin{multline}
\calQ'_3(\zeta_3) \calQ'_2(\zeta_2) = - ((\tr \circ (\rho_{32})_{\zeta_3, \zeta_2}) \otimes \id)(\calR(t \otimes 1)) (1 - q^{(h'_0 + h'_1 - 2 h'_2)/3})^{-1} \\
= - ((\tr \circ (\rho_{32})_{\zeta_3, \zeta_2}) \otimes \id)(\calR(t \otimes 1)) (1 - q^{- h'_2})^{-1}. \label{q1q2}
\end{multline}
The last equation follows from the relation
\begin{equation*}
h'_0 + h'_1 + h'_2 = 0.
\end{equation*}

When $\zeta_3 = q^{2/s} \zeta_2 = \zeta$ the $\uqbp$-module $(W_{32})_{\zeta_3, \zeta_2} = (W_{32})_{\zeta, q^{-2/s} \zeta}$ has a $\uqbp$-submodule formed by the basis vectors $v_n$ with $n_3 > 0$. This submodule is isomorphic to the shifted $\uqbp$-module $(W_{32})_{q^{-2/s} \zeta, \zeta}[\xi]$, where
\begin{equation}
\xi(h_0) = 1, \qquad \xi(h_1) = 1, \qquad \xi(h_2) = - 2. \label{xi}
\end{equation}
It is evident that the quotient module $(W_{32})_{\zeta, q^{-2/s} \zeta} / (W_{32})_{q^{-2/s} \zeta, \zeta}$ is isomorphic to the $\uqbp$-module defined by the relations
\begin{gather*}
q^{\nu h_0} v_n = q^{\nu (n_1 + 2 n_2 + 1)} v_n, \quad q^{\nu h_1} v_n = q^{\nu (- 2 n_1 - n_2 + 1)} v_n, \quad
q^{\nu h_2} v_n = q^{\nu (n_1 - n_2 - 2)} v_n, \\
e_0 v_n = q^{- n_1} v_{n + \varepsilon_2}, \\
e_1 v_n = \zeta^s \kappa_q^{-1} q^{- n_1 - 1} [n_1]_q v_{n - \varepsilon_1}, \quad
e_2 v_n = q^{n_1 - n_2 + 1} [n_2]_q v_{n + \varepsilon_1 - \varepsilon_2},
\end{gather*}
where $n$ denotes the pair $(n_1, n_2) \in \bbZ_+^2$. Introduce a new basis formed by the vectors
\begin{equation*}
w_{(n_1, n_2)} = c_1^{n_1} c_2^{n_2} v_{(n_2, n_1)},
\end{equation*}
where
\begin{equation*}
c_1 = r_s^{- s_0} q^{s_0/s} \zeta^{-s_0}, \qquad c_2 = r_s^{s_1} q^{1 - s_1/s} \zeta^{-s_0 - s_2}.
\end{equation*}
For this basis we have
\begin{gather*}
q^{\nu h_0} w_n = q^{\nu (2 n_1 + n_2 + 1)} w_n, \quad q^{\nu h_1} w_n = q^{\nu (- n_1 - 2 n_2 + 1)} w_n, \quad
q^{\nu h_2} w_n = q^{\nu (- n_1 + n_2 - 2)} w_n, \\
e_0 w_n = \widetilde \zeta^{s_0} q^{- n_2} w_{n + \varepsilon_1}, \\*
e_1 w_n = \widetilde \zeta^{s_1} \kappa_q^{-1} q^{- n_2} [n_2]_q w_{n - \varepsilon_2}, \quad
e_2 w_n = - \widetilde \zeta^{s_2} q^{- n_1 + n_2 + 1} [n_1]_q w_{n - \varepsilon_1 + \varepsilon_2},
\end{gather*}
where\footnote{Recall that we denote by $r_s$ some fixed $s$th root of $-1$.}
\begin{equation*}
\widetilde \zeta = r_s q^{- 1/s} \zeta.
\end{equation*}
Thus, the quotient module $(W_{32})_{\zeta, q^{-2/s} \zeta} / (W_{32})_{q^{-2/s} \zeta, \zeta}$ is isomorphic to the $\uqbp$-module $(\oW_1)_{\widetilde \zeta}[\xi]$, where $\xi$ is again given by relations (\ref{xi}). We see that
\begin{multline*}
((\tr \circ (\rho_{32})_{\zeta, q^{-2/s} \zeta}) \otimes \id)(\calR(t \otimes 1))
= ((\tr \circ (\rho_{32})_{q^{-2/s} \zeta, \zeta}[\xi]) \otimes \id)(\calR(t \otimes 1)) \\+ (\tr \circ (\orho_1)_{r_s q^{-1/s} \zeta}[\xi]) \otimes \id)(\calR(t \otimes 1)).
\end{multline*}
Remembering equation (\ref{q1q2}), we obtain\footnote{Recall that our universal $Q$-operators depends on $\zeta$ via $\zeta^s$. Therefore, the expression $\ocalQ{}'_1(r_s \zeta)$ does not depend on the choice of $r_s$. Recall also that ${\orho}_1 = \rho \circ \tau$.}
\begin{equation*}
\calQ'_3(q^{1/s} \zeta) \calQ'_2(q^{-1/s} \zeta) - \calQ'_3(q^{-1/s} \zeta) \calQ'_2(q^{1/s} \zeta) q^{- h'_2} = {} - \ocalQ'_1(r_s \zeta) \, q^{- h'_2} (1 - q^{- h'_2})^{-1}.
\end{equation*}
The relations for other pairs of the universal $Q$-operators $\calQ{}'_i(\zeta)$ can be be obtained from this relation with the help of equation (\ref{qsq}). Further, using (\ref{bqtq}), we come to the equation 
\begin{equation*}
\ocalQ'_1(q^{1/s} \zeta) \ocalQ'_2(q^{-1/s} \zeta) - \ocalQ'_1(q^{-1/s} \zeta) \ocalQ'_2(q^{1/s} \zeta) q^{- h'_1} = {} - \calQ'_3(r_s \zeta) \, q^{- h'_1} (1 - q^{- h'_1})^{-1}.
\end{equation*}
The relations for other pairs of the universal $Q$-operators $\ocalQ{}'_i(\zeta)$ can be be obtained from this relation with the help of equation (\ref{qsq}).

Now we redefine the universal $Q$-operators as
\begin{equation}
\calQ_i(\zeta) = \zeta^{\calD_i} \, \calQ'_i(\zeta), \qquad \ocalQ_i(\zeta) = \zeta^{- \calD_i} \, \ocalQ'_i(r_s \zeta), \label{rdq}
\end{equation}
where
\begin{gather*}
\calD_1 = (h'_0 - h'_1) \, s / 6, \qquad \calD_2 = (h'_1 - h'_2) \, s / 6, \qquad \calD_3 = (h'_2 - h'_0) \, s / 6.
\end{gather*}
It is worth to note here that
\begin{equation*}
\calD_1 + \calD_2 + \calD_3 = 0.
\end{equation*}
The commutativity properties of the universal $Q$-operators $\calQ_i(\zeta)$, $\ocalQ_i(\zeta)$ are the same as of $\calQ'_i(\zeta)$, $\ocalQ'_i(\zeta)$. After the above redefinition, the functional relations for the pairs of the universal $Q$-operators take the universal determinant form
\begin{align*}
\calC_i \ocalQ_i(\zeta) & = \calQ_j(q^{1/s} \zeta) \calQ_k(q^{- 1/s} \zeta) - \calQ_j(q^{-1/s} \zeta) \calQ_k(q^{1/s} \zeta), \\
\calC_i \calQ_i(\zeta) & = \ocalQ_j(q^{- 1/s} \zeta) \ocalQ_k(q^{1/s} \zeta) - \ocalQ_j(q^{1/s} \zeta) \ocalQ_k(q^{- 1/s} \zeta),
\end{align*}
where
\begin{equation*}
\calC_i = q^{- \calD_i / s} (q^{2 \calD_j / s} - q^{2 \calD_k / s})^{-1},
\end{equation*}
and $(i,\, j, \, k)$ is a cyclic permutation of $(1, \, 2, \, 3)$.

It is easy to see that
\begin{equation*}
\calD_{i+1} = \sigma(\calD_i)|_{\phi \to \sigma(\phi)}.
\end{equation*}
Therefore, one has
\begin{equation}
\calC_{i+1} = \sigma(\calC_i)|_{\phi \to \sigma(\phi)}. \label{cipo}
\end{equation}
Here and below we assume that the definition of $\calC_i$ and $\calD_i$ is extended to arbitrary integer values of the index $i$ periodically with the period $3$.
It follows from relation (\ref{cipo}) and equation (\ref{qsq}) that
\begin{equation}
\calQ_{i + 1}(\zeta) = \sigma(\calQ_i(\zeta))|_{\phi \to \sigma(\phi)}, \qquad \ocalQ{}_{i + 1}(\zeta) = \sigma(\ocalQ{}_i(\zeta))|_{\phi \to \sigma(\phi)}. \label{qsqa}
\end{equation}
One can also verify that
\begin{equation*}
\calD_i = - \tau(\calD_{- i + 1})|_{\phi \to \tau(\phi)}
\end{equation*}
and
\begin{equation}
\calC_i = \tau(\calC_{- i + 1})|_{\phi \to \tau(\phi)}. \label{ctc}
\end{equation}
This relation, together with equation (\ref{bqtq}), leads to
\begin{equation}
\ocalQ{}_i(\zeta) = \tau(\calQ_{- i + 1}(r_s \zeta))|_{\phi \to \tau(\phi)}. \label{qtq}
\end{equation}

\subsection{\texorpdfstring{Product of three universal $Q$-operators}{Product of three universal Q-operators}} \label{s:ptqo}

A convenient way to analyse the product of three universal $Q$-operators is to start with the product of $\calQ'_3(\zeta_3)$, $\calQ'_2(\zeta_2)$ and $\calQ'_1(\zeta_1)$. To construct $\calQ'_3(\zeta_3)$ and $\calQ'_2(\zeta_2)$ we use (\ref{qp3}) and (\ref{qp2}), and for $\calQ'_1(\zeta_1)$ the equation
\begin{equation*}
\calQ'_1(\zeta_1) =  ((\tr \circ (\rho^{\scriptscriptstyle --}_1)_{\zeta_1}) \otimes \id)(\calR (t \otimes 1)).
\end{equation*}
Similarly as for the case of two universal $Q$-opertsors, one can show~\cite{BooGoeKluNirRaz14a} that
\begin{equation*}
\calQ'_3(\zeta_3) \calQ'_2(\zeta_2) \calQ'_1(\zeta_1) = - ((\tr \circ ((\rho^{\scriptscriptstyle ++}_3)_{\zeta_3} \otimes_\Delta (\rho^{\scriptscriptstyle -+}_2)_{\zeta_2} \otimes_\Delta (\rho^{\scriptscriptstyle --}_1)_{\zeta_1})) \otimes \id) (\calR (t \otimes 1)).
\end{equation*}
Hence, we have to analyse the tensor product of the representations $(\rho^{\scriptscriptstyle ++}_3)_{\zeta_3}$, $(\rho^{\scriptscriptstyle -+}_2)_{\zeta_2}$ and $(\rho^{\scriptscriptstyle --}_1)_{\zeta_1}$.

It is demonstrated in appendix \ref{a:tprhorhorho} that the $\uqbp$-module
\begin{equation*}
W_{\zeta_3, \zeta_2, \zeta_1} = (W^{\scriptscriptstyle ++}_3)_{\zeta_3} \otimes (W^{\scriptscriptstyle -+}_2)_{\zeta_2} \otimes (W^{\scriptscriptstyle --}_1)_{\zeta_1}
\end{equation*}
has a basis $\{w^k_n\}$ such that
\begin{align}
q^{\nu h_0} w^k_n & = q^{\nu(n_1 + 2 n_2 + n_3 + k_1 + k_2 + 2 k_3 + 4)} w^k_n, \label{th0v} \\*
q^{\nu h_1} w^k_n & = q^{\nu(-2 n_1 - n_2 + n_3 + k_1 - 2 k_2 - k_3 - 2)} w^k_n, \\*
q^{\nu h_2} w^k_n & = q^{\nu(n_1 - n_2 - 2 n_3 - 2 k_1 + k_2 - k_3 - 2)} w^k_n, \\
e_0 w^k_n & = \zeta^{s_0} q^{- \lambda_1 - \lambda_3 - n_3} w^k_{n + \varepsilon_2},  \label{te0v} \\
e_1 w^k_n & = \zeta^{s_1 - s} \kappa_q^{-1} q^{\lambda_1 + \lambda_2}(q^{- n_1 - n_2 + n_3 + 1} \zeta_2^s - q^{n_1 + n_2 - n_3 + 1} \zeta_1^s) [n_1]_q w^k_{n - \varepsilon_1} \notag \\*
& - \zeta^{s_1} q^{\mu_1 - n_2 + n_3 + 2} [n_2]_q w^k_{n - \varepsilon_2 + \varepsilon_3} - \zeta_1^{s_1} q^{2 n_1 + n_2 - n_3 - k_1 + k_3} [k_2]_q w^{k - \varepsilon_2}_n \notag \\
& - \zeta^{s_1 - s_2} \zeta_1^{s_2} \kappa_q q^{\mu_1 + \mu_2 - 2 n_2 + n_3 + 2 k_1 + k_2 - k_3 + 4} [n_1]_q [k_3]_q w^{k + \varepsilon_2 - \varepsilon_3}_{n - \varepsilon_1 + \varepsilon_3} \notag \\
& + \zeta^{s_1 - s_2} \zeta_2^{s_2} \kappa_q q^{\mu_1 + \mu_2 - 2 n_2 + n_3 + 1} [n_1]_q [k_1]_q w^{k - \varepsilon_1}_{n - \varepsilon_1 + \varepsilon_3} \notag \\
& - \zeta^{-s_2} \zeta_1^{s_1 + s_2} \kappa_q q^{\mu_2 + 2 n_1 + n_2 - n_3 + k_1 + 1} [n_1]_q [k_3]_q w^{k - \varepsilon_3}_{n - \varepsilon_1 + \varepsilon_2}, \\
e_2 w^k_n & = \zeta^{s_2 - s} \kappa_q^{-1} q^{\lambda_2 + \lambda_3} (q^{- n_3 - 1} \zeta_3^s - q^{n_3 - 1} \zeta_2^s) [n_3]_q w^k_{n - \varepsilon_3} \notag \\
& + \zeta^{s_2} q^{- \mu_2 + 2 n_3} [n_2]_q w^k_{n + \varepsilon_1 - \varepsilon_2} - \zeta_2^{s_2} q^{n_1 - n_2 + 2 n_3} [k_1]_q w^{k - \varepsilon_1}_n \notag \\
& + \zeta_1^{s_2} q^{n_1 - n_2 + 2 n_3 + 2 k_1 + k_2 - k_3 + 3} [k_3]_q w^{k + \varepsilon_2 - \varepsilon_3}_n. \label{te2v}
\end{align}
Here $n$ and $k$ mean triples $(n_1, n_2, n_3)$ and $(k_1, k_2, k_3)$ of non-negative integers. Introduce for the triples $k$ the colexicographic total order assuming that $k' \le k$ if and only if $k'_3 \le k_3$, or $k'_3 = k_3$ and $k'_2 \le k_2$, or $k'_3 = k_3$, $k'_2 = k_2$ and $k'_1 \le k_1$. We see that there is an increasing filtration on $W_{\zeta_3, \zeta_2, \zeta_1}$ formed by the $\uqbp$-submodules
\begin{equation*}
(W_{\zeta_3, \zeta_2, \zeta_1})_k = \bigoplus_{\ell \le k} \bigoplus_n \bbC w_n^\ell.
\end{equation*}
Putting
\begin{equation*}
\zeta_1 = q^{- 2(\lambda_1 + 1) / s} \zeta, \qquad \zeta_2 = q^{- 2 \lambda_2 / s} \zeta, \qquad \zeta_3 = q^{- 2(\lambda_3 - 1) / s} \zeta,
\end{equation*}
one discovers that in this case
\begin{equation*}
(W_{\zeta_3, \zeta_2 \zeta_1})_k / \bigcup_{\ell < k} (W_{\zeta_3, \zeta_2, \zeta_1})_\ell \cong (\widetilde V^\lambda)_\zeta[\xi_k],
\end{equation*}
where
\begin{gather*}
\xi_k(h_0) = \mu_1 + \mu_2 + k_1 + k_2 + 2 k_3 + 4, \\
\xi_k(h_1) = - \mu_1 + k_1 - 2 k_2 - k_3 - 2, \qquad \xi_k(h_2) = - \mu_2 - 2 k_1 + k_2 - k_3 - 2.
\end{gather*}
Now, simple calculations lead to the following result
\begin{equation}
\calC \, \widetilde \calT^\lambda(\zeta) = \calQ_1(q^{- 2 (\lambda_1 + 1)/s} \zeta) \calQ_2(q^{- 2 \lambda_2/s} \zeta) \calQ_3(q^{- 2 (\lambda_3 - 1)/s} \zeta), \label{ctt}
\end{equation}
where
\begin{equation*}
\calC = \calC_1 \calC_2 \calC_3 = (q^{2\calD_1/s} - q^{2\calD_2/s})^{-1} (q^{2\calD_2/s} - q^{2\calD_3/s})^{-1} (q^{2\calD_3/s} - q^{2\calD_1/s})^{-1}.
\end{equation*}
It is instructive to rewrite (\ref{ctt}) as
\begin{equation}
\calC \, \widetilde \calT^\lambda(\zeta) = \calQ_1(q^{- 2 (\lambda + \rho)_1/s} \zeta) \calQ_2(q^{- 2 (\lambda + \rho)_2/s} \zeta) \calQ_3(q^{- 2 (\lambda + \rho)_3/s} \zeta), \label{tqqq}
\end{equation}
where $\rho = \alpha_1 + \alpha_2$.

It follows from equation (\ref{ctc}) that
\begin{equation*}
\calC = \tau(\calC)|_{\phi \to \tau(\phi)}.
\end{equation*}
Using this relation, equation (\ref{ttt}) and equation (\ref{qtq}), we find
\begin{equation*}
\calC \, \widetilde{\ocalT}{}'^{(- \lambda_3, \, - \lambda_2, \, - \lambda_1)}(r_s \zeta) = \ocalQ_1(q^{2 (\lambda + \rho)_1/s} \zeta) \ocalQ_2(q^{2 (\lambda + \rho)_2/s} \zeta) \ocalQ_3(q^{2 (\lambda + \rho)_3/s} \zeta).
\end{equation*}
Hence, if we define
\begin{equation*}
\widetilde{\ocalT}{}^{(\lambda_1, \, \lambda_2, \, \lambda_3)}_i(\zeta) = \widetilde{\ocalT}{}'^{(- \lambda_3, \, - \lambda_2, \, - \lambda_1)}_i(r_s \zeta),
\end{equation*}
we obtain
\begin{equation}
\calC \, \widetilde{\ocalT}{}^\lambda(\zeta) = \ocalQ_1(q^{2 (\lambda + \rho)_1/s} \zeta) \ocalQ_2(q^{2 (\lambda + \rho)_2/s} \zeta) \ocalQ_3(q^{2 (\lambda + \rho)_3/s} \zeta). \label{btbqbqbq}
\end{equation}

Consider an automorphism of $\uqgliii$, similar to the automorphism $\tau$ of $\uqlsliii$, which we also denote by $\tau$. This automorphism is given by the equations 
\begin{align}
& \tau(q^{\nu G_1}) = q^{- \nu G_3}, &&  \tau(q^{\nu G_2}) = q^{- \nu G_2}, &&  \tau(q^{\nu G_3}) = q^{- \nu G_1}, \label{tga} \\
& \tau(E_1) = E_2, && \tau(E_2) = E_1, &
& \tau(F_1) = F_2, && \tau(F_2) = F_1. \label{tgb}
\end{align}
It is easy to understand that there are the isomorphisms of representations
\begin{equation*}
\widetilde \pi^{(\lambda_1, \, \lambda_2, \, \lambda_3)} \circ \tau \cong \widetilde \pi^{(- \lambda_3, \, - \lambda_2, \, - \lambda_1)}, \qquad \pi^{(\lambda_1, \, \lambda_2, \, \lambda_3)} \circ \tau \cong \pi^{(- \lambda_3, \, - \lambda_2, \, - \lambda_1)}.
\end{equation*}

Now, defining the universal monodromy operators $\ocalM$
\begin{equation}
\ocalM_i(\zeta) = ((\ovarphi_i)_{r_s \zeta} \otimes \id)(\calR), \qquad \widetilde{\ocalM}{}_i^\lambda(\zeta) = ((\widetilde{\ovarphi}{}_i^\lambda)_{r_s \zeta} \otimes \id)(\calR), \label{bmiz}
\end{equation}
where\footnote{Here the first appearance of $\tau$ means the automorphism of $\uqgliii$ defined by equations (\ref{tga}), (\ref{tgb}), and the second one the automorphism of $\uqlsliii$ defined by (\ref{taue})--(\ref{tauh}).} 
\begin{equation*}
\ovarphi_i = \tau \circ \varphi \circ \tau \circ \sigma^{- i + 1}
\end{equation*}
and $\widetilde{\ovarphi}{}_i^\lambda = \widetilde \pi^\lambda \circ \ovarphi_i$, we see that
\begin{equation*}
\widetilde \ocalT{}^\lambda_i(\zeta) = (\tr^\lambda \otimes \id)(\ocalM{}_i(\zeta) (({\ovarphi}_i)_{r_s \zeta}(t) \otimes 1)) = (\tr \otimes \id)(\widetilde \ocalM{}^\lambda_i(\zeta) ((\widetilde {\ovarphi}_i^\lambda)_{r_s \zeta}(t) \otimes 1)).
\end{equation*}
For the finite dimensional representations $\pi^\lambda$ we define
\begin{equation*}
\ocalT{}^{(\lambda_1, \, \lambda_2, \, \lambda_3)}_i(\zeta) = \ocalT{}'^{(- \lambda_3, \, - \lambda_2, \, - \lambda_1)}_i(r_s \zeta).
\end{equation*}
These universal transfer operators are expressed via the universal monodromy operators $\ocalM_i(\zeta)$ or
\begin{equation*}
\ocalM{}_i^\lambda(\zeta) = ((\ovarphi{}_i^\lambda)_{r_s \zeta} \otimes \id)(\calR),
\end{equation*}
where $\ovarphi{}_i^\lambda = \pi^\lambda \circ \ovarphi_i$, as
\begin{equation*}
\ocalT{}^\lambda_i(\zeta) = (\tr^\lambda \otimes \id)(\ocalM{}_i(\zeta) ((\ovarphi_i)_{r_s \zeta}(t) \otimes 1)) = (\tr \otimes \id)(\ocalM{}^\lambda_i(\zeta) ((\ovarphi_i^\lambda)_{r_s \zeta}(t) \otimes 1)).
\end{equation*}
More explicitly, one can write
\begin{align*}
& \widetilde{\ocalT}{}_i^\lambda(\zeta) = ((\widetilde \tr{}^\lambda \circ (\ovarphi_i)_{r_s \zeta}) \otimes \id)(\calR (t \otimes 1)) = ((\tr \circ (\widetilde {\ovarphi}{}_i^\lambda)_{r_s \zeta}) \otimes \id)(\calR (t \otimes 1)), \\
& \ocalT{}_i^\lambda(\zeta) = ((\tr{}^\lambda \circ (\ovarphi_i)_{r_s \zeta}) \otimes \id)(\calR (t \otimes 1)) = ((\tr \circ (\ovarphi{}_i^\lambda)_{r_s \zeta}) \otimes \id)(\calR (t \otimes 1)).
\end{align*}

Using relation (\ref{twtt}) which follows from the Berstein--Gelfand--Gelfand resolution, we come to the determinant representations
\begin{align}
& \calC \, \calT^\lambda(\zeta) = \det \left( \calQ_i(q^{- 2 (\lambda + \rho)_j / s} \zeta) \right)_{i,\, j = 1,2,3}, \label{tdq1} \\
& \calC \, \ocalT^\lambda(\zeta) = \det \left( \ocalQ_i(q^{2 (\lambda + \rho)_j /s } \zeta) \right)_{i, \, j = 1,2,3}. \label{tdq2}
\end{align}
Note that if any two of the components of $\lambda + \rho$ coincide, then $\calT^\lambda(\zeta) = 0$ and $\ocalT^\lambda(\zeta) = 0$.

We use relations (\ref{tdq1}) and (\ref{tdq2}) to define $\calT^\lambda(\zeta)$ and $\ocalT^\lambda(\zeta)$ for arbitrary $\lambda \in \gothg^*$. It is useful to have in mind that with such definition
\begin{equation}
\calT^{p(\lambda + \rho) - \rho}(\zeta) = \sgn(p) \calT^\lambda(\zeta), \qquad \ocalT^{p(\lambda + \rho) - \rho}(\zeta) = \sgn(p) \ocalT^\lambda(\zeta) \label{**}
\end{equation}
for any permutation $p \in \mathrm S_3$.

In fact the universal transfer operators $\widetilde \calT_i^\lambda(\zeta)$ and $\widetilde \ocalT{}_i^\lambda(\zeta)$ are not independent. In particular, using (\ref{tst}), one obtains from (\ref{tqqq}) and (\ref{btbqbqbq}) the equations
\begin{align*}
& \widetilde \calT_2^{(\lambda_1, \, \lambda_2, \, \lambda_3)}(\zeta) = \widetilde \calT_1^{(\lambda_3 - 2, \, \lambda_1 + 1, \, \lambda_2 + 1)}(\zeta), && \widetilde \calT_3^{(\lambda_1, \, \lambda_2, \, \lambda_3)}(\zeta) = \widetilde \calT_1^{(\lambda_2 - 1, \, \lambda_3 - 1, \, \lambda_1 + 2)}(\zeta), \\
& \widetilde{\ocalT}{}_2^{(\lambda_1, \, \lambda_2, \, \lambda_3)}(\zeta) = \widetilde{\ocalT}{}_1^{(\lambda_3 - 2, \, \lambda_1 + 1, \, \lambda_2 + 1)}(\zeta), && \widetilde{ \ocalT}{}_3^{(\lambda_1, \, \lambda_2, \, \lambda_3)}(\zeta) = \widetilde{\ocalT}{}_1^{(\lambda_2 - 1, \, \lambda_3 - 1, \, \lambda_1 + 2)}(\zeta).
\end{align*}
The same is true for the universal transfer operators $\calT^\lambda_i(\zeta)$ and $\ocalT{}^\lambda_i(\zeta)$. Therefore, we use below only the universal transfer operators $\widetilde \calT^\lambda(\zeta)$, $\widetilde{\ocalT}{}^\lambda(\zeta)$ and $\calT^\lambda(\zeta)$, $\ocalT{}^\lambda(\zeta)$. For the operators  $\widetilde \calT^\lambda(\zeta)$ and $\widetilde{\ocalT}{}^\lambda(\zeta)$ we also have
\begin{align*}
& \widetilde \calT^{(\lambda_1 + \nu, \, \lambda_2 + \nu, \, \lambda_3 + \nu)}(q^{2 \nu / s} \zeta) = \widetilde \calT^{(\lambda_1, \, \lambda_2, \, \lambda_3)}(\zeta), \\
& \widetilde \ocalT{}^{(\lambda_1 + \nu, \, \lambda_2 + \nu, \, \lambda_3 + \nu)}(q^{- 2 \nu / s} \zeta) = \widetilde \ocalT{}^{(\lambda_1, \, \lambda_2, \, \lambda_3)}(\zeta),
\end{align*}
where $\nu$ is an arbitrary complex number. The same relations are valid for the operators $\calT^\lambda(\zeta)$ and $\ocalT{}^\lambda(\zeta)$:
\begin{align}
& \calT^{(\lambda_1 + \nu, \, \lambda_2 + \nu, \, \lambda_3 + \nu)}(q^{2 \nu / s} \zeta) = \calT^{(\lambda_1, \, \lambda_2, \, \lambda_3)}(\zeta), \label{tls} \\
& \ocalT{}^{(\lambda_1 + \nu, \, \lambda_2 + \nu, \, \lambda_3 + \nu)}(q^{- 2 \nu / s} \zeta) = \ocalT{}^{(\lambda_1, \, \lambda_2, \, \lambda_3)}(\zeta). \label{btls}
\end{align}

\subsection{\texorpdfstring{$TQ$- and $TT$-relations}{TQ- and TT-relations}}

Let $\lambda_j$, $j = 1, \ldots, 4$, be arbitrary complex numbers. Define three four dimensional row-vectors $\calP_i(\zeta)$, $i = 1,2,3$, by the equation
\begin{equation*}
(\calP_i(\zeta))_j = \calQ_i(q^{-2 \lambda_j / s} \zeta), \qquad j = 1, \ldots, 4,
\end{equation*}
and construct three four-by-four matrices whose first three rows are the vectors $\calP_i(\zeta)$ and the last vector is the vector $\calP_k(\zeta)$ where $k$ is $1$, $2$ or $3$. The determinant of any of these matrices certainly vanish. Expanding it over the last row and taking into account (\ref{tdq1}), written as
\begin{equation}
\calC \, \calT^{\lambda - \rho}(\zeta) = \det \left( \calQ_i(q^{- 2 \lambda_j / s} \zeta) \right)_{i,\, j = 1,2,3}, \label{ctlmr}
\end{equation}
we obtain the equation
\begin{multline*}
\calT^{(\lambda_1 - 1, \, \lambda_2, \, \lambda_3 + 1)}(\zeta) \calQ_k(q^{- 2 \lambda_4 / s} \zeta) - \calT^{(\lambda_1 - 1, \, \lambda_2, \, \lambda_4 + 1)}(\zeta) \calQ_k(q^{- 2 \lambda_3 / s} \zeta) \\
+ \calT^{(\lambda_1 - 1, \, \lambda_3, \, \lambda_4 + 1)}(\zeta) \calQ_k(q^{- 2 \lambda_2 / s} \zeta) - \calT^{(\lambda_2 - 1, \, \lambda_3, \, \lambda_4 + 1)}(\zeta) \calQ_k(q^{- 2 \lambda_1 / s} \zeta) = 0.
\end{multline*}
We call this equation the universal $TQ$-relation. Assuming that
\begin{equation*}
\lambda_1 = 2, \qquad \lambda_2 = 1, \qquad \lambda_3 = 0, \qquad \lambda_4 = -1,
\end{equation*}
we obtain the relation
\begin{multline}
\calT^{(1, \, 1, \, 1)}(\zeta) \calQ_k(q^{2 / s} \zeta) - \calT^{(1, \, 1, \, 0)}(\zeta) \calQ_k(\zeta) \\
+ \calT^{(1, \, 0, \, 0)}(\zeta) \calQ_k(q^{- 2 / s} \zeta) - \calT^{(0, \, 0, \, 0)}(\zeta) \calQ_k(q^{- 4 / s} \zeta) = 0. \label{tqi}
\end{multline}
It follows from the structure of the universal $R$-matrix, see, for example, the paper \cite{TolKho92}, that $\calT^{(0, \, 0, \, 0)}(\zeta) = 1$. Therefore, as follows from (\ref{tls}), we have
\begin{equation*}
\calT^{(\nu, \, \nu, \, \nu)}(\zeta) = 1
\end{equation*}
for any $\nu \in \bbC$. This property leads to a simpler form of (\ref{tqi}):
\begin{equation}
\calQ_k(q^{2 / s} \zeta) - \calT^{(1, \, 1, \, 0)}(\zeta) \calQ_k(\zeta)
+ \calT^{(1, \, 0, \, 0)}(\zeta) \calQ_k(q^{- 2 / s} \zeta) -  \calQ_k(q^{- 4 / s} \zeta) = 0. \label{tq}
\end{equation}
In a similar way we obtain
\begin{equation}
\ocalQ_k(q^{- 2 / s} \zeta) - \ocalT^{(1, \, 1, \, 0)}(\zeta) \ocalQ_k(\zeta) + \ocalT^{(1, \, 0, \, 0)}(\zeta) \ocalQ_k(q^{2 / s} \zeta) -  \ocalQ_k(q^{4 / s} \zeta) = 0. \label{btq}
\end{equation}
Equations (\ref{tq}) and (\ref{btq}) are analogues of the famous Baxter's $TQ$-relations for the case under consideration. For the first time they were obtained by Pronko and Stroganov from the nested Bethe ansatz equations \cite{ProStr00}.

It is possible to obtain relations containing only $\calT^{(1, \, 0, \, 0)}(\zeta)$ or $\calT^{(1, \, 1, \, 0)}(\zeta)$, or their barred analogues. Here we follow the paper \cite{BazFraLukMenSta11}. First remind the Jacobi identity for determinants, see, for example, the book \cite{Hir04}. Let $D$ be the determinant of some square matrix. Denote by $D \left[ \begin{array}{c} i \\ j \end{array} \right]$ the determinant of the same matrix with the $i$th row and the $j$th column removed, and by $D \left[ \begin{array}{cc} i & k \\ j & \ell \end{array} \right]$ the determinant of that matrix with the $i$th and $k$th rows and the $j$th and $\ell$th columns removed. The Jacobi identity looks as
\begin{equation}
D \left[ \begin{array}{cc} i & j \\ i & j \end{array} \right] D = D \left[ \begin{array}{c} i \\ i \end{array} \right] D \left[ \begin{array}{c} j \\ j \end{array} \right] - D \left[ \begin{array}{c} i \\ j \end{array} \right] D \left[ \begin{array}{c} j \\ i \end{array} \right]. \label{ji}
\end{equation}
For $k = 1, 2, 3$ define
\begin{equation*}
\calH^k(\zeta_1, ..., \zeta_k) = \det (\calQ_i(\zeta_j))_{i, j = 1, \ldots, k} \, ,
\end{equation*}
and assume that $\calH^0 = 1$. Introduce the matrix
\begin{equation}
\left( \begin{array}{cccc}
0 & 1 & 0 & 0 \\
\calQ_1(\zeta_1) & \calQ_1(\zeta_2) & \calQ_1(\zeta_3) & \calQ_1(\zeta_4) \\
\calQ_2(\zeta_1) & \calQ_2(\zeta_2) & \calQ_2(\zeta_3) & \calQ_2(\zeta_4) \\
\calQ_3(\zeta_1) & \calQ_3(\zeta_2) & \calQ_3(\zeta_3) & \calQ_3(\zeta_4)
\end{array}
\right), \label{mq}
\end{equation}
and apply to it the Jacobi identity (\ref{ji}) with $i = 1$ and $j = 4$. This gives the relation
\begin{equation}
\calH^3(\zeta_1, \zeta_3, \zeta_4) \calH^2(\zeta_2, \zeta_3) = \calH^3(\zeta_2, \zeta_3, \zeta_4) \calH^2(\zeta_1, \zeta_3) + \calH^3(\zeta_1, \zeta_2, \zeta_3) \calH^2(\zeta_3, \zeta_4). \label{jia}
\end{equation}
Now apply the Jacobi identity (\ref{ji}) with $i = 1$ and $j = 3$ to the matrix obtained from the matrix (\ref{mq}) by removing the last row and the last column. We come to the relation
\begin{equation*}
\calH^2(\zeta_1, \zeta_3) \calH^1(\zeta_2) = \calH^2(\zeta_2, \zeta_3) \calH^1(\zeta_1) + \calH^2(\zeta_1, \zeta_2) \calH^1(\zeta_3).
\end{equation*}
Using this equation to exclude $\calH^2(\zeta_1, \zeta_3)$ from (\ref{jia}), we obtain
\begin{multline*}
\calH^3(\zeta_1, \zeta_3, \zeta_4) \calH^2(\zeta_2, \zeta_3) \calH^1(\zeta_2) = \calH^3(\zeta_2, \zeta_3, \zeta_4) \calH^2(\zeta_2, \zeta_3) \calH^1(\zeta_1) \\
+ \calH^3(\zeta_2, \zeta_3, \zeta_4) \calH^2(\zeta_1, \zeta_2) \calH^1(\zeta_3) + \calH^3(\zeta_1, \zeta_2, \zeta_3) \calH^2(\zeta_3, \zeta_4) \calH^1(\zeta_2).
\end{multline*}
Put in the above equation
\begin{equation*}
\zeta_1 = q^{- 4/s} \zeta, \qquad \zeta_2 = q^{- 2/s} \zeta, \qquad \zeta_3 = \zeta, \qquad \zeta_4 = q^{2/s} \zeta,
\end{equation*}
and take into account that
\begin{gather*}
\calH^3(q^{- 2(\lambda_1 + 1)/s} \zeta, \, q^{- 2\lambda_2/s} \zeta, \, q^{- 2(\lambda_3 - 1)/s} \zeta) = \calC \calT^\lambda(\zeta), \\
\calH^2(q^{-1/s} \zeta, \, q^{1/s} \zeta) = - \calC_3 \ocalQ_3(\zeta), \qquad \calH^1(\zeta) = \calQ_1(\zeta).
\end{gather*}
The resulting equation is
\begin{multline*}
\calT^{(1,0,0)}(\zeta) \calQ_1(q^{- 2/s}\zeta) {\ocalQ}_3(q^{- 1 / s} \zeta) =   \calQ_1(q^{-4/s}\zeta) {\ocalQ}_3(q^{- 1 /s} \zeta) \\[.3em]
+ \calQ_1(\zeta) {\ocalQ}_3(q^{- 3/s}\zeta) + \calQ_1(q^{- 2/s}\zeta) {\ocalQ}_3(q^{1/s}\zeta).
\end{multline*}
In fact, one can demonstrate that
\begin{multline}
\calT^{(1,0,0)}(\zeta) \calQ_i(q^{- 2/s}\zeta) {\ocalQ}_j(q^{- 1 / s} \zeta) =   \calQ_i(q^{-4/s}\zeta) {\ocalQ}_j(q^{- 1 /s} \zeta) \\[.3em]
+ \calQ_i(\zeta) {\ocalQ}_j(q^{- 3/s}\zeta) + \calQ_i(q^{- 2/s}\zeta) {\ocalQ}_j(q^{1/s}\zeta) \label{t100qq}
\end{multline}
for all $i \ne j$. Here one also uses (\ref{tsta}) and (\ref{qsqa}). Similarly, one obtains
\begin{multline}
\calT^{(1,1,0)}(\zeta) \calQ_i(\zeta) {\ocalQ}_j(q^{-1/s}\zeta)
= \calQ_i(q^{2/s}\zeta) {\ocalQ}_j(q^{-1/s}\zeta) \\[.3em] + \calQ_i(q^{-2/s}\zeta) {\ocalQ}_j(q^{1/s}\zeta) + \calQ_i(\zeta) 
{\ocalQ}_j(q^{-3/s}\zeta) \label{t110qq}
\end{multline}
again for all $i \ne j$.

Proceed now to $TT$-relations. Define the row-vector $\calP_4(\zeta)$ by the equation
\begin{equation*}
(\calP_4(\zeta))_j = \det \left( \begin{array}{ccc}
\calQ_1 (q^{-2 \lambda_j / s} \zeta) & \calQ_1 (q^{-2 \lambda_5 / s} \zeta) & \calQ_1 (q^{-2 \lambda_6 / s} \zeta) \\
\calQ_2 (q^{-2 \lambda_j / s} \zeta) & \calQ_2 (q^{-2 \lambda_5 / s} \zeta) & \calQ_2 (q^{-2 \lambda_6 / s} \zeta) \\
\calQ_3 (q^{-2 \lambda_j / s} \zeta) & \calQ_3 (q^{-2 \lambda_5 / s} \zeta) & \calQ_3 (q^{-2 \lambda_6 / s} \zeta) \\
\end{array} \right), \qquad j = 1, \ldots, 4,
\end{equation*}
where $\lambda_5$ and $\lambda_6$ are two additional complex numbers. It is easy to see that this vector is a linear combination of the vectors $\calP_1(\zeta)$, $\calP_2(\zeta)$ and $\calP_3(\zeta)$. Therefore, the determinant of the four-by-four matrix constructed from the vectors $\calP_i(\zeta)$, $i = 1, \ldots, 4$, vanishes.  Expanding it over the last row and again taking into account (\ref{ctlmr}), we come to the equation
\begin{multline}
\calT^{(\lambda_1 - 1, \, \lambda_2, \, \lambda_3 + 1)}(\zeta) \calT^{(\lambda_4 - 1, \, \lambda_5, \, \lambda_6 + 1)}(\zeta) \\- \calT^{(\lambda_1 - 1, \, \lambda_2, \, \lambda_4 + 1)}(\zeta) \calT^{(\lambda_3 - 1, \, \lambda_5, \, \lambda_6 + 1)}(\zeta) 
+ \calT^{(\lambda_1 - 1, \, \lambda_3, \, \lambda_4 + 1)}(\zeta) \calT^{(\lambda_2 - 1, \, \lambda_5, \, \lambda_6 + 1)}(\zeta) \\- \calT^{(\lambda_2 - 1, \, \lambda_3, \, \lambda_4 + 1)}(\zeta) \calT^{(\lambda_1 - 1, \, \lambda_5, \, \lambda_6 + 1)}(\zeta) = 0, \label{utt}
\end{multline}
which we call the universal $TT$-relation. Putting in (\ref{utt})
\begin{equation*}
\lambda_1 = \ell + 2, \qquad \lambda_2 = \ell + 1, \qquad \lambda_3 = 1, \qquad \lambda_4 = 0, \qquad \lambda_5 = 0, \qquad \lambda_6 = -1,
\end{equation*}
where $\ell$ is a positive integer, we obtain
\begin{equation}
\calT^{(\ell - 1, \, 0, \, 0)}(q^{-2/s} \zeta) \calT^{(\ell + 1, \, 0, \, 0)}(\zeta) = \calT^{(\ell, \, 0, \, 0)}(q^{-2/s} \zeta) \calT^{(\ell, \, 0, \, 0)}(\zeta) - \calT^{(\ell, \, \ell, \, 0)}(q^{-2/s} \zeta). \label{fr1}
\end{equation}
From the other hand the substitution
\begin{equation*}
\lambda_1 = \ell + 2, \qquad \lambda_2 = \ell, \qquad \lambda_3 = 0, \qquad \lambda_4 = \ell + 1, \qquad \lambda_5 = \ell + 1, \qquad \lambda_6 = -1,
\end{equation*}
where again $\ell$ is a positive integer, gives
\begin{equation}
\calT^{(\ell - 1, \, \ell - 1, \, 0)}(q^{-2/s} \zeta) \calT^{(\ell + 1, \, \ell + 1, \, 0)}(\zeta) = \calT^{(\ell, \, \ell, \, 0)}(q^{-2/s} \zeta) \calT^{(\ell, \, \ell, \, 0)}(\zeta) - \calT^{(\ell, \, 0, \, 0)}(\zeta). \label{fr2}
\end{equation}
Here the first relation of (\ref{**}) is used. Equations (\ref{fr1}) and (\ref{fr2}) are usually called the fusion relations, see \cite{KluPea92, KunNakSuz94, BazHibKho02}. They allow one to express the universal transfer operators $\calT^{(\ell, \, 0, \, 0)}(\zeta)$ and $\calT^{(\ell, \, \ell, 0)}(\zeta)$ via $\calT^{(1, \, 0, \, 0)}(\zeta)$ and $\calT^{(1, \, 1, \, 0)}(\zeta)$.

Further, for two positive integers $\ell_1$ and $\ell_2$, such that $\ell_1 \ge \ell_2$, putting in (\ref{utt})
\begin{equation*}
\lambda_1 = \ell_1 + 2, \qquad \lambda_2 = \ell_2 + 1, \qquad \lambda_3 = 1, \qquad \lambda_4 = 0, \qquad \lambda_5 = 0, \qquad \lambda_6 = -1,
\end{equation*}
we obtain
\begin{equation}
\calT^{(\ell_1, \, \ell_2, \, 0)}(\zeta) = \calT^{(\ell_1, \, 0, \, 0)}(\zeta) \calT^{(\ell_2, \, 0, \, 0)}(q^{2/s} \zeta) - \calT^{(\ell_1 + 1, \, 0, \, 0)}(q^{2/s} \zeta) \calT^{(\ell_2 - 1, \, 0, \, 0)}(\zeta). \label{fr3}
\end{equation}
Certainly, relation (\ref{fr1}) is a partial case of (\ref{fr3}). Thus, we see that $\calT^{(\ell_1, \, \ell_2, \, 0)}(\zeta)$ can be expressed via $\calT^{(\ell, \, 0, \, 0)}(\zeta)$ and, therefore, via $\calT^{(1, \, 0, \, 0)}(\zeta)$ and $\calT^{(1, \, 1, \, 0)}(\zeta)$. The explicit expression is given by the quantum Jacobi--Trudi identity \cite{BazRes90, Che87, BazHibKho02}
\begin{equation}
\calT^{(\ell_1, \, \ell_2, \, 0)}(\zeta) = \det \left( \calE_{\ell^t_i - i + j}(q^{- 2(j - 1)/s} \zeta) \right)_{1 \le i, \, j \le \ell_1}, \label{jt}
\end{equation}
where
\begin{equation*}
\calE_0(\zeta) = \calE_3(\zeta) = 1, \qquad \calE_1(\zeta) = \calT^{(1, \, 0, \, 0)}(\zeta), \qquad \calE_2(\zeta) = \calT^{(1, \, 1, \, 0)}(\zeta),
\end{equation*}
and $\calE_k(\zeta) = 0$ for $k < 0$ and $k > 3$. The numbers $\ell^t_i$ are defined as\footnote{One associates with the numbers $\ell_1$ and $\ell_2$ the Young diagram with the rows of length  $\ell_1$ and $\ell_2$. The numbers $\ell^t_i$ describe the rows of the transposed diagram.}
\begin{equation*}
\ell^t_i = 2, \quad 1 \le i \le \ell_2, \qquad \ell^t_i = 1, \quad \ell_2 < i \le \ell_1.
\end{equation*}
To prove identity (\ref{jt}) we put in (\ref{utt})
\begin{equation*}
\lambda_1 = 2, \qquad \lambda_2 = 1, \qquad \lambda_3 = 0, \qquad \lambda_4 = -1, \qquad \lambda_5 = \ell_1 + 1, \qquad \lambda_6 = \ell_2,
\end{equation*}
use the first relation of (\ref{**}), and come to the relation
\begin{multline}
\calT^{(\ell_1, \, \ell_2, \, 0)}(\zeta) = \calT^{(1, \, 1, \, 0)}(\zeta) \calT^{(\ell_1 - 1, \, \ell_2 - 1, \, 0)}(q^{-2/s} \zeta) \\
- \calT^{(1, \, 0, \, 0)}(\zeta) \calT^{(\ell_1 - 2, \, \ell_2 - 2, \, 0)}(q^{-4/s} \zeta) + \calT^{(\ell_1 - 3, \, \ell_2 - 3, \, 0)}(q^{-6/s} \zeta). \label{pjt1}
\end{multline}
It is worth to write a few partial cases of this relation in the form
\begin{align}
\calT^{(\ell, \, 2, \, 0)}(\zeta) & = \calT^{(1, \, 1, \, 0)}(\zeta) \calT^{(\ell - 1, \, 1, \, 0)}(q^{-2/s} \zeta) - \calT^{(1, \, 0, \, 0)}(\zeta) \calT^{(\ell - 2, \, 0, \, 0)}(q^{-4/s} \zeta), \label{pjt2} \\
\calT^{(\ell, \, 1, \, 0)}(\zeta) & = \calT^{(1, \, 1, \, 0)}(\zeta) \calT^{(\ell - 1, \, 0, \, 0)}(q^{-2/s} \zeta) - \calT^{(\ell - 2, \, 0, \, 0)}(q^{-4/s} \zeta), \label{pjt3} \\
\calT^{(\ell, \, 0, \, 0)}(\zeta) & = \calT^{(1, \, 0, \, 0)}(\zeta) \calT^{(\ell - 1, \, 0, \, 0)}(q^{-2/s} \zeta) - \calT^{(\ell - 1, \, 1, \, 0)}(q^{-2/s} \zeta). \label{pjt4}
\end{align}
Here we again use the first relation of (\ref{**}). As (\ref{fr1})--(\ref{fr3}), relations (\ref{pjt1})--(\ref{pjt4}) allow us to express $\calT^{(\ell_1, \, \ell_2, \, 0)}(\zeta)$ via $\calT^{(1, \, 0, \, 0)}(\zeta)$ and $\calT^{(1, \, 1, \, 0)}(\zeta)$. It is not difficult to see that (\ref{jt}) satisfies (\ref{pjt1})--(\ref{pjt4}). Thus, the identity (\ref{jt}) is true.

The $TT$-relations for the barred universal transfer operators are obtained by the change $q$ to $q^{-1}$ in the relations for the unbarred ones. 

A last remark is in order. It is clear that the action of the mappings $\varphi^{(\lambda_1, \, \lambda_2, \, \lambda_3)}$ and $ \ovarphi^{(\lambda_1, \, \lambda_2, \, \lambda_3)}$ on the generators of $\uqbp$ is the same except their action on the generator $e_0$. Using equations (\ref{pi100a})--(\ref{pi100d}) and (\ref{pi110a})--(\ref{pi110d}), we obtain the following relations
\begin{equation*}
(\ovarphi^{(1, \, 0, \, 0)})_\zeta(e_0) = - q^3 (\varphi^{(1, \, 0, \, 0)})_\zeta(e_0), \qquad (\ovarphi^{(1, \, 1, \, 0)})_\zeta(e_0) = - q (\varphi^{(1, \, 1, \, 0)})_\zeta(e_0).
\end{equation*}
Taking into account the discussion given at the end of section \ref{sss:uto}, we see that
\begin{equation}
\ocalT^{(1,\, 0, \, 0)}(\zeta) = \calT^{(1, \, 0, \, 0)}(q^{3/s} \zeta), \qquad \ocalT^{(1, \, 1, \, 0)}(\zeta) = \calT^{(1, \, 1, \, 0)}(q^{1/s} \zeta). \label{octct}
\end{equation}
Using these equations and the quantum Jacobi--Trudi identity (\ref{jt}) one can easily prove that
\begin{equation}
\ocalT^{(\ell, \, 0, \, 0)}(\zeta) = \calT^{(\ell, \, 0, \, 0)}(q^{(2 \ell + 1)/s} \zeta), \qquad \ocalT^{(\ell, \, \ell, \, 0)}(\zeta) = \calT^{(\ell, \, \ell, \, 0)}(q^{(2 \ell - 1)/s} \zeta). \label{bttot}
\end{equation}
The quantum Jacobi--Trudi identity for $\ocalT^{(\ell_1, \, \ell_2, \, 0)}(\zeta)$ and equation (\ref{octct}) allow us to express $\ocalT^{(\ell_1, \, \ell_2, \, 0)}(\zeta)$ via $\calT^{(1,\, 0, \, 0)}(\zeta)$ and $\calT^{(1,\, 1, \, 0)}(\zeta)$.

\section{Spin chain}

\subsection{Integrability objects}

Let us start with a systematization of our zoo of universal integrability objects. There are two types of them. An object of the first type is constructed using some homomorphism from $\uqlsliii$ to some algebra $A$ and the corresponding power of the automorphism $\sigma$. These are the universal monodromy operators and the universal $L$-operators. We denote a general representative of this type as $\calX_i(\zeta)$. It is an element of $A \otimes \uqlsliii$.  More concretely, we have a family of homomorphisms $\omega_i$ such that
\begin{equation*}
\omega_{i + 1} = \omega_i \circ \sigma^{-1},
\end{equation*}
and define $\calX_i(\zeta)$ as
\begin{equation*}
\calX_i(\zeta) = ((\omega_i)_\zeta \otimes \id)(\calR).
\end{equation*}
The objects of the second type are constructed from the objects of the first type by using some trace $\tr_A$ on the algebra $A$. These are the universal transfer operators and the universal $Q$-operators. We denote a general representative of the second type as $\calY_i(\zeta)$. It is an element of $\uqlsliii$. We assume that
\begin{equation}
\calY_i(\zeta) = (\tr_A \otimes \id)(\calX_i(\zeta) ((\omega_i)_\zeta(t) \otimes 1)), \label{cycx}
\end{equation}
where $t \in \uqlsliii$ is a twist element. In fact, due to the redefinition (\ref{rdq}), it is not quite so for the connection of the universal $L$-operators and the universal $Q$-operators. However, the necessary modification of (\ref{cycx}) and of the further relations for the corresponding integrability objects can be performed easily, see also the discussion below. 

A concrete integrable model is defined by a choice of a representation for the second factor of the tensor product $\uqlsliii \otimes \uqlsliii$. For the case of a spin chain one fixes a finite dimensional representation $\psi$ on the vector space $U$ and defines the operators
\begin{gather}
X_i(\zeta | \eta_1, \ldots, \eta_n) = (\id \otimes  (\psi_{\eta_1} \otimes_{\Delta^{\mathrm{op}}} \cdots \otimes_{\Delta^{\mathrm{op}}} \psi_{\eta_n}))(\calX_i(\zeta)), \label{dxi} \\
Y_i(\zeta | \eta_1, \ldots, \eta_n) =(\psi_{\eta_1} \otimes_{\Delta^{\mathrm{op}}} \cdots \otimes_{\Delta^{\mathrm{op}}} \psi_{\eta_n})(\calY_i(\zeta)), \label{dyi}
\end{gather}
associated with a chain of length $n$. Here, $\eta_1$, $\ldots$, $\eta_n$ are the spectral parameters associated with the sites of the chain. It is clear that the operator $X_i(\zeta | \eta_1, \ldots, \eta_n)$ is an element of $A \otimes \End(U)^{\otimes n}$ and  $Y_i(\zeta | \eta_1, \ldots, \eta_n)$ is an element of $\End(U)^{\otimes n}$. It follows from (\ref{cycx}) that
\begin{equation*}
Y_i(\zeta | \eta_1, \ldots, \eta_n) = (\tr_A \otimes \id)(X_i(\zeta | \eta_1, \ldots, \eta_n) ((\omega_i)_\zeta(t) \otimes 1)).
\end{equation*}

For all cases, except the case of $Q$-operators $Q_i(\zeta | \eta_1, \ldots, \eta_n)$ and $\oQ_i(\zeta | \eta_1, \ldots, \eta_n)$, using equations (\ref{ggd}) and (\ref{ggr}), one can demonstrate that
\begin{equation}
X_i(\zeta \nu | \eta_1 \nu, \ldots \eta_n \nu) = X_i(\zeta | \eta_1, \ldots, \eta_n), \qquad Y_i(\zeta \nu | \eta_1 \nu, \ldots \eta_n \nu) = Y_i(\zeta | \eta_1, \ldots, \eta_n) \label{xinu}
\end{equation}
for any $\nu \in \bbC^\times$. In particular, for the operators
\begin{align*}
& Q'_i(\zeta | \eta_1, \ldots, \eta_n) = (\psi_{\eta_1} \otimes_{\Delta^{\mathrm{op}}} \cdots  \otimes_{\Delta^{\mathrm{op}}} \psi_{\eta_n})(\calQ'_i(\zeta)), \\
& \oQ'_i(\zeta | \eta_1, \ldots, \eta_n) = (\psi_{\eta_1} \otimes_{\Delta^{\mathrm{op}}} \cdots \otimes_{\Delta^{\mathrm{op}}} \psi_{\eta_n})(\ocalQ'_i(\zeta))
\end{align*}
we have
\begin{equation*}
Q'_i(\zeta \nu | \eta_1 \nu, \ldots \eta_n \nu) = Q'_i(\zeta | \eta_1, \ldots, \eta_n), \qquad \oQ{}'_i(\zeta \nu | \eta_1 \nu, \ldots \eta_n \nu) = \oQ{}'_i(\zeta | \eta_1, \ldots, \eta_n).
\end{equation*}
However, if we define the $Q$-operators $Q_i(\zeta | \eta_1, \ldots \eta_n)$ and $\oQ_i(\zeta | \eta_1, \ldots \eta_n)$ with the help of equation (\ref{dyi}) we will not obtain objects satisfying the second relation of (\ref{xinu}). Nevertheless, the corresponding functional relations have the simplest form namely in terms of the $Q$-operators $Q_i(\zeta | \eta_1, \ldots \eta_n)$ and $\oQ_i(\zeta | \eta_1, \ldots \eta_n)$.

Now introduce the matrix
\begin{equation*}
\bbX_i(\zeta | \eta_1, \ldots, \eta_n) = ((\bbX_i(\zeta | \eta_1, \ldots, \eta_n))_{k_1 \ldots k_n | \ell_1 \ldots \ell_n}),
\end{equation*}
where $(\bbX_i(\zeta | \eta_1, \ldots, \eta_n))_{k_1 \ldots k_n | \ell_1 \ldots \ell_n}$ are the elements of $A$ defined by the equation
\begin{equation*}
X_i(\zeta | \eta_1, \ldots, \eta_n) = \sum_{k_1, \ldots, k_n = 1}^3 \sum_{\ell_1, \ldots, \ell_n = 1}^3 (\bbX_i(\zeta | \eta_1, \ldots, \eta_n))_{k_1 \ldots k_n | \ell_1 \ldots \ell_n} \otimes (E_{k_1 \ell_1} \otimes \cdots \otimes  E_{k_n \ell_n}).
\end{equation*}
Here $E_{k \ell}$ are the basis elements of $\End(U)$ associated with some basis of $U$, see, for example, appendix \ref{a:dr}. Similarly, we define the matrix
\begin{equation*}
\bbY_i(\zeta | \eta_1, \ldots, \eta_n) = ((\bbY_i(\zeta | \eta_1, \ldots, \eta_n))_{k_1 \ldots k_n | \ell_1 \ldots \ell_n}),
\end{equation*}
where $(\bbY_i(\zeta | \eta_1, \ldots, \eta_n))_{k_1 \ldots k_n | \ell_1 \ldots \ell_n}$ are the complex numbers defined by the equation
\begin{equation*}
Y_i(\zeta | \eta_1, \ldots, \eta_n) = \sum_{k_1, \ldots, k_n = 1}^3 \sum_{\ell_1, \ldots, \ell_n = 1}^3 (\bbY_i(\zeta | \eta_1, \ldots, \eta_n))_{k_1 \ldots k_n | \ell_1 \ldots \ell_n} E_{k_1 \ell_1} \otimes \cdots \otimes  E_{k_n \ell_n}.
\end{equation*}
For the one-site case we introduce the matrices $\bbX(\zeta)$ and $\bbY(\zeta)$ depending on one spectral parameter. It is clear that the matrices $\bbX_i(\zeta | \eta_1, \ldots, \eta_n)$ and $\bbY_i(\zeta | \eta_1, \ldots, \eta_n)$ contain the same information as the operators $X_i(\zeta | \eta_1, \ldots, \eta_n)$ and $Y_i(\zeta | \eta_1, \ldots, \eta_n)$. However, for concrete physical applications it is more convenient to work with the matrices.

Below we take as $\psi$ the representation $\varphi^{(1, \, 0, \, 0)}$. The universal integrability objects $\calX_i(\zeta)$ satisfy the following relation
\begin{equation*}
\calX_{i + 1}(\zeta) = (\id \otimes \sigma)(\calX_i(\zeta))|_{s \to \sigma(s)}.
\end{equation*}
Hence, we have
\begin{equation*}
X_{i + 1}(\zeta) = (\id \otimes (\varphi^{(1, \, 0, \, 0)} \circ \sigma))(\calX_i(\zeta))|_{s \to \sigma(s)}.
\end{equation*}
Using equations (\ref{vp100a})--(\ref{vp100d}), we conclude that for any $a \in \uqlsliii$ one obtains
\begin{equation*}
(\varphi^{(1,\, 0, \, 0)} \circ \sigma)(a) = O \bigl( \varphi^{(1, \, 0, \,0)}(a) \bigr) O^{-1},
\end{equation*}
where the endomorphism $O$ has the matrix form
\begin{equation*}
\bbO = \left( \begin{array}{ccc}
0 & 0 & q \\
1 & 0 & 0 \\
0 & 1 & 0
\end{array} \right).
\end{equation*}
Thus, the relation
\begin{equation}
\bbX_{i + 1}(\zeta) = \bbO \, \bbX_i(\zeta) \bbO^{-1} \, |_{s \to \sigma(s)} \label{xipo}
\end{equation}
is valid. For a chain of an arbitrary length we obtain
\begin{equation*}
\bbX_{i + 1}(\zeta | \eta_1, \ldots, \eta_n) = (\underbrace{\bbO \times \cdots \times \bbO}_n) \, \bbX_i(\zeta | \eta_1, \ldots, \eta_n) (\underbrace{\bbO \times \cdots \times \bbO}_n)^{-1} \, |_{s \to \sigma(s)}.
\end{equation*}
In the same way we determine that
\begin{equation*}
\bbY_{i + 1}(\zeta | \eta_1, \ldots, \eta_n) = (\underbrace{\bbO \times \cdots \times \bbO}_n) \, \bbY_i(\zeta | \eta_1, \ldots, \eta_n) (\underbrace{\bbO \times \cdots \times \bbO}_n)^{-1} \, |_{\phi \to \sigma(\phi)}.
\end{equation*}

The explicit form of the matrices $\bbM(\zeta)$ and $\obbM{}'_i(\zeta)$ were found in the paper \cite{Raz13}.\footnote{The matrix $\obbM{}'_i(\zeta)$ in \cite{Raz13} is denoted as $\obbM{}_i(\zeta)$.} As follows from (\ref{bmiz}), the monodromy operator $\obbM{}_i(\zeta)$ can be obtained from $\obbM{}'_i(\zeta)$ with the help of the relation
\begin{equation*}
\obbM{}_i(\zeta) = \tau(\obbM{}'_i(r_s \zeta)),
\end{equation*}
where the automorphism $\tau$ of $\uqgliii$ defined by equations (\ref{tga}) and (\ref{tgb}) is applied to the matrix entries.

There is another method to find the expression for $\obbM{}_i(\zeta)$. Using equation (\ref{bmtom}), we see that
\begin{equation*}
\oM{}'_i(\zeta) = ((\varphi_{- i + 2})_\zeta \otimes \ovarphi'^{(1, \, 0, \, 0)})(\calR)|_{s \to \tau(s)}.
\end{equation*}
Then, taking into account (\ref{oeqd}), we have
\begin{equation*}
\oM{}'_i(\zeta) = (1 \otimes P)(((\varphi_{- i + 2})_{r_s^{-3} q^{-3/s} \zeta} \otimes \varphi^{(1, \, 0, \, 0) * S^{-1}})(\calR))(1 \otimes P^{-1})|_{s \to \tau(s)},
\end{equation*}
and equation (\ref{mvdp}) gives
\begin{equation*}
\obbM{}'_i(\zeta) = \bbP ((\bbM_{- i + 2}(r_s^{-3} q^{- 3/s} \zeta))^{-1})^t \bbP^{-1} |_{s \to \tau(s)},
\end{equation*}
where the matrix $\bbP$ is given by equation (\ref{bbp}).

The explicit expressions for various $L$-operators were obtained in the paper \cite{BooGoeKluNirRaz10}. However, in the present paper we use different definition of $q$-oscillators. Therefore, we give here the following expressions for two $L$-operators:
\begin{align}
& \bbL'(\zeta) = \rme^{f_3(\zeta^s)} \notag \\
& \hspace{1.5em} \times \left( \begin{array}{ccc}
q^{N_1 + N_2} - \zeta^s q^{- N_1 - N_2 - 2} & \zeta^{s - s_1} b_1 q^{- 2 N_1 - N_2} & \zeta^{s -s_1 - s_2} b_2 q^{- 2 N_2 + 1} \\[.3em]
\zeta^{s_1} \kappa_q b_1^\dagger q^{N_1} & q^{- N_1} & 0 \\[.3em]
\zeta^{s_1 + s_2} \kappa_q b_2^\dagger q^{- N_1 + N_2 - 1} & - \zeta^{s_2} \kappa_q b_1 b_2^\dagger q^{- 2 N_1 + N_2 + 1} & q^{- N_2}
\end{array} \right), \label{bbl} \\
& \obbL{}'(\zeta) = \rme^{f_3(- q \zeta^s) + f_3(- q^{-1} \zeta^s)} \notag \\
& \hspace{3.em} \times \left( \begin{array}{ccc}
q^{ - N_1 - N_2} & - \zeta^{s - s_1} \kappa_q b_2^\dagger q^{N_2 + 1}  & \zeta^{s - s_1 - s_2} \kappa_q b_1^\dagger q^{N_1 - N_2 + 1} \\[.3em]
\zeta^{s_1} b_2 q^{- N_1 - 2 N_2} & q^{N_2} + \zeta^s q^{- N_2 - 1} & \zeta^{s - s_2} \kappa_q b_1^\dagger b_2 q^{N_1 - 2 N_2 + 1} \\[.3em]
- \zeta^{s_1 + s_2} b_1 q^{-2 N_1} &  - \zeta^{s_2} \kappa_q b_1 b_2^\dagger q^{- N_1 + 2 N_2 + 1}  & q^{N_1} + \zeta^s q^{- N_1 - 1}
\end{array} \right), \label{obbl}
\end{align}
where the function
\begin{equation*}
f_3(\zeta) = \sum_{k = 1}^\infty \frac{1}{q^{2k} + 1 + q^{-2k}} \, \frac{\zeta^k}{k}
\end{equation*}
satisfies the defining relation
\begin{equation*}
f_3(q^2\zeta) + f_3(\zeta) + f_3(q^{-2}\zeta) = - \log(1 - \zeta).
\end{equation*}
The expressions for all operators $\bbL'_i(\zeta)$ and $\obbL{}'_i(\zeta)$ can be obtained with the help of equation (\ref{xipo}). In a similar way as in the case of the monodromy matrices one obtains
\begin{equation*}
\obbL{}'_i(\zeta) = \bbP ((\bbL'_{- i + 1}(r_s^{-3} q^{- 3/s} \zeta))^{-1})^t \bbP^{-1} |_{s \to \tau(s)}.
\end{equation*}

\subsection{Functional relations}

For the transfer operators corresponding to the finite dimensional representations $\pi^\lambda$ of $\uqgliii$ on the auxiliary space we have the expressions
\begin{gather*}
T_i^\lambda(\zeta | \eta_1, \ldots, \eta_n) = \tr^\lambda (M_i(\zeta | \eta_1, \ldots, \eta_n) ((\varphi_i)_\zeta(t) \otimes 1)), \\
\oT_i^\lambda(\zeta | \eta_1, \ldots, \eta_n) = \tr^\lambda (\oM_i(\zeta | \eta_1, \ldots, \eta_n) ((\ovarphi_i)_{r_s \zeta}(t) \otimes 1)).
\end{gather*}
One can demonstrate, in particular, that
\begin{gather*}
\bbT^\lambda(\zeta | \eta_1, \ldots, \eta_n) = \tr^\lambda ((\bbM(\zeta \eta_1^{-1}) \boxtimes \cdots \boxtimes \bbM(\zeta \eta_n^{-1})) \, q^{ - \Phi_1 G_1 - \Phi_2 G_2 - \Phi_3 G_3}), \\
\obbT^\lambda(\zeta | \eta_1, \ldots, \eta_n) = \tr^\lambda ((\obbM(\zeta \eta_1^{-1}) \boxtimes \cdots \boxtimes \obbM(\zeta \eta_n^{-1})) \, q^{- \Phi_1 G_1 - \Phi_2 G_2 - \Phi_3 G_3}),
\end{gather*}
where $\boxtimes$ denotes the natural generalization of the Kronecker product to the case of matrices with noncommuting entries, see, for example, the paper \cite{BooGoeKluNirRaz14a}, and
\begin{equation*}
\Phi_1 = (\phi_0 - \phi_1)/3, \qquad \Phi_2 = (\phi_1 - \phi_2)/3, \qquad \Phi_3 = (\phi_2 - \phi_0)/3.
\end{equation*}
Similar expression can be written for the transfer operators corresponding to the infinite dimensional representations $\widetilde \pi^\lambda$ of $\uqgliii$.

Further, we obtain
\begin{align*}
& \bbQ_i(\zeta | \eta_1, \ldots, \eta_n) = \zeta^{s \Phi_i / 2} \tr^{\scriptscriptstyle ++} \big((\zeta^{\bbD_i}\bbL'_i(\zeta \eta_1^{-1}) \boxtimes \cdots \boxtimes \zeta^{\bbD_i} \bbL'_i(\zeta \eta_n^{-1})) \rho_i(t) \big), \\
& \obbQ_i(\zeta | \eta_1, \ldots, \eta_n) = \zeta^{- s \Phi_i / 2} \tr^{\scriptscriptstyle ++} \big((\zeta^{-\bbD_i}\obbL'_i(r_s \zeta \eta_1^{-1}) \boxtimes \cdots \boxtimes \zeta^{-\bbD_i} \obbL'_i(r_s \zeta \eta_n^{-1 })) \orho_i(t) \big),
\end{align*}
where
\begin{gather}
\zeta^{\bbD_1} = \left( \begin{array}{ccc}
\zeta^{-s/3} \\
& \zeta^{s/6} \\
& & \zeta^{s/6}
\end{array} \right), \qquad \zeta^{\bbD_2} = \left( \begin{array}{ccc}
\zeta^{s/6} \\
& \zeta^{- s/3} \\
& & \zeta^{s/6}
\end{array} \right), \label{zda} \\
\zeta^{\bbD_3} = \left( \begin{array}{ccc}
\zeta^{s/6} \\
& \zeta^{s/6} \\
& & \zeta^{- s/3}
\end{array} \right) \label{zdb}
\end{gather}
and $\zeta^{- \bbD_i}$ is the inverse of $\zeta^{\bbD_i}$. It is easy to determine that
\begin{gather*}
\rho_1(t) = q^{- (\Phi_1 - \Phi_2) N_1 + (\Phi_3 - \Phi_1) N_2}, \qquad \rho_2(t) = q^{- (\Phi_2 - \Phi_3) N_1 + (\Phi_1 - \Phi_2) N_2}, \\*
\rho_3(t) = q^{- (\Phi_3 - \Phi_1) N_1 + (\Phi_2 - \Phi_3) N_2}, \\
\orho_1(t) = q^{(\Phi_1 - \Phi_3) N_1 - (\Phi_2 - \Phi_1) N_2}, \qquad \orho_2(t) = q^{(\Phi_2 - \Phi_1) N_1 - (\Phi_3 - \Phi_2) N_2}, \\* 
\orho_3(t) = q^{(\Phi_3 - \Phi_2) N_1 - (\Phi_1 - \Phi_3) N_2}.
\end{gather*}

The functional relations for integrability objects corresponding to some choice of the representation of $\uqlsliii$ on the quantum space are easily obtained from the functional relations for the corresponding universal integrability objects by the simple substitution. However, it is important for applications to have integrability objects with a simple analytical structure with respect to the spectral parameter $\zeta$. This is not the case for the objects introduced above. Therefore, we introduce the transfer operators $\bbT^{\mathrm p \lambda}_i(\zeta | \eta_1, \ldots, \eta_n)$ and $\obbT^{\mathrm p \lambda}_i(\zeta | \eta_1, \ldots, \eta_n)$ being Laurent polynomials in $\zeta^{s / 2}$ and related to the transfer operators $\bbT^\lambda_i(\zeta | \eta_1, \ldots, \eta_n)$ and $\obbT{}^\lambda_i(\zeta | \eta_1, \ldots, \eta_n)$ as
\begin{align*}
& \bbT^\lambda_i(\zeta | \eta_1, \ldots, \eta_n) = q^{- (\lambda_1 + \lambda_2 + \lambda_3) n / 3} \prod_{k = 1}^n (\zeta \eta_k^{-1})^{s / 2} \\*
& \hspace{4em} {} \times \prod_{k = 1}^n \rme^{f_3(q^{- 2(\lambda_1 + 1)} (\zeta \eta_k^{-1})^s) + f_3(q^{- 2\lambda_2} (\zeta \eta_k^{-1})^s) + f_3(q^{- 2(\lambda_3 - 1)} (\zeta \eta_k^{-1})^s)} \, \bbT^{\mathrm p \lambda}_i(\zeta | \eta_1, \ldots, \eta_n), \\
& \obbT{}_i^\lambda(\zeta | \eta_1, \ldots, \eta_n) = q^{2(\lambda_1 + \lambda_2 + \lambda_3) n / 3} \prod_{k = 1}^n (\zeta \eta_k^{-1})^s \\*
& \hspace{2em} {} \times \prod_{k = 1}^n \rme^{f_3(q^{2(\lambda_1 + 1) - 1} (\zeta \eta_k^{-1})^s) + f_3(q^{2\lambda_2 - 1} (\zeta \eta_k^{-1})^s) + f_3(q^{2(\lambda_3 - 1) - 1} (\zeta \eta_k^{-1})^s)} \\*
& \hspace{4em} {} \times \prod_{k = 1}^n \rme^{f_3(q^{2(\lambda_1 + 1) + 1} (\zeta \eta_k^{-1})^s) + f_3(q^{2\lambda_2 + 1} (\zeta \eta_k^{-1})^s) + f_3(q^{2(\lambda_3 - 1) + 1} (\zeta \eta_k^{-1})^s)} \obbT{}_i^{\mathrm p \lambda}(\zeta | \eta_1, \ldots, \eta_n).
\end{align*}
To prove that we really have Laurent polynomials one can use the explicit expressions for the monodromy operators $\bbM(\zeta)$ and $\obbM'(\zeta)$ from the paper \cite{Raz13}.

Similarly, we introduce the $Q$-operators $\bbQ^{\mathrm p}_i (\zeta | \eta_1, \ldots, \eta_n)$ and $\obbQ^{\mathrm p}_i (\zeta | \eta_1, \ldots, \eta_n)$ being Laurent polynomials in $\zeta^{s / 2}$ and related to the initial $Q$-operators $\bbQ_i (\zeta | \eta_1, \ldots, \eta_n)$ and $\obbQ_i (\zeta | \eta_1, \ldots, \eta_n)$ as
\begin{align*}
& \bbQ_i(\zeta | \eta_1, \ldots, \eta_n) = \zeta^{s \Phi_i / 2 + n s / 6} \prod_{k = 1}^n \rme^{f_3((\zeta \eta_k^{-1})^s)} \, \bbQ^{\mathrm p}_i (\zeta | \eta_1, \ldots, \eta_n), \\
& \obbQ{}_i(\zeta | \eta_1, \ldots, \eta_n) =  \zeta^{- s \Phi_i / 2 + n s / 3} \prod_{k = 1}^n \rme^{f_3(q^{-1} (\zeta \eta_k^{-1})^s) + f_3(q (\zeta \eta_k^{-1})^s)} \, \obbQ{}^{\mathrm p}_i (\zeta | \eta_1, \ldots, \eta_n).
\end{align*}
The required polynomiality follows from the explicit form (\ref{bbl}) and (\ref{obbl}) of the $L$-opera\-tors $\bbL'(\zeta)$ and $\obbL'(\zeta)$ and the explicit form (\ref{zda}) and (\ref{zdb}) of the matrices $\zeta^{\bbD_i}$.   

In terms of polynomial objects the $TQ$-relation (\ref{tq}) takes the form 
\begin{multline*}
q^{\Phi_i} \prod_{k = 1}^n b(q (\zeta \eta_k^{-1})^{- s / 2}) \, \bbQ_i^{\mathrm p} (q^{2 / s} \zeta) \\- \bbT^{\mathrm p (1, \, 1, \, 0)}(\zeta) \, \bbQ_i^{\mathrm p} (\zeta) 
+ q^{- \Phi_i} \bbT^{\mathrm p (1, \, 0, \, 0)}(\zeta) \, \bbQ_i^{\mathrm p} (q^{- 2 / s} \zeta) \\ - q^{- 2 \Phi_i} \prod_{k = 1}^n b((\zeta \eta_k^{-1})^{- s / 2}) \, \bbQ_i^{\mathrm p} (q^{- 4 / s} \zeta) = 0,
\end{multline*}
while instead of (\ref{btq}) we have
\begin{multline*}
q^{\Phi_i} \prod_{k = 1}^n \big[ b(q^{-3/2}(\zeta \eta_k^{-1})^{- s / 2}) b(q^{-1/2}(\zeta \eta_k^{-1})^{- s / 2} ) \big] \, \obbQ{}_i^{\mathrm p} (q^{- 2 / s} \zeta) \\
- \obbT^{\mathrm p (1, \, 1, \, 0)}(\zeta) \, \obbQ{}_i^{\mathrm p} (\zeta) 
+ q^{- \Phi_i} \obbT^{\mathrm p (1, \, 0, \, 0)}(\zeta) \, \obbQ_i^{\mathrm p} (q^{2 / s} \zeta) \\ - q^{- 2 \Phi_i} \prod_{k = 1}^n \big[ b(q^{- 1/2} (\zeta \eta_k^{-1})^{- s / 2}) b(q^{1/2} (\zeta \eta_k^{-1})^{- s / 2}) \big] \, \obbQ_i^{\mathrm p} (q^{4 / s} \zeta) = 0.
\end{multline*}
Here and below we use the notation
\begin{equation*}
b(\zeta) = \zeta - \zeta^{-1},
\end{equation*}
and skip the explicit dependence of the transfer operators and $Q$-operators on the spectral parameters $\eta_1$, $\ldots$, $\eta_n$. For relations (\ref{t100qq}) and (\ref{t110qq}) we obtain 
\begin{align*}
& \bbT^{\mathrm p(1,0,0)}(\zeta) \bbQ^{\mathrm p}_i(q^{-2/s} \zeta) \obbQ{}^{\mathrm p}_j(q^{-1/s} \zeta) \\
& \hspace{6em} {} = q^{-\Phi_i} \prod_{k = 1}^n b((\zeta \eta_k^{-1})^{-s/2}) 
\bbQ^{\mathrm p}_i(q^{-4/s} \zeta) \obbQ{}^{\mathrm p}_j(q^{-1/s} \zeta) \\
& \hspace{6em} {} + q^{\Phi_i + \Phi_j} \prod_{k = 1}^n b((\zeta \eta_k^{-1})^{-s/2}) \bbQ^{\mathrm p}_i(\zeta) \obbQ{}^{\mathrm p}_j(q^{-3/s} \zeta) \\
& \hspace{15em} {} + q^{-\Phi_j} \prod_{k = 1}^n b(q(\zeta \eta_k^{-1})^{-s/2}) \bbQ^{\mathrm p}_i(q^{-2/s} \zeta) \obbQ{}^{\mathrm p}_j(q^{1/s} \zeta)  
\end{align*}
and
\begin{align*}
& \bbT^{\mathrm p(1,1,0)}(\zeta) \bbQ^{\mathrm p}_i(\zeta) \obbQ{}^{\mathrm p}_j(q^{-1/s}\zeta) \\
& \hspace{7em} {} = q^{\Phi_i} \prod_{k=1}^n b(q(\zeta\eta_k^{-1})^{-s/2}) 
\bbQ^{\mathrm p}_i(q^{2/s}\zeta) \obbQ{}^{\mathrm p}_j(q^{-1/s}\zeta) \\
& \hspace{7em} {} + q^{-\Phi_i-\Phi_j}
\prod_{k=1}^n b(q(\zeta\eta_k^{-1})^{-s/2}) 
\bbQ^{\mathrm p}_i(q^{-2/s}\zeta) \obbQ{}^{\mathrm p}_j(q^{1/s}\zeta) \\
& \hspace{17em} {} + q^{\Phi_j} 
\prod_{k=1}^n b((\zeta\eta_k^{-1})^{-s/2}) 
\bbQ^{\mathrm p}_i(\zeta) \obbQ{}^{\mathrm p}_j(q^{-3/s}\zeta).  
\end{align*}

The $TT$-relation (\ref{fr3}) and its barred analogue look like 
\begin{align*}
& \prod_{k = 1}^n b(q^{-1} (\zeta \eta_k^{-1})^{- s / 2}) \bbT^{\mathrm p (\ell_1, \, \ell_2, \, 0)}(\zeta) \\*
& \hspace{6em} {} = \bbT^{\mathrm p (\ell_1, \, 0, \, 0)}(\zeta) \bbT^{\mathrm p (\ell_2, \, 0, \, 0)}(q^{2/s} \zeta) - \bbT^{\mathrm p (\ell_1 + 1, \, 0, \, 0)}(q^{2/s} \zeta) \bbT^{\mathrm p (\ell_2 - 1, \, 0, \, 0)}(\zeta), \\
&\prod_{k = 1}^n b(q^{3/2} (\zeta \eta_k^{-1})^{- s / 2})  b(q^{1/2} (\zeta \eta_k^{-1})^{- s / 2}) \obbT^{\mathrm p (\ell_1, \, \ell_2, \, 0)}(\zeta) \\
& \hspace{5em} = \obbT^{\mathrm p (\ell_1, \, 0, \, 0)}(\zeta) \obbT^{\mathrm p (\ell_2, \, 0, \, 0)}(q^{- 2/s} \zeta) - \obbT^{\mathrm p (\ell_1 + 1, \, 0, \, 0)}(q^{- 2/s} \zeta) \obbT^{\mathrm p (\ell_2 - 1, \, 0, \, 0)}(\zeta).
\end{align*}
As the final example we give the expression for the $TT$-relation (\ref{fr2}) and its barred analogue:
\begin{align*}
& \bbT^{\mathrm p (\ell - 1, \, \ell - 1, \, 0)}(q^{-2/s} \zeta) \bbT^{\mathrm p (\ell + 1, \, \ell + 1, \, 0)}(\zeta) = \bbT^{\mathrm p (\ell, \, \ell, \, 0)}(q^{-2/s} \zeta) \bbT^{\mathrm p (\ell, \, \ell, \, 0)}(\zeta) \\* 
& \hspace{20em} {} - \prod_{k = 1}^n b(q^{\ell + 1} (\zeta \eta_k^{-1})^{- s/2}) \bbT^{\mathrm p (\ell, \, 0, \, 0)}(\zeta), \\
& \obbT^{\mathrm p (\ell - 1, \, \ell - 1, \, 0)}(q^{2/s} \zeta) \obbT^{\mathrm p (\ell + 1, \, \ell + 1, \, 0)}(\zeta) = \obbT^{\mathrm p (\ell, \, \ell, \, 0)}(q^{2/s} \zeta) \obbT^{\mathrm p (\ell, \, \ell, \, 0)}(\zeta) \\* 
& \hspace{9em} - \prod_{k = 1}^n b(q^{- \ell - 3/2} (\zeta \eta_k^{-1})^{- s/2}) b(q^{- \ell - 1/2} (\zeta \eta_k^{-1})^{- s/2}) \obbT^{\mathrm p (\ell, \, 0, \, 0)}(\zeta).
\end{align*}
It is not difficult to obtain the expressions for all other functional relations in terms of polynomial objects. It is also worth to note that relations (\ref{bttot}) take the form
\begin{align*}
& \obbT^{(\ell, \, 0, \, 0)}(\zeta) = \prod_{k = 1}^n b(q^{1/2} (\zeta \eta_k^{-1})^{-s/2}) \bbT^{(\ell, \, 0, \, 0)}(q^{(2 \ell + 1)/s} \zeta), \\
& \obbT^{(\ell, \, \ell, \, 0)}(\zeta) = \prod_{k = 1}^n b(q^{- (2 \ell + 1)/2} (\zeta \eta_k^{-1})^{-s/2}) \bbT^{(\ell, \, \ell, \, 0)}(q^{(2 \ell - 1)/s} \zeta).
\end{align*}

\section{Conclusions}

We gave a detailed construction of the universal integrability objects related to the integrable systems associated with the quantum group $\uqlsliii$. The full proof of the functional relations in the form independent of the representation of the quantum group on the quantum space was presented.  Generalizations to the case of the quantum group $\mathrm{U}_q(\mathcal L(\mathfrak{sl}_N))$ and quantum supergroup $\mathrm{U}_q(\mathcal L(\mathfrak{gl}_{N | M})$ are discussed in the papers \cite{Koj08} and \cite{Tsu13b}, respectively, however, mostly without proofs. However, the proofs for the case of the quantum supergroup $\mathrm{U}_q(\mathcal L(\mathfrak{sl}_{2 | 1}))$ are given in \cite{BazTsu08}. A discussion of the functional relations for the systems related to $\uqlslii$ with $q$ being a root of unity can be found in \cite{Kor05a} and references therein.

Note that, in fact, we define and enumerate the $Q$-operators with the help of the elements of the automorphisms group of the Dynkin diagram of the affine Lie algebra $\widehat{\calL}(\mathfrak{sl}_3)$. The similar approach is used for the case of the integrable systems related to the quantum group  $\mathrm{U}_q(\mathcal L(\mathfrak{sl}_N))$ in the paper \cite{Koj08}. In the paper \cite{Tsu10} Tsuboi suggested defining and enumerating $Q$-operators with the help of Hasse diagrams. This approach was used then in   \cite{BazFraLukMenSta11, FraLukMenSta11, KazLeuTsu12, AleKazLeuTsuZab13, Tsu13a, FraLukMenSta13, Tsu13b}.\footnote{We are sorry for the inevitable incompleteness of the list.} We will not discuss here what is the ``right'' way, and only note that for the case considered in the present paper both approaches are equivalent.

The next remark is on $TT$-relations. Usually one considers only three-term fusion relations of the type (\ref{fr1}) and (\ref{fr2}), see, for example, \cite{KluPea92, KunNakSuz94}. We demonstrated that in our case the four-term $TT$-relations of the type (\ref{pjt1}) are also useful, in particular, to prove the quantum Jacobi--Trudi identity. It is evident from our construction that in the case of $\mathrm{U}_q(\mathcal L(\mathfrak{sl}_N))$ one has in general $(N+1)$-term $TT$-relations.

Among other approaches to the derivation of the $TQ$-relations for the ``deformed'' case we would like to mention the approach based on the factorization of the Yang--Baxter operators, see \cite{ChiDerKarKir13} and references therein, and the approach based on the concept of $q$-cha\-rac\-ters~\cite{FreRes99} developed in the paper \cite{FreHer13}. It would be interesting to generalize the approach based on formulas for group characters \cite{KazLeuTsu12, AleKazLeuTsuZab13} to the ``deformed'' case.

\vskip .5em

{\em Acknowledgements.\/} This work was supported in part by the DFG grant KL \hbox{645/10-1} and by the Volkswagen Foundation. Kh.S.N. and A.V.R. were supported in part by the RFBR grants \#~13-01-00217 and \#~14-01-91335.

\appendix

\section{\texorpdfstring{Verma modules and representations of $\uqgliii$}{Verma modules and representations of Uq(gl3)}} \label{a:vmr}

Let $\lambda$ be an element of $\gothg^*$, where $\gothg$ is the standard Cartan subalgebra of $\gothgl_3$ with the standard basis formed by the elements $G_i = E_{ii}$, $i = 1, 2, 3$, see section \ref{ss:qguqslii}. We identify $\lambda$ with the set of its components $(\lambda_1, \lambda_2, \lambda_3)$ with respect to the dual basis of the basis $\{G_i\}$. In particular, for the simple roots of $\gothgl_3$ we have the identification
\begin{equation*}
\alpha_1 = (1, \, -1, \, 0), \qquad \alpha_2 = (0, \, 1, \, -1).
\end{equation*}
The Verma $\uqgliii$-module $\widetilde V^\lambda$  is the $\uqgliii$-module with the highest weight vector $v_0$ satisfying the relations
\begin{gather}
q^{\nu G_1} v_0 = q^{\nu \lambda_1} v_0, \qquad q^{\nu G_2} v_0 = q^{\nu \lambda_2} v_0, \qquad q^{\nu G_3} v_0 = q^{\nu \lambda_3} v_0, \label{hv0} \\
E_1 v_0 = 0, \qquad E_2 v_0 = 0. \label{ev0}
\end{gather}
It is convenient to denote
\begin{equation}
\mu_1 = \lambda_1 - \lambda_2, \qquad \mu_2 = \lambda_2 - \lambda_3 \label{ml}
\end{equation}
so that we have
\begin{equation*}
q^{\nu H_1} v_0 = q^{\nu \mu_1} v_0, \qquad q^{\nu H_2} v_0 = q^{\nu \mu_2} v_0,
\end{equation*}
where $H_1$ and $H_2$ are the standard Cartan generators of the standard Cartan subalgebra of the Lie algebra $\gothsl_3$ given by equation (\ref{hk}).

The vectors
\begin{equation}
v_n = F_1^{n_1} F_3^{n_2} F_2^{n_3} v_0, \label{vn}
\end{equation}
where $n_1, n_2, n_3 \in \bbZ_+$ and $n = (n_1, n_2, n_3)$, form a basis of $\widetilde V^\lambda$. Let us describe the action of the generators of $\uqgliii$ on the elements of this basis.

Using (\ref{alphah}) and the second relation of (\ref{xexf}), we obtain
\begin{equation*}
q^{\nu G_i} F_j = q^{- \nu c_{i j}} F_j \, q^{\nu G_i}, \qquad \nu \in \bbC.
\end{equation*}
Now, taking into account (\ref{hv0}), we see that
\begin{equation*}
q^{\nu G_1} v_n = q^{\nu(\lambda_1 - n_1 - n_2)} v_n, \quad q^{\nu G_2} v_n = q^{\nu(\lambda_2 + n_1 - n_3)} v_n, \quad q^{\nu G_3} v_n = q^{\nu(\lambda_3 + n_2 + n_3)} v_n.
\end{equation*}

The action of $F_1$ is very simple. Namely, it directly follows from (\ref{vn}) that\footnote{For $k \in \bbZ$ we use the notation $n + k \varepsilon_1 = (n_1 + k, n_2, n_3)$, $n + k \varepsilon_2 = (n_1, n_2 + k, n_3)$ and $n + k \varepsilon_3 = (n_1, n_2, n_3 + k)$.}
\begin{equation*}
F_1 v_n = v_{n + \varepsilon_1}.
\end{equation*}
To find the action of $F_2$, we determine from the second relation of (\ref{e3f3}) and the first relation of (\ref{f3f}) that
\begin{equation*}
F_2^{\mathstrut} F_1^{n_1} = q^{n_1} F_1^{n_1} F_2^{\mathstrut} + [n_1]_q F_1^{n_1 - 1} F_3^{\mathstrut},
\end{equation*}
and from the second relation of (\ref{f3f}) that
\begin{equation*}
F_2^{\mathstrut} F_3^{n_2} = q^{-n_2} F_3^{n_2} F_2^{\mathstrut}.
\end{equation*}
These two relations give
\begin{equation*}
F_2 v_n =  q^{n_1 - n_2} v_{n + \varepsilon_3} + [n_1]_q v_{n - \varepsilon_1 + \varepsilon_2}.
\end{equation*}
For the action of $F_3$, using (\ref{f3f}), we obtain
\begin{equation*}
F_3 v_n = q^{- n_1} v_{n + \varepsilon_2}.
\end{equation*}

It follows from (\ref{ef}) and from the second relation of (\ref{xexf}) that
\begin{equation*}
E_1^{\mathstrut} F_1^{n_1} = F_1^{n_1} E_1^{\mathstrut} + [n_1]_q F_1^{n_1 - 1} [H_1 - n_1 + 1]_q,
\end{equation*}
while the first relation of (\ref{f3e}) and the second relation of (\ref{f3f}) give
\begin{equation*}
E_1^{\mathstrut} F_3^{n_2} = F_3^{n_2} E_1^{\mathstrut} - q^{- n_2 + 2} [n_2]_q F_3^{n_2 - 1} F_2^{\mathstrut} q^{H_1}.
\end{equation*}
Therefore, one has
\begin{equation}
E_1 v_n = [\mu_1 - n_1 - n_2 + n_3 + 1]_q [n_1]_q v_{n - \varepsilon_1} -
 q^{\mu_1 - n_2 + n_3 + 2} [n_2]_q v_{n - \varepsilon_2 + \varepsilon_3}. \label{e1vn}
\end{equation}
Similarly, the second relation of (\ref{f3e}) and the first relation of (\ref{f3f}) give
\begin{equation*}
E_2^{\mathstrut} F_3^{n_2} = F_3^{n_2} E_2^{\mathstrut} + [n_2]_q F_1^{} F_3^{n_2 - 1} q^{-H_2}.
\end{equation*}
while it follows from (\ref{ef}) and from the second relation of (\ref{xexf}) that
\begin{equation*}
E_2^{\mathstrut} F_2^{n_3} = F_2^{n_3} E_2^{\mathstrut} + [n_3]_q F_2^{n_3 - 1} [H_2 - n_3 + 1]_q,
\end{equation*}
Therefore, one has
\begin{equation}
E_2 v_n = [\mu_2 - n_3 + 1]_q \, [n_3]_q v_{n - \varepsilon_3} + q^{- \mu_2 + 2 n_3} [n_2]_q v_{n + \varepsilon_1 - \varepsilon_2}. \label{e2vn}
\end{equation}
Finally, from the definition of $E_3$, and from relations (\ref{e1vn}) and (\ref{e2vn}) we obtain
\begin{multline*}
E_3 v_n = q^{n_1} [\mu_1 + \mu_2 - n_1 - n_2 - n_3 + 1]_q [n_2]_q v_{n - \varepsilon_2} \\
- q^{-\mu_1 + n_1 + n_2 - n_3 - 1} [\mu_2 - n_3 + 1]_q [n_1]_q [n_3]_q v_{n - \varepsilon_1 - \varepsilon_3}.
\end{multline*}

We denote the representation of $\uqgliii$ corresponding to the module $\widetilde V^\lambda$ by $\widetilde \pi^\lambda$. When the numbers $\mu_1$ and $\mu_2$ defined by (\ref{ml}) are non-negative integers, the module $\widetilde V^\lambda$ has a maximal submodule such that the corresponding quotient module is finite dimensional. We denote this finite dimensional $\uqgliii$-module by $V^\lambda$ and the corresponding representation by $\pi^\lambda$.

\section{\texorpdfstring{Tensor product of representations $(\rho^{\scriptscriptstyle ++}_3)_{\zeta_3}$ and $(\rho^{\scriptscriptstyle -+}_2)_{\zeta_2}$}{Tensor product of representations (rho++3)zeta3 and (rho-+2)zeta2}}
\label{a:tprhorho}

\subsection{Basis}

Consider the $\uqbp$-module $(W_3^{\scriptscriptstyle ++})_{\zeta_3} \otimes (W_2^{\scriptscriptstyle -+})_{\zeta_2}$ corresponding to the representation $(\rho^{\scriptscriptstyle ++}_3)_{\zeta_3} \otimes_\Delta (\rho^{\scriptscriptstyle -+}_2)_{\zeta_2}$. By definition, we have
\begin{align*}
q^{\nu h_0} v &= q^{\nu(2 N_{A1} + N_{A2} - N_{B1} + N_{B2})} v, \\
q^{\nu h_1} v &= q^{\nu(- N_{A1} + N_{A2} - N_{B1} - 2 N_{B2})} v, \\
q^{\nu h_2} v &= q^{\nu(- N_{A1} - 2 N_{A2} + 2 N_{B1} + N_{B2})} v, \\
e_0 v &= \zeta_3^{s_0} b^\dagger_{A1} q^{- N_{A2}} v - \zeta_2^{s_0} b^{}_{B1} b^\dagger_{B2} q^{- 2N_{A1} - N_{A2} - N_{B1} + N_{B2} + 1} v, \\
e_1 v &= - \zeta_3^{s_1} b^{}_{A1} b^\dagger_{A2} q^{- N_{A1} + N_{A2} + 1} v + \zeta_2^{s_1} \kappa_q^{-1} b^{}_{B2} q^{N_{A1} - N_{A2} - N_{B2}} v, \\
e_2 v &= \zeta_3^{s_2} \kappa_q^{-1} b^{}_{A2} q^{- N_{A2}} v + \zeta_2^{s_2} b^\dagger_{B1} q^{N_{A1} + 2 N_{A2} - N_{B2}} v,
\end{align*}
where we mark the $q$-oscillators related to the first and second factors by $A$ and $B$. Below in this appendix $q^{\nu h_i}$ and $e_i$, $i = 0, 1, 2$, denote the operators given by the above equations.

The $q$-oscillators act on the module $(W_3^{\scriptscriptstyle ++})_{\zeta_3} \otimes (W_2^{\scriptscriptstyle -+})_{\zeta_2}$ as on the module $W^{\scriptscriptstyle +} \otimes W^{\scriptscriptstyle +} \otimes W^{\scriptscriptstyle -} \otimes W^{\scriptscriptstyle +}$. In particular, in accordance with our choice of the representations, we have\footnote{Here and below in this appendix $v_0$ means $v_0 \otimes v_0 \otimes v_0 \otimes v_0$.}
\begin{equation*}
b^{}_{A1} v_0 = 0, \qquad b^{}_{A2} v_0 = 0, \qquad b^\dagger_{B1} v_0 = 0, \qquad b^{}_{B2} v_0 = 0.
\end{equation*}
This gives
\begin{equation}
e_1 v_0 = 0, \qquad e_2 v_0 = 0. \label{ev}
\end{equation}
These are the equations which we want to have choosing the representations for $q$-oscil\-lators.

Introduce operators $x_0$, $x_1$ and $x_2$ defined as
\begin{gather}
x_0 = e_1 e_2 - q^{-1} e_2 e_1, \qquad x_1 = e_2 e_0 - q^{-1} e_0 e_2, \label{x0x1} \\
x_2 = e_0 e_1 - q^{-1} e_1 e_0. \label{x2}
\end{gather}
Useful properties of these operators are
\begin{gather}
e_0 x_1 = q^{-1} x_1 e_0 , \qquad e_1 x_2 = q^{-1} x_2 e_1, \qquad e_2 x_ = q^{-1} x_0 e_2 , \label{ex1} \\
e_0 x_2  = q x_2 e_0, \qquad e_1 x_0 = q x_0 e_1, \qquad e_2 x_1 = q x_1 e_2. \label{ex2}
\end{gather}
In fact, they are a direct consequence of the Serre relations.

As it is clear from what follows, it is convenient to consider the basis
\begin{equation*}
w^k_n = x_1^{n_1} e_0^{n_2} x_2^{n_3} b_{B1}^k v_0.
\end{equation*}
Here $n_1$, $n_2$, $n_3$ and $k$ are nonnegative integers, and $n$ in the left hand side means the triple $n_1$, $n_2$, $n_3$.

\subsection{\texorpdfstring{Action of $q^{\nu h_0}$, $q^{\nu h_1}$ and $q^{\nu h_2}$}{Action of qnuh0, qnuh1 and qnuh2}}

It is easy to determine that
\begin{gather*}
q^{\nu h_0} w^k_n = q^{\nu (n_1 + 2 n_2 + n_3 + k + 1)} w^k_n, \qquad
q^{\nu h_1} w^k_n = q^{\nu (- 2 n_1 - n_2 + n_3 + k + 1)} w^k_n, \\*
q^{\nu h_2} w^k_n = q^{\nu (n_1 - n_2 - 2 n_3 - 2 k - 2)} w^k_n.
\end{gather*}

\subsection{\texorpdfstring{Action of $e_0$}{Action of e0}}

The action of the operator $e_0$ is very simple. Indeed, the first equation of (\ref{ex1}) immediately gives
\begin{equation*}
e_0 w_n^k = q^{- n_1} w_{n + \varepsilon_2}^k.
\end{equation*}

\subsection{\texorpdfstring{Action of $e_1$}{Action of e1}}

The idea which is used here is to move $e_1$ to $v_0$ and to use the first equation of (\ref{ev}). To this end we first define the operator $y_1$ as
\begin{equation}
y_1 = e_1 x_1 - x_1 e_1. \label{y1}
\end{equation}
One can verify that
\begin{equation*}
y_1 x_1 = q^{-2} x_1 y_1,
\end{equation*}
and we have
\begin{equation*}
e^{}_1 x_1^{n_1} = x_1^{n_1} e^{}_1 + q^{- n_1 + 1} [n_1]_q x_1^{n_1 - 1} y^{}_1.
\end{equation*}
To move $e_1$ through $e_0$ we use equation (\ref{x2}) and the first equation of (\ref{ex2}) to obtain
\begin{equation*}
e_1^{} e_0^{n_2} = q^{n_2} e_0^{n_2} e^{}_1 - q [n_2]_q e_0^{n_2 - 1} x_2^{}.
\end{equation*}
The second equation of (\ref{ex1}) and the equation
\begin{equation*}
e^{}_1 b_{B1}^k v_0 = 0
\end{equation*}
allow us to terminate the motion of $e_1$. Note that in the process of the motion an additional operator $y_1$ appeared. Now we move it using the equations
\begin{equation*}
y_1 e_0 = e_0 y_1
\end{equation*}
and
\begin{equation}
y_1 x_2 = q^2 x_2 y_1 \label{y1x2}
\end{equation}
which give
\begin{equation*}
y^{}_1 e_0^{n_2} = e_0^{n_2} y^{}_1
\end{equation*}
and
\begin{equation*}
y^{}_1 x_2^{n_3} = q^{2 n_3} x_2^{n_3} y^{}_1.
\end{equation*}
The equation
\begin{equation*}
y_1 b_{B1}^k v_0 = \zeta_2^s \kappa_q^{-1} b_{B1}^k v_0 + \zeta_2^{s_2} \kappa_q q [k]_q x_2 b_{B1}^{k - 1} v_0
\end{equation*}
allows us to terminate the motion of $y_1$. Here we use the relations
\begin{equation*}
b(b^\dagger)^k = q^k (b^\dagger)^k b + [k]_q (b^\dagger)^{k-1} q^{-N}, \qquad
b^\dagger b^k = q^{-k} b^k b^\dagger - q^{-1} [k]_q b^{k-1} q^{-N}.
\end{equation*}
The result of the movements is the equation
\begin{multline*}
e_1 w^k_n = - q [n_2]_q w^k_{n - \varepsilon_2 + \varepsilon_3} + \zeta_2^s \kappa_q^{-1} q^{- n_1 + 2 n_3 + 1} [n_1]_q w^k_{n - \varepsilon_1} \\
+ \zeta_2^{s_2} \kappa_q q^{- n_1 + 2 n_3 + 2} [n_1]_q [k]_q w^{k - 1}_{n -\varepsilon_1 + \varepsilon_3}.
\end{multline*}

\subsection{\texorpdfstring{Action of $e_2$}{Action of e2}}

It follows from the third equality of (\ref{ex2}) that
\begin{equation*}
e^{}_2 x_1^{n_1} = q^{n_1} x_1^{n_1} e^{}_2.
\end{equation*}
The second equation of (\ref{x0x1}) and the first equation of (\ref{ex1}) give
\begin{equation*}
e^{}_2 e_0^{n_2} = q^{-n_2} e_0^{n_2} e_2 + q^{- n_2 + 1} [n_2]_q x_1 e_0^{n_2 - 1}.
\end{equation*}
Let us denote by $y_2$ the operator acting as
\begin{equation}
y_2 = e_2 x_2 - x_2 e_2. \label{y2}
\end{equation}
One can make sure that
\begin{equation*}
y_2 = {} - y_0 - y_1,
\end{equation*}
where $y_1$ is defined by (\ref{y1}) and the operator $y_0$ is given by the equation
\begin{equation}
y_0 = e_0 x_0 - x_0 e_0. \label{y0}
\end{equation}
The equation
\begin{equation*}
y_0 x_2 = q^{-2} x_2 y_0
\end{equation*}
together with (\ref{y1x2}) gives
\begin{equation*}
e^{}_2 x_2^{n_3} = x_2^{n_3} e^{}_2 - q^{- n_3 + 1} [n_3]_q x_2^{n_3 - 1} y_0 - q^{n_3 - 1} [n_3]_q x_2^{n_3 - 1} y_1.
\end{equation*}
The motion of $e_2$ is terminated by the relation
\begin{equation*}
e_2 b_{B1}^k v_0 = - \zeta_2^{s_2} [k]_q b_{B1}^{k - 1} v_0.
\end{equation*}
After all, using the equations
\begin{align*}
y_0 b_{B1}^k v_0 &= - \zeta_3^s \kappa_q^{-1} q^{-2} b_{B1}^k v_0, \\
y_1 b_{B1}^k v_0 &= \zeta_2^s \kappa_q^{-1} b_{B1}^k v_0 + \zeta_2^{s_2} \kappa_q q [k]_q x_2 b_{B1}^{k - 1} v_0,
\end{align*}
we come to the final result
\begin{multline*}
e_2 w^k_n = (q^{- n_3} \zeta_3^s - q^{n_3} \zeta_2^s) \kappa_q^{-1} q^{n_1 - n_2 - 1} [n_3]_q w^k_{n - \varepsilon_3} \\
+ q^{n_1 - n_2 + 1} [n_2]_q w^k_{n + \varepsilon_1 - \varepsilon_2} - \zeta_2^{s_2} q^{n_1 - n_2 + 2 n_3}[k]_q w^{k - 1}_n.
\end{multline*}

\section{\texorpdfstring{Tensor product of representations $(\rho^{\scriptscriptstyle ++}_3)_{\zeta_3}$, $(\rho^{\scriptscriptstyle -+}_2)_{\zeta_2}$\\ and $(\rho^{\scriptscriptstyle --}_1)_{\zeta_1}$}{Tensor product of representations (rho++3)zeta3, (rho-+2)zeta2 and (rho--1)zeta1}}
\label{a:tprhorhorho}

\subsection{Basis}

Consider the $\uqbp$-module $(W_3^{\scriptscriptstyle ++})_{\zeta_3} \otimes (W_2^{\scriptscriptstyle -+})_{\zeta_2} \otimes (W_1^{\scriptscriptstyle --})_{\zeta_1}$ corresponding to the representation $(\rho^{\scriptscriptstyle ++}_3)_{\zeta_3} \otimes_\Delta (\rho^{\scriptscriptstyle -+}_2)_{\zeta_2} \otimes_\Delta (\rho_1^{\scriptscriptstyle --})_{\zeta_1}$. By definition, we have
\begin{align*}
q^{\nu h_0} v &= q^{\nu(2 N_{A_1} + N_{A_2} - N_{B_1} + N_{B_2} - N_{C_1} - 2 N_{C_2})} v, \\
q^{\nu h_1} v &= q^{\nu(- N_{A_1} + N_{A_2} - N_{B_1} - 2 N_{B_2} + 2 N_{C_1} + N_{C_2})} v, \\
q^{\nu h_2} v &= q^{\nu(- N_{A_1} - 2 N_{A_2} + 2 N_{B_1} + N_{B_2} - N_{C_1} + N_{C_2})} v, \\
e_0 v &= \zeta_3^{s_0} b^\dagger_{A1} q^{- N_{A2}} v - \zeta_2^{s_0} b^{}_{B1} b^\dagger_{B2} q^{- 2N_{A1} - N_{A2} - N_{B1} + N_{B2} + 1} v \\
& \hspace{10em} {} + \zeta_1^{s_0}\kappa_q^{-1} b_{C2} q^{- 2 N_{A1} - N_{A2} + N_{B1} - N_{B2} - N_{C2}} v, \\
e_1 v &= - \zeta_3^{s_1} b^{}_{A1} b^\dagger_{A2} q^{- N_{A1} + N_{A2} + 1} v + \zeta_2^{s_1} \kappa_q^{-1} b^{}_{B2} q^{N_{A1} - N_{A2} - N_{B2}} v \\
& \hspace{10em} {} + \zeta_1^{s_1} b_{C1}^\dagger q^{N_{A1} - N_{A2} + N_{B1} + 2 N_{B2} - N_{C2}} v, \\
e_2 v &= \zeta_3^{s_2} \kappa_q^{-1} b^{}_{A2} q^{- N_{A2}} v + \zeta_2^{s_2} b^\dagger_{B1} q^{N_{A1} + 2 N_{A2} - N_{B2}} v \\*
& \hspace{10em} {} - \zeta_1^{s_2} b_{C1} b_{C2}^\dagger q^{N_{A1} + 2 N_{A2} - 2 N_{B1} - N_{B2} - N_{C1} + N_{C2} + 1} v,
\end{align*}
where we mark the $q$-oscillators related to the first, second and third factors by $A$, $B$ and $C$. Below in this appendix $q^{\nu h_i}$ and $e_i$, $i = 0, 1, 2$, denote the operators given by the above equations.

Denote by $v_0$ the tensor product of the vacuum vectors for the chosen representations of $\Osc_q$. We have
\begin{equation*}
b^{\mathstrut}_{A1} v_0 = 0, \quad b^{\mathstrut}_{A2} v_0 = 0, \quad b^\dagger_{B1} v_0 = 0, \quad b^{\mathstrut}_{B2} v_0 = 0, \quad b^\dagger_{C1} v_0 = 0, \quad b^\dagger_{C2} v_0 = 0. 
\end{equation*}
As for the case of the tensor product of two representations, we obtain
\begin{equation*}
e_1 v_0 = 0, \qquad e_2 v_0 = 0.
\end{equation*}
Define the operators $x_0$, $x_1$ and $x_2$ whose action on the representation space is described by equations (\ref{x0x1}) and (\ref{x2}), and construct the basis formed by the vectors
\begin{equation*}
w_n^k = x_1^{n_1} e_0^{n_2} x_2^{n_3} b^{k_1}_{B1} b^{k_2}_{C1} b^{k_3}_{C2} v_0.
\end{equation*}
Here $n$ and $k$ mean triples $(n_1, n_2, n_3)$ and $(k_1, k_2, k_3)$ of non-negative integers. Note that in the case under consideration, equations (\ref{ex1}) and (\ref{ex2}) remain valid. Let us show that the action of the generators of $\uqlsliii$ is given by the formulas
\begin{align}
q^{\nu h_0} w^k_n & = q^{\nu(n_1 + 2 n_2 + n_3 + k_1 + k_2 + 2 k_3 + 4)} w^k_n, \label{tqh0} \\*
q^{\nu h_1} w^k_n & = q^{\nu(-2 n_1 - n_2 + n_3 + k_1 - 2 k_2 - k_3 - 2)} w^k_n, \label{tqh1} \\*
q^{\nu h_2} w^k_n & = q^{\nu(n_1 - n_2 - 2 n_3 - 2 k_1 + k_2 - k_3 - 2)} w^k_n, \label{tqh2} \\*
e_0 w^k_n & = q^{- n_1} w^k_{n + \varepsilon_2},  \label{te0} \\
e_1 w^k_n & = \kappa_q^{-1} q^{n_2 + n_3} (q^{- n_1 - n_2 + n_3 + 1} \zeta_2^s - q^{n_1 + n_2 - n_3 + 1} \zeta_1^s) [n_1]_q w^k_{n - \varepsilon_1} \notag \\*
& - q [n_2]_q w^k_{n - \varepsilon_2 + \varepsilon_3} - \zeta_1^{s_1} q^{2 n_1 + n_2 - n_3 - k_1 + k_3} [k_2]_q w^{k - \varepsilon_2}_n \notag \\*
& - \zeta_1^{s_2} \kappa_q q^{- n_1 + 2 n_3 + 2 k_1 + k_2 - k_3 + 5} [n_1]_q [k_3]_q w^{k + \varepsilon_2 - \varepsilon_3}_{n - \varepsilon_1 + \varepsilon_3} \notag \\*
& + \zeta_2^{s_2} \kappa_q q^{- n_1 + 2 n_3 + 2} [n_1]_q [k_1]_q w^{k - \varepsilon_1}_{n - \varepsilon_1 + \varepsilon_3} \notag \\*
& - \zeta_1^{s_1 + s_2} \kappa_q q^{n_1 + 2 n_2 + n_3 + k_1 + 2} [n_1]_q [k_3]_q w^{k - \varepsilon_3}_{n - \varepsilon_1 + \varepsilon_2}, \label{te1} \\
e_2 w^k_n & = \kappa_q^{-1} q^{n_1 - n_2} (q^{- n_3 - 1} \zeta_3^s - q^{n_3 - 1} \zeta_2^s) [n_3]_q w^k_{n - \varepsilon_3} \notag \\
& + q^{n_1 - n_2 + 1} [n_2]_q w^k_{n + \varepsilon_1 - \varepsilon_2} - \zeta_2^{s_2} q^{n_1 - n_2 + 2 n_3} [k_1]_q w^{k - \varepsilon_1}_n \notag \\
& + \zeta_1^{s_2} q^{n_1 - n_2 + 2 n_3 + 2 k_1 + k_2 - k_3 + 3} [k_3]_q w^{k + \varepsilon_2 - \varepsilon_3}_n. \label{te2}
\end{align}

\subsection{\texorpdfstring{Action of $q^{\nu h_0}$, $q^{\nu h_1}$ and $q^{\nu h_2}$}{Action of qnuh0, qnuh1 and qnuh2}}

Equations (\ref{tqh0})--(\ref{tqh2}) are simple consequences of the defining relations of the algebra $\Osc_q$.

\subsection{\texorpdfstring{Action of $e_0$}{Action of e0}}

The first equation of (\ref{ex1}) gives (\ref{te0}).

\subsection{\texorpdfstring{Action of $e_1$}{Action of e1}}

First, define the operators $y_0$, $y_1$ and $y_2$ by equations (\ref{y1}), (\ref{y2}) and (\ref{y0}). It appears convenient to represent them as
\begin{equation*}
y_0 = z_1 - z_3, \qquad y_1 = z_2 - z_1, \qquad y_2 = z_3 - z_2.
\end{equation*}
In fact, we define
\begin{align*}
z_3 = - \zeta_1^{s_0} \zeta_2^{s_1} \zeta_3^{s_2} & \kappa_q^{-1} q \, b_{A2} b_{B2} b_{C2} q^{- N_{A1} - 3 N_{A2} + N_{B1} - 2 N_{B2} - N_{C2}}  \\*
& {} - \zeta_2^{s_0 + s_1} \zeta_3^{s_2} \kappa_q^{-1} b_{A2} b_{B1} q^{- N_{A1} - 3 N_{A2} - N_{B1} - N_{B2}} \\*
& {} - \zeta_1^{s_0} \zeta_3^{s_1 + s_2} \kappa_q^{-1} b_{A1} b_{C2} q^{- 3 N_{A1} - 2 N_{A2} + N_{B1} - N_{B2} - N_{C2}} \\*
& {} + \zeta_2^{s_0} \zeta_3^{s_1 + s_2} q \, b_{A1} b_{B1} b_{B2}^\dagger q^{- 3 N_{A1} - 2 N_{A2} - N_{B1} + N_{B2}} + \zeta_3^s \kappa_q^{-1} q^{-2} q^{- 2 N_{A1} - 2 N_{A2}},
\end{align*}
and then assume that
\begin{equation*}
z_1 = y_0 + z_3, \qquad z_2 = - y_2 + z_3.
\end{equation*}
The operators $z_i$, $i=1,2,3$, can actually be read off, for instance, from an appropriate product of three different $L$-operators, just in the spirit of techniques of \cite{BooJimMiwSmiTak09}.

One can demonstrate that
\begin{equation*}
z_1 x_1 = q^2 x_1 z_1, \qquad z_2 x_1 = q^{-2} x_1 z_2,
\end{equation*}
and we have
\begin{gather*}
e_1^{\mathstrut} x_1^{n_1} = x_1^{n_1} e_1^{\mathstrut} + q^{- n_1 + 1} [n_1]^{\mathstrut}_q x_1^{n_1 - 1} z_1^{\mathstrut} - q^{n_1 - 1} [n_1]^{\mathstrut}_q x_1^{n_1 - 1} z_2^{\mathstrut}.
\end{gather*}
It follows from definition (\ref{x2}) and the first equation of (\ref{ex2}) that
\begin{equation*}
e_1^{\mathstrut} e_0^{n_2} = q^{n_2} e_0^{n_2} e_1^{\mathstrut} - q [n]_q^{\mathstrut} e_0^{n_2 - 1} x_2^{\mathstrut},
\end{equation*}
while the second equation of (\ref{ex1}) gives
\begin{equation*}
e_1^{\mathstrut} x_2^{n_3} = q^{- n_3} x_2^{n_3} e_1^{\mathstrut}.
\end{equation*}
One can determine that
\begin{equation*}
z_1 e_0 = q^2 e_0 z_1, \qquad z_2 e_0 = e_0 z_2,
\end{equation*}
and, therefore,
\begin{equation*}
z^{\mathstrut}_1 e_0^{n_2} = q^{2 n_2} e_0^{n_2} z_1^{\mathstrut}, \qquad z^{\mathstrut}_2 e_0^{n_2} = e_0^{n_2} z_2^{\mathstrut}.
\end{equation*}
Define now the operator $r_3$ as
\begin{equation*}
r_3 = z_1 x_2 - q^2 x_2 z_1.
\end{equation*}
It follows from the equation
\begin{equation*}
r_3 x_2 = x_2 r_3
\end{equation*}
that
\begin{equation*}
z_1^{\mathstrut} x_2^{n_3} = q^{2n_3}_{\mathstrut} x_2^{n_3} z_1^{\mathstrut} + q^{n_3 - 1} [n_3]_q^{\mathstrut} x_2^{n_3 - 1} r_3.
\end{equation*}
Now we denote
\begin{equation*}
\varphi_n = x_1^{n_1} e_0^{n_2} x_2^{n_3}
\end{equation*}
and, using the equation
\begin{equation*}
z_2 x_2 = q^2 x_2 z_2,
\end{equation*}
obtain
\begin{multline}
e_1 \varphi_n = q^{n_2 - n_3} \varphi_n e_1 - q [n_2]_q \varphi_{n - \varepsilon_2 + \varepsilon_3} - q^{n_1 + 2 n_2 + 2 n_3 - 1} [n_1]_q \varphi_{n - \varepsilon_1} z_1 \\+ q^{- n_1 + 2 n_3 + 1} [n_1]_q \varphi_{n - \varepsilon_1} z_2 - q^{n_1 + 2 n_2 + n_3 - 2} [n_1]_q [n_3]_q \varphi_{n - \varepsilon_1 - \varepsilon_3} r_3. \label{e1phin_3}
\end{multline}
Define also the operator
\begin{equation*}
p_3 = x_1 x_2 - q^3 x_2 x_1.
\end{equation*}
Using the equation
\begin{equation*}
p_3 x_2 = q x_2 p_3,
\end{equation*}
we find
\begin{equation*}
x_1^{\mathstrut} x_2^{n_3} = q^{3 n_3} x_2^{n_3} x_1^{\mathstrut} + q^{2(n_3 - 1)} [n_3]_q^{\mathstrut} x_2^{n_3 - 1} p_3^{\mathstrut}.
\end{equation*}
Now, having in mind that
\begin{equation*}
w_n^k = \varphi_n^{\mathstrut} w_0^k
\end{equation*}
and using equation (\ref{e1phin_3}) together with the equations
\begin{align*}
& e_1 w_0^k = - \zeta_1^{s_1} q^{- k_1 + k_3} [k_2]_q w_0^{k - \varepsilon_2}, \\
& z_1 w^k_0 = \zeta_1^s \kappa_q^{-1} q^2 w^k_0 + \zeta_1^{s_1} \kappa_q q^{- k_1 + k_3 + 1} [k_2]_q x_1 w^{k - \varepsilon_2}_0 + \zeta_1^{s_1 + s_2} \kappa_q q^{k_1 + 3} [k_3]_q e_0 
w_0^{k - \varepsilon_3}, \\
& z_2 w^k_0 = \zeta_2^s \kappa_q^{-1} w^k_0 + \zeta_2^{s_2} \kappa_q q [k_1] x_2 w_0^{k - \varepsilon_1} - \zeta_1^{s_2} \kappa_q q^{2 k_1 + k_2 - k_3 + 4} [k_3]_q x_2 w_0^{k + \varepsilon_2 - \varepsilon_3},
\end{align*}
one can obtain some expression for $e_1 w_n^k$. Comparing it with (\ref{te1}) we see that equation (\ref{te1}) is true if the equation
\begin{equation*}
r_3 w_0^k = \zeta_1^{s_1} \kappa_q q^{- k_1 + k_3} [k_2]_q p_3 w_0^{k - \varepsilon_2} - \zeta_1^{s} q^3 x_2 w_0^{k}
\end{equation*}
is true. The validity of this equation can be verified by direct calculation using the explicit forms of the operators $r_3$ and $p_3$.

\subsection{\texorpdfstring{Action of $e_2$}{Action of e2}}

The proof is actually the same as for the case of $e_1$. It is based on the equation
\begin{multline*}
e_2 \varphi_n = q^{n_1 - n_2} \varphi_n e_2 + q^{n_1 - n_2 + 1} [n_2]_q \varphi_{n + \varepsilon_1 - \varepsilon_2} \\- q^{n_1 - n_2 + n_3 - 1} [n_3]_q \varphi_{n - \varepsilon_3} z_2 + q^{n_1 - n_2 - n_3 + 1} [n_3]_q \varphi_{n - \varepsilon_3} z_3
\end{multline*}
and the relations
\begin{align*}
& e_2 w_0^k = - \zeta_2^{s_2} [k_1]_q w_0^{k - \varepsilon_1} + \zeta_1^{s_2} q^{2 k_1 + k_2 - k_3 + 3} [k_3]_q w_0^{k + \varepsilon_2 - \varepsilon_3}, \\
& z_3 w_0^k = \zeta_3^s \kappa_q^{-1} q^{-2} w_0^k.
\end{align*}

\subsection{Another basis}

For any $\xi \in \widetilde \gothh^*$, we use as the basis of the $\uqbp$-module $(\widetilde V^\lambda)_\zeta[\xi]$ the basis $\{v_n\}$ of the underlying $\uqgliii$-module $\widetilde V^\lambda$ defined by equation (\ref{vn}). It is easy to show that the action of the generators $F_1$, $F_2$ and $F_3$ used to construct this basis can be expressed via the action of the generators $e_0$, $e_1$ and $e_2$ in the following way:
\begin{align}
& F_1 v = \zeta^{- s_0 - s_2} q^{2(\lambda_1 + \lambda_2 + \lambda_3) / 3} (e_2 e_0 - q^{-1} e_0 e_2) \, q^{(h_2 - \xi(h_2) - h_0 + \xi(h_0)) / 3} v, \label{f1vev} \\
& F_2 v = \zeta^{- s_0 - s_1} q^{2(\lambda_1 + \lambda_2 + \lambda_3) / 3} (e_0 e_1 - q^{-1} \, e_1 e_0) \, q^{(h_0 - \xi(h_0) - h_1 + \xi(h_1)) / 3} v, \label{f2vev} \\
& F_3 v = \zeta^{- s_0} q^{2(\lambda_1 + \lambda_2 + \lambda_3) / 3} e_0 \, q^{(h_1 - \xi(h_1) - h_2 + \xi(h_2)) / 3} v. \label{f3vev}
\end{align}
We expect the appearance of the submodule isomorphic to $(\widetilde V^\lambda)_\zeta[\xi]$ for some $\lambda \in \widetilde \gothh^*$, $\zeta \in \bbC^\times$ and $\xi \in \widetilde \gothh^*$. It is also natural to expect that the vacuum vector of the tensor product of the representations under consideration coincides with the vector $v_0$ of $(\widetilde V^\lambda)_\zeta[\xi]$. It follows from equations (\ref{tqh0})--(\ref{tqh2}) that we should assume that
\begin{equation}
\xi(h_0) = \mu_1 + \mu_2 + 4, \qquad \xi(h_1) = - \mu_1 - 2, \qquad \xi(h_2) = - \mu_2 - 2. \label{xif}
\end{equation}
Being inspired by equations (\ref{f1vev})--(\ref{f3vev}) and taking into account (\ref{xif}), we introduce the operators
\begin{align*}
& \widetilde x_1 v = \zeta^{- s_0 - s_2} q^{\lambda_1 + \lambda_2 + 2} (e_2 e_0 - q^{-1} e_0 e_2) \, q^{(h_2 - h_0) / 3} v, \\
& \widetilde x_2 v = \zeta^{- s_0 - s_1} q^{\lambda_2 + \lambda_3 - 2} (e_0 e_1 - q^{-1} \, e_1 e_0) \, q^{(h_0 - h_1) / 3} v, \\
& \widetilde e_0 v = \zeta^{- s_0} q^{\lambda_1 + \lambda_3} e_0 \, q^{(h_1 - h_2) / 3} v,
\end{align*}
which are analogues of the generators $F_1$, $F_2$ and $F_3$ acting on the module $(\widetilde V^\lambda)_\zeta[\xi]$. Now we define the basis $\{w_n^k\}$ formed by the vectors
\begin{equation*}
w_n^k = \widetilde x_1^{\, n_1} \widetilde e_0^{\, n_2} \widetilde x_2^{\, n_3} b_{B1}^{k_1} b_{C1}^{k_2} b_{C2}^{k_3} v_0.
\end{equation*}
It is clear that the action of $q^{\nu h_0}$, $q^{\nu h_1}$ and $q^{\nu h_2}$ is again given by (\ref{tqh0}), (\ref{tqh1}) and (\ref{tqh2}). The formulas for the action of $e_0$, $e_1$ and $e_2$ can be easily obtained from (\ref{te0}), (\ref{te1}) and (\ref{te2}) and have the form
\begin{align*}
e_0 w^k_n & = \zeta^{s_0} q^{- \lambda_1 - \lambda_3 - n_3 - k_1 + k_2} w^k_{n + \varepsilon_2}, \\
e_1 w^k_n & = \zeta^{s_1 - s} \kappa_q^{-1} q^{\lambda_1 + \lambda_2 - k_1 - k_3}(q^{- n_1 - n_2 + n_3 + 1} \zeta_2^s - q^{n_1 + n_2 - n_3 + 1} \zeta_1^s) [n_1]_q w^k_{n - \varepsilon_1} \\*
& - \zeta^{s_1} q^{\mu_1 - n_2 + n_3 + k_1 - 2 k_2 - k_3 + 2} [n_2]_q w^k_{n - \varepsilon_2 + \varepsilon_3} - \zeta_1^{s_1} q^{2 n_1 - k_1 + k_3} [k_2]_q w^{k - \varepsilon_2}_n \\*
& - \zeta^{s_1 - s_2} \zeta_1^{s_2} \kappa_q q^{\mu_1 + \mu_2 - n_1 - n_2 + n_3 + k_1 - 3 k_3 + 5} [n_1]_q [k_3]_q w^{k + \varepsilon_2 - \varepsilon_3}_{n - \varepsilon_1 + \varepsilon_3} \\*
& + \zeta^{s_1 - s_2} \zeta_2^{s_2} \kappa_q q^{\mu_1 + \mu_2 - n_1 - n_2 + n_3 - k_1 - k_2 -2 k_3 + 2} [n_1]_q [k_1]_q w^{k - \varepsilon_1}_{n - \varepsilon_1 + \varepsilon_3} \\*
& - \zeta^{-s_2} \zeta_1^{s_1 + s_2} \kappa_q q^{\mu_2 + n_1 + n_2 - k_1 + k_2 - k_3 + 2} [n_1]_q [k_3]_q w^{k - \varepsilon_3}_{n - \varepsilon_1 + \varepsilon_2}, \\
e_2 w^k_n & = \zeta^{s_2 - s} \kappa_q^{-1} q^{\lambda_2 + \lambda_3 + k_2 + k_3} (q^{- n_3 - 1} \zeta_3^s - q^{n_3 - 1} \zeta_2^s) [n_3]_q w^k_{n  - \varepsilon_3} \\
& + \zeta^{s_2} q^{- \mu_2 + 2 n_3 + 2 k_1 - k_2 + k_3} [n_2]_q w^k_{n + \varepsilon_1 - \varepsilon_2} - \zeta_2^{s_2} q^{2 n_3} [k_1]_q w^{k - \varepsilon_1}_n \\
& + \zeta_1^{s_2} q^{2 n_3 + 2 k_1 + k_2 - k_3 + 3} [k_3]_q w^{k + \varepsilon_2 - \varepsilon_3}_n.
\end{align*}
After the substitution
\begin{equation*}
w_n^k \to q^{- (k_1 + k_3) n_1 + (k_1 - k_2) n_2 + (k_2 + k_3) n_3} w_n^k
\end{equation*}
we come to relations (\ref{th0v})--(\ref{te2v}).

\section{Dual representations and monodromy operators} \label{a:dr}

Let $A$ be a quantum group, and $\calR$ its universal $R$-matrix. A monodromy operator corresponding to the choice of the representation $\varphi$ for the auxiliary space and the representation $\psi$ for the quantum space is defined as
\begin{equation*}
M_{\varphi, \psi}(\zeta | \eta) = (\varphi_\zeta \otimes \psi_\eta)(\calR),
\end{equation*}
where $\zeta$ and $\eta$ are spectral parameters, and the mappings $\varphi_\zeta$ and $\psi_\eta$ are defined with the help of equation (\ref{vpz}). In fact, it is often convenient to consider the case where $\varphi$ is a homomorphism from the quantum group to some algebra $B$, see, for example, section~\ref{ss:gr}. Certainly, a representation is a partial case of such homomorphism.

Dual of a representation of $A$ can be defined with the help of any antiautomorphism of $A$. Here we use for this end the inverse of the antipode $S$. If $\psi$ is a representation of $A$ on the vector space $U$, the dual representation $\psi^{* S^{-1}}$ is defined by the equation
\begin{equation*}
\langle \psi^{* S^{-1}}(a) v^*, v \rangle = \langle v^*, \psi(S^{-1}(a)) v \rangle
\end{equation*}
for any $v \in U$ and $v^* \in U^*$. If the mapping $*: \End(U) \to \End(U^*)$ is defined as\footnote{Usually, one denotes $*(M)$ as $M^*$. It is more convenient for our consideration to use the former notation.}
\begin{equation*}
\langle *(M) v^*, v \rangle = \langle v^*, M v \rangle,
\end{equation*}
we can write
\begin{equation*}
\psi^{* S^{-1}} = * \circ \psi \circ S^{-1}.
\end{equation*}
Now, using the relation \cite[p. 124]{ChaPre94}
\begin{equation*}
(\id \otimes S^{-1})(\calR) = \calR^{-1},
\end{equation*}
we obtain the equation
\begin{equation}
M_{\varphi, \psi^{*S^{-1}}}(\zeta) = (\id \otimes *)((M_{\varphi, \psi}(\zeta))^{-1}). \label{meim}
\end{equation}

Let $\{e_i\}$ be a basis of $U$ and $\{E_{ij}\}$ the corresponding basis of $\End(U)$ defined by the relation
\begin{equation*}
E_{ij} \, e_k = e_i \, \delta_{jk}.
\end{equation*}
For any endomorphism $M \in \End(U)$ we have
\begin{equation*}
M e_j = \sum_i e_i M_{ij}.
\end{equation*}
One can verify that
\begin{equation*}
M = \sum_{i, j}  M_{i j} E_{i j}.
\end{equation*}
Let $\{e^*_i\}$ be the dual basis satisfying the relation
\begin{equation*}
\langle e_i^*, e_j \rangle = \delta_{ij},
\end{equation*}
and $\{E^*_{ij}\}$ the corresponding basis of $\End(U^*)$, so that
\begin{equation*}
E^*_{ij} \, e^*_k = e^*_i \, \delta_{jk}.
\end{equation*}
It is not difficult to determine that
\begin{equation}
*(M)_{ij} = M_{ji}, \label{smij}
\end{equation} 
where the numbers $*(M)_{ij}$ are defined by the equation
\begin{equation*}
*(M) = \sum_{i, j} *(M)_{i j} E^*_{i j}. 
\end{equation*}
It follows from equations (\ref{meim}) and (\ref{smij}) that for the matrices
\begin{equation*}
\bbM_{\varphi, \psi}(\zeta) = ((\bbM_{\varphi, \psi}(\zeta))_{ij}), \qquad \bbM_{\varphi, \psi^{* S^{-1}}}(\zeta) = ((\bbM_{\varphi, \psi^{* S^{-1}}}(\zeta))_{i j}),
\end{equation*}
where the quantities $(\bbM_{\varphi, \psi}(\zeta))_{ij}$ and $(\bbM_{\varphi, \psi^{S^{-1}}}(\zeta))_{i j}$ are defined by the relations
\begin{equation*}
M_{\varphi, \psi}(\zeta) = \sum_{i, j} (\bbM_{\varphi, \psi}(\zeta))_{i j} \, E_{ij}, \qquad M_{\varphi, \psi^{* S^{-1}}}(\zeta) = \sum_{i, j} (\bbM_{\varphi, \psi^{* S^{-1}}}(\zeta))_{i j} \, E^*_{ij},
\end{equation*}
one has
\begin{equation}
\bbM_{\varphi, \psi^{* S^{-1}}}(\zeta) = ((\bbM_{\varphi, \psi}(\zeta))^{-1})^t. \label{mvdp}
\end{equation}

\section{Finite dimensional representations}

In this appendix we give explicit formulas for the finite dimensional representations of the quantum groups $\uqgliii$ and $\uqlsliii$ used in the main text.

The representation of $\uqgliii$ with the highest weight $(1, \, 0, \, 0)$ is three dimensional and one has
\begin{gather}
\pi^{(1, \, 0, \, 0)}(q^{\nu G_1}) = q^\nu E_{11} + E_{22} + E_{33}, \qquad \pi^{(1, \, 0, \, 0)}(q^{\nu G_2}) = E_{11} + q^\nu E_{22} + E_{33}, \label{pi100a}  \\
\pi^{(1, \, 0, \, 0)}(q^{\nu G_3}) = E_{11} + E_{22} + q^\nu E_{33}, \\
\pi^{(1, \, 0, \, 0)}(E_1) = E_{12}, \qquad \pi^{(1, \, 0, \, 0)}(E_2) = E_{23}, \\
\pi^{(1, \, 0, \, 0)}(F_1) = E_{21}, \qquad \pi^{(1, \, 0, \, 0)}(F_2) = E_{32}. \label{pi100d}
\end{gather}
It is useful to have in mind that
\begin{equation*}
\pi^{(1, \, 0, \, 0)}(E_3) = E_{13}, \qquad \pi^{(1, \, 0, \, 0)}(F_3) = E_{31}.
\end{equation*}
The representation of $\uqgliii$ with the highest weight $(1, \, 1, \, 0)$ is also three dimensional and with the corresponding choice of the basis one can determine that
\begin{gather}
\pi^{(1, \, 1, \, 0)}(q^{\nu G_1}) = q^\nu E_{11} + q^\nu E_{22} + E_{33}, \qquad \pi^{(1, \, 1, \, 0)}(q^{\nu G_2}) = q^\nu E_{11} + E_{22} + q^\nu E_{33}, \label{pi110a} \\ \pi^{(1, \, 1, \, 0)}(q^{\nu G_3}) = E_{11} + q^\nu E_{22} + q^\nu E_{33}, \\
\pi^{(1, \, 1, \, 0)}(E_1) = E_{23}, \qquad \pi^{(1, \, 1, \, 0)}(E_2) = E_{12}, \\
\pi^{(1, \, 1, \, 0)}(F_1) = E_{32}, \qquad \pi^{(1, \, 1, \, 0)}(F_2) = E_{21}. \label{pi110d}
\end{gather}
Here one has
\begin{equation*}
\pi^{(1, \, 1, \, 0)}(E_3) = - q^{-1} E_{13}, \qquad \pi^{(1, \, 1, \, 0)}(F_3) = - q E_{31}.
\end{equation*}

Using equations (\ref{jha})--(\ref{jhc}) describing the Jimbo's homomorphism and equations (\ref{pi100a})--(\ref{pi100d}), for the representation $\varphi^{(1, \, 0, \, 0)}$ of $\uqlsliii$ we obtain
\begin{align}
& \varphi^{(1, \, 0, \, 0)}(q^{\nu h_0}) = q^{- \nu} E_{11} + E_{22} + q^\nu E_{33}, \label{vp100a} \\
& \varphi^{(1, \, 0, \, 0)}(q^{\nu h_1}) = q^\nu E_{11} + q^{- \nu} E_{22} + E_{33}, \\
& \varphi^{(1, \, 0, \, 0)}(q^{\nu h_2}) = E_{11} + q^\nu E_{22} + q^{- \nu} E_{33},
\end{align}
\vspace{-1.3em}
\begin{align}
& \varphi^{(1, \, 0, \, 0)}(e_0) = q^{-1} E_{31}, && \varphi^{(1, \, 0, \, 0)}(e_1) = E_{12}, && \varphi^{(1, \, 0, \, 0)}(e_2) = E_{23}, \\
& \varphi^{(1, \, 0, \, 0)}(f_0) = q E_{13}, && \varphi^{(1, \, 0, \, 0)}(f_1) = E_{21}, && \varphi^{(1, \, 0, \, 0)}(f_2) = E_{32}. \label{vp100d}
\end{align}
It is easy to see that for the representation $\ovarphi'^{(1, \, 0, \, 0)}$ of $\uqlsliii$ we have
\begin{align}
& \ovarphi'^{(1, \, 0, \, 0)}(q^{\nu h_0}) = q^{- \nu} E_{11} + E_{22} + q^\nu E_{33}, \label{vpp100a} \\
& \ovarphi'^{(1, \, 0, \, 0)}(q^{\nu h_1}) = E_{11} + q^{\nu} E_{22} + q^{- \nu} E_{33}, \\
& \ovarphi'^{(1, \, 0, \, 0)}(q^{\nu h_2}) = q^\nu E_{11} + q^{- \nu} E_{22} + E_{33},
\end{align}
\vspace{-1.3em}
\begin{align}
& \ovarphi'^{(1, \, 0, \, 0)}(e_0) = q^{-1} E_{31}, && \ovarphi'^{(1, \, 0, \, 0)}(e_1) = E_{23}, && \ovarphi'^{(1, \, 0, \, 0)}(e_2) = E_{12}, \\
& \ovarphi'^{(1, \, 0, \, 0)}(f_0) = q E_{13}, && \ovarphi'^{(1, \, 0, \, 0)}(f_1) = E_{32}, && \ovarphi'^{(1, \, 0, \, 0)}(f_2) = E_{21}. \label{vpp100d}
\end{align}

Now consider the dual representation $\varphi^{(1, \, 0, \, 0)* S^{-1}}$. Using the basis of $(\bbC^3)^*$ dual for the standard basis of $\bbC^3$ we come to the equations
\begin{align*}
& \varphi^{(1, \, 0, \, 0)* S^{-1}} (q^{\nu h_0}) = q^\nu E_{11} + E_{22} + q^{-\nu} E_{33}, \\
& \varphi^{(1, \, 0, \, 0)* S^{-1}} (q^{\nu h_1}) = q^{- \nu} E_{11} + q^{\nu} E_{22} + E_{33}, \\
& \varphi^{(1, \, 0, \, 0)* S^{-1}} (q^{\nu h_2}) = E_{11} + q^{- \nu} E_{22} + q^\nu E_{33},
\end{align*}
\vspace{-1.3em}
\begin{align*}
& \varphi^{(1, \, 0, \, 0)* S^{-1}} (e_0) = - q^{-2} E_{13}, && \varphi^{(1, \, 0, \, 0)* S^{-1}} (e_1) = - q^{-1} E_{21}, && \varphi^{(1, \, 0, \, 0)* S^{-1}} (e_2) = - q^{-1} E_{32}, \\ & \varphi^{(1, \, 0, \, 0)* S^{-1}} (f_0) = - q^2 E_{31}, && \varphi^{(1, \, 0, \, 0)* S^{-1}} (f_1) = - q E_{12}, && \varphi^{(1, \, 0, \, 0)* S^{-1}} (f_2) = - q E_{23}
\end{align*}
Comparing these equations with (\ref{vpp100a})-(\ref{vpp100d}), we conclude that
\begin{equation}
(\ovarphi'^{(1, \, 0, \, 0)})_\zeta (a) = P (\varphi^{(1, \, 0, \, 0)* S^{-1}})_{r_s^3 q^{3/s} \zeta}(a) P^{-1}, \label{oeqd}
\end{equation}
where the endomorphism $P$ has the skew-diagonal matrix form:
\begin{equation}
\bbP = \left( \begin{array}{crc}
0 & 0 & r_s^{-3 s_2} q^{1 - 3 s_2/s} \\
0 & -1 & 0 \\
r_s^{3 s_1} q^{- 1 + 3 s_1/s} & 0 & 0
\end{array} \right). \label{bbp}
\end{equation}
Thus, the representation $(\ovarphi'^{(1, \, 0, \, 0)})_\zeta$ is equivalent to the representation $(\varphi^{(1, \, 0, \, 0)* S^{-1}})_{r_s^3 q^{3/s} \zeta}$.

\bibliographystyle{amsrusunsrt}

\bibliography{IntegrableSystems}

\end{document}